\documentclass[prd,aps,twocolumn,a4paper,showkeys,nofootinbib, superscriptaddress]{revtex4-2}

\usepackage{acronym}
\usepackage[T1]{fontenc}
\usepackage[utf8]{inputenc}
\usepackage{xcolor}
\usepackage{amssymb}
\usepackage{amsmath}
\usepackage{physics}
\usepackage{comment}
\usepackage{bigints}
\usepackage{amsmath}
\usepackage{graphicx}% Include figure files
\usepackage{soul}
\usepackage{ulem}
\usepackage[breaklinks,colorlinks,citecolor=blue]{hyperref}
\usepackage[all]{hypcap}

\begin{document}
\title{Detectability of QCD phase transitions in binary neutron star mergers:\\ 
Bayesian inference with the next generation gravitational wave detectors}

\author{Aviral Prakash}
\affiliation{Institute for Gravitation \& the Cosmos, The Pennsylvania State University, University Park PA 16802, USA}
\affiliation{Department of Physics, The Pennsylvania State University, University Park PA 16802, USA}

\author{Ish Gupta}
\affiliation{Institute for Gravitation \& the Cosmos, The Pennsylvania State University, University Park PA 16802, USA}
\affiliation{Department of Physics, The Pennsylvania State University, University Park PA 16802, USA}

\author{Matteo Breschi}
\affiliation{Theoretisch-Physikalisches Institut, Friedrich-Schiller-Universit{\"a}t Jena, 07743 Jena, Germany}
\affiliation{Scuola Internazionale Superiore di Studi Avanzati (SISSA), 34136 Trieste, Italy}
\affiliation{Istituto Nazionale di Fisica Nucleare (INFN), Sezione di Trieste, 34127 Trieste, Italy}

\author{Rahul Kashyap}
\affiliation{Institute for Gravitation \& the Cosmos, The Pennsylvania State University, University Park PA 16802, USA}
\affiliation{Department of Physics, The Pennsylvania State University, University Park PA 16802, USA}

\author{David Radice}
\thanks{Alfred P.~Sloan Fellow}
\affiliation{Institute for Gravitation \& the Cosmos, The Pennsylvania State University, University Park PA 16802, USA}
\affiliation{Department of Physics, The Pennsylvania State University, University Park PA 16802, USA}
\affiliation{Department of Astronomy \& Astrophysics,  The Pennsylvania State University, University Park PA 16802, USA}

\author{Sebastiano Bernuzzi}
\affiliation{Theoretisch-Physikalisches Institut, Friedrich-Schiller-Universit{\"a}t Jena, 07743 Jena, Germany}

\author{Domenico Logoteta}
\affiliation{Dipartimento di Fisica, Universit\`{a} di Pisa, Largo B.  Pontecorvo, 3 I-56127 Pisa, Italy}
\affiliation{INFN, Sezione di Pisa, Largo B. Pontecorvo, 3 I-56127 Pisa, Italy}

\author{B.S. Sathyaprakash}
\affiliation{Institute for Gravitation \& the Cosmos, The Pennsylvania State University, University Park PA 16802, USA}
\affiliation{Department of Physics, The Pennsylvania State University, University Park PA 16802, USA}
\affiliation{Department of Astronomy \& Astrophysics,  The Pennsylvania State University, University Park PA 16802, USA}

\date{\today}

\begin{abstract}
	We study the detectability of postmerger QCD phase transitions in neutron star binaries with next-generation gravitational-wave detectors Cosmic Explorer and Einstein Telescope. We perform numerical relativity simulations of neutron star mergers with equations of state that include a quark deconfinement phase transition through either a Gibbs or Maxwell construction. These are followed by Bayesian parameter estimation of the associated gravitational-wave signals using the \texttt{NRPMw} waveform model, with priors inferred from the analysis of the inspiral signal. We assess the ability of the model to measure the postmerger peak frequency $f_2^{\rm peak}$ and identify aspects that should be improved in the model. We show that, even at postmerger signal to noise ratios as low as 10, the model can distinguish (at the 90\% level) $f_2^{\rm peak}$ between binaries with and without a phase transition in most cases. Phase-transition induced deviations in the $f_2^{\rm peak}$ from the predictions of equation-of-state insensitive relations can also be detected if they exceed $1.6\,\sigma$. Our results suggest that next-generation gravitational wave detectors can measure phase transition effects in binary neutron star mergers. However, unless the phase transition is ``strong'', disentangling it from other hadronic physics uncertainties will require significant theory improvements.
\end{abstract}

\maketitle

%%%%%%%%%%%%%%%%%%%%%%%%%%%%%%%%%%%
%%%%%%% INTRODUCTION %%%%%%%%%%%%%%
%%%%%%%%%%%%%%%%%%%%%%%%%%%%%%%%%%%

\section{\label{sec:level1}Introduction}

%%%%%%%%%GW170817%%%%%%%%%%%%

The discoveries of the gravitational wave (GW) event GW170817 \cite{LIGOScientific:2017vwq} from a merger of two neutron stars, the associated short gamma ray burst GRB170817A and the optical transient AT2017gfo \cite{LIGOScientific:2017ync}, revitalized the field of multimessenger astronomy. It is now possible to probe high-energy astrophysical phenomena through their GW signatures in addition to electromagnetic radiation. The emitted GW spectra from a merger of two neutron stars spans a broad range of frequencies. GWs from an inspiral (at frequencies $\lesssim 10^3\ \mathrm{Hz}$) signal provide a wealth of information about the intrinsic properties of a binary such as its component masses and tidal deformabilities. On the other hand, postmerger GW emission (at frequencies $\gtrsim 10^3\ \mathrm{Hz}$) can inform us about the dynamically evolving merger remnant. No postmerger signal from GW170817 was detected thereby leaving to speculation the fate of the remnant. We encourage the reader to refer to refs. \cite{Radice:2020ddv, Bernuzzi:2020tgt} for recent reviews.

%%%%%%%%% Physical Processes %%%%%%%%%%%%
With the upcoming generation of GW detectors like the Einstein Telescope (ET) \cite{Punturo:2010zz, Hild:2010id} or the Cosmic Explorer (CE) \cite{LIGOScientific:2016wof, Reitze:2019iox, Evans:2021gyd, Evans:2023euw}, it is expected that the postmerger phase of evolution will be within reach of detector sensitivities \cite{Branchesi:2023mws, Gupta:2023lga}. This would imply observational constraints on the physical processes in neutron star mergers, particularly the ones arising in the postmerger. The postmerger emission is characterized by GWs emitted in the kilohertz regime from the dynamically ($\mathcal{O}\sim10^{-3}\ {\rm s}$) changing quadrupolar moment of the merger remnant. Changes in the quadrupolar moment depend strongly on the underlying equation of state (EOS) which describes the thermodynamic equilibrium state of matter in the neutron star bulk. EOSs may involve a multitude of physical processes like temperature dependent effects \cite{Perego:2019adq, Hammond:2021vtv, Blacker:2023onp, Most:2021ktk, Fields:2023bhs}, neutrino interactions and microphysics \cite{Radice:2016dwd, Radice:2018pdn, Radice:2021jtw, Schianchi:2023uky, Foucart:2022bth, Radice:2023zlw, Zappa:2022rpd, Loffredo:2022prq, Camilletti:2022jms, Most:2021zvc, Most:2022yhe, Combi:2022nhg, George:2020veu, Siegel:2017jug, Martin:2017dhc, Fujibayashi:2020qda, Grohs:2022fyq, Grohs:2023pgq, Richers:2019grc}, appearance of hyperons \cite{Sekiguchi:2011mc, Radice:2016rys}, and high-density phase transitions \cite{Sekiguchi:2011mc, Radice:2016rys, Most:2018eaw, Most:2019onn, Bauswein:2018bma, Bauswein:2020ggy, Blacker:2020nlq, Blacker:2023onp, Weih:2019xvw, Prakash:2021wpz, Liebling:2020dhf, Kedia:2022nns, Mathews:2022ria, Huang:2022mqp, Fujimoto:2022xhv, Tootle:2022pvd, Demircik:2022uol, Espino:2023llj, Guo:2023som, Haque:2022dsc} which can leave imprints on the postmerger emission. Additionally, magnetic fields and magnetohydrodynamic turbulence \cite{Ciolfi:2020cpf, Ciolfi:2017uak, Radice:2017zta, Shibata:2017xht, Margalit:2022rde} may influence the postmerger emission by redistributing the angular momentum in the remnant.

%%%%%%%%% NR Works on QCD Phase Transitions %%%%%%%%%%%%

In recent years, there has been a significant impetus in understanding the behavior of supranuclear ($> 2.7\times 10^{14}\;\mathrm{g\;cm}^{-3}$) matter expected to be realized in and around the core of heavy neutron stars, neutron star merger remnants or core collapse supernovae. Processes like a possible phase transition to deconfined quark matter or the appearance of hyperons have garnered particular interest in reference to binary neutron star (BNS) mergers as they are expected to influence the postmerger GW emission from a merger remnant which in turn can provide excellent test beds for probing strongly-interacting matter. Modelling efforts in this direction typically involve comparing GW emission from a nucleonic EOS to that computed from an EOS that has additional degrees of freedom. In this regard, the works by Sekiguchi et al.~\cite{Sekiguchi:2011mc} and Radice et al.~\cite{Radice:2016rys} explored the appearance of hyperons in a BNS merger and reported on their effects on the postmerger GW signal, i.e., a compactification of the merger remnant leading to shorter postmerger signals as compared to models without hyperons.

Most et al.~\cite{Most:2018eaw, Most:2019onn} considered a first order phase transition to deconfined quarks and obtained similar results for the postmerger GW emission along with a small de-phasing. The works by Bauswein et al.~\cite{Bauswein:2018bma, Bauswein:2020ggy} identified large shifts (30-121 Hz) in the postmerger peak frequency (which we call $f_2^{\mathrm{peak}}$ in this work) of their quark models as compared to their hadronic models. They claimed that sufficiently large shifts in $f_2^{\mathrm{peak}}$, breaking the degeneracy of EOS-insensitive relations, could be a tell-tale sign of first order phase transitions. Extending this work, Blacker et al.~\cite{Blacker:2020nlq} attempted to constrain the onset density of such phase transitions. In another work Blacker et al.~\cite{Blacker:2023onp} disentangled and explored the thermodynamics of deconfined quark matter with respect to BNS mergers. Weih et al.~\cite{Weih:2019xvw} reported on double-peaked frequency spectra as a signature of a delayed phase transition that resulted in a metastable hypermassive neutron star (HMNS). Studies by Prakash et al.~\cite{Prakash:2021wpz} however, found no smoking-gun evidences of GW signatures and observed shifts in postmerger peak frequency that were degenerate with other hadronic EOSs. They also computed potential electromagnetic signatures of these phase transitions. Liebling et al.~\cite{Liebling:2020dhf} computed similar postmerger GW signatures and observed changes in the magnetic field topology in the bulk of the star. In contrast to modelling $1^{\rm{st}}$ order phase transitions, refs. \cite{Kedia:2022nns, Mathews:2022ria, Huang:2022mqp} explored such deconfinement processes via a quark-hadron crossover (QHC) by constraining the $f_2^{\mathrm{peak}}$ and chirp frequencies. In this regard, Fujimoto et al.~\cite{Fujimoto:2022xhv} have compared GW signatures arising from a $1^{\rm{st}}$ order phase transition with those from a QHC and show the results from the QHC scenario to be consistent with electromagnetic counterparts observed from GW170817. 

More recently, there have been efforts \cite{Tootle:2022pvd, Demircik:2022uol} to employ the novel holographic V-QCD framework to construct EOSs with a deconfinement phase transition and compute their GW signals. Consistent with previous works, an early collapse for softer EOSs is observed. Espino et al.~\cite{Espino:2023llj}, for the first time, investigated multi-modal signatures of deconfinement phase transitions and reported on a weakening of the one-armed spiral instability that increased with the strength of the phase transition. Guo et al.~\citep{Guo:2023som} contrasted the GW signatures between EOSs that modelled such phase transitions via a Maxwell's construction, a Gibb's construction and a QHC and showed that lower phase transition densities lead to more compact remnants that collapse into a black hole. In a parallel study, Haque et al.~\cite{Haque:2022dsc} varied the onset density of the phase transition and examined its impact on the postmerger GW frequency.

%%%%%%%%%% QCD PT Estimation from Postmerger %%%%%%%%%%%%%

Both pre-merger (late inspiral) and postmerger phases of a BNS evolution can provide useful information with reference to phase transitions to deconfined quarks. Extensive efforts by several groups have gone into modelling the postmerger GW emission \cite{Hotokezaka:2013iia, Bauswein:2015vxa, Bose:2017jvk, Easter:2020ifj, Soultanis:2021oia, Tsang:2019esi, Breschi:2019srl, Breschi:2022xnc}. Chatziioannou et al.~\cite{Chatziioannou:2017ixj} and Wijngaarden et al.~\cite{Wijngaarden:2022sah} employ model independent inference via $\tt{Bayeswave}$ to resconstruct the postmerger signals while using NR calibrated compact binary coalescence templates for the inspiral. While this kind of a hybrid model-agnostic approach does indeed offer more flexibility towards modelling particular waveform morphologies as compared to analytical models, an absence of a model implies no way for a likelihood computation and hence a comparison using Bayes' factors or odd's ratios to other approaches cannot be made.
% DR
On the other hand, Tsang et al.~\cite{Tsang:2019esi} and Easter et al.~\cite{Easter:2020ifj} employed damped sinusoidal models to describe the postmerger emission. Breschi et al.~\cite{Breschi:2019srl, Breschi:2022xnc} constructed analytic models of postmerger emission which were calibrated by numerical relativity simulations. Subsequently, these models were employed in refs. \cite{Breschi:2019srl, Breschi:2022ens, Breschi:2023mdj} to potentially detect EOS softening via the production of $\Lambda$ hyperons. In particular, Breschi et al.~\cite{Breschi:2022ens} recovered differences in the postmerger peak frequency and remnant lifetimes to constrain the said effects in a BNS merger.

%%%%%%%%%% QCD PT Estimation from Inspiral %%%%%%%%%%%%%

To complement the above mentioned postmerger studies, there have also been several efforts to constrain nuclear properties of high-density matter using the late inspiral phase of a binary merger \cite{Chatziioannou:2017ixj, Han:2018mtj, Sieniawska:2018zzj, Raithel:2022efm, Essick:2023fso}. In particular, Mondal et al.~\cite{Mondal:2023gbf} employed a phenomenological meta-modelling approach to the EOSs and constrained QCD phase transitions via measurements of tidal parameters. Essick et al.~\cite{Essick:2023fso} constructed non-parametric representations of EOSs and attempted to infer an onset of QCD phase transitions from the EOS itself. Raithel et al.~\citep{Raithel:2022efm} have examined the impact of phase transition on an inference of tidal deformability using inspiral GW signals and have found degeneracies between the EOS with phase transition and that with hadrons while keeping the tidal deformability constant. Raithel et al.~\cite{Raithel:2022aee} also present an interesting case of ``tidal deformability doppelgängers'' where they employ quark EOSs with differences in pressure at nuclear saturation but which predict tidal parameters consistent with that of exclusively hadronic EOSs. Pang et al.~\citep{Pang:2020ilf} have computed Bayes factors of binary mergers with and without a phase transition while also considering the strength of phase transitions as a paremeter for Bayesian inference. 

While most of the works discussed above remark that such deconfinement phase transitions (and EOS softening effects in general) are potentially detectable, the refs.~\cite{Huang:2022mqp, Bauswein:2015vxa, Easter:2020ifj, Wijngaarden:2022sah, Breschi:2022ens, Breschi:2023mdj, Breschi:2019srl, Essick:2023fso, Mondal:2023gbf, Pang:2020ilf} pave a concrete path in defining an observational strategy to observe their effects with kilohertz gravitational waves. 

%%%%%%% Quasi Universal Relations %%%%%%%%%

It has been shown from NR simulations of neutron star binaries \cite{Bauswein:2011tp, Hotokezaka:2013iia, Bernuzzi:2014kca, Bernuzzi:2015rla, Rezzolla:2016nxn, Zappa:2017xba, Bauswein:2012ya, Liebling:2020dhf,Lioutas:2021jbl} that there exists a correlation between the $f_2^{\mathrm{peak}}$ frequency of the postmerger and an inspiral property of the binary, e.g. a suitable combination of tidal parameters from the inspiral, the radius of a neutron star of a fixed mass or the compactness of a neutron star. Such relations are insensitive to the EOS and are also referred to as quasi universal relations (QURs). Indeed, such relations have been employed to construct analytical waveform models \cite{Breschi:2019srl, Breschi:2022xnc}. Several works \cite{Bauswein:2018bma, Bauswein:2020ggy} claim that a violation of a universal relation between $f_2^{\rm peak}$ and the tidal deformability of a $1.35 M_{\odot}$ neutron star ($\Lambda_{1.35}$) can be taken to be a smoking-gun evidence of QCD phase transitions. Wijngaarden et al.~\cite{Wijngaarden:2022sah} even demonstrate that Bayesian error estimates for a joint detection of $f_2^{\mathrm{peak}}$ and $\tilde{\Lambda}$ at sufficiently high signal to noise ratios (SNRs) can be distinguished from the established QURs. At the same time, Breschi et al. in ref.~\cite{Breschi:2023mdj} perform a pre-postmerger consistency test and show that a breaking of an EOS insensitive relation between $f_2^{\rm peak}$ and tidal polarizability $\kappa_2^T$ to a given confidence level cannot be taken to be a confident signature of the softening of the EOS. In this work, we place our calculations in the context of previous findings by applying error estimates from Bayesian inference to NR simulations.

%%%%%%%% Organization of paper %%%%%%%%%%

We utilize Bayesian inference on the inspiral and postmerger signals to recover estimates on tidal properties and postmerger spectra respectively. We then use these estimates in reference to the universal relation by Breschi et al.~\cite{Breschi:2022xnc} to show a potential detectability of QCD phase transitions at postmerger SNRs as low as 10. To this aim, we employ composition-dependent, finite-temperature EOSs describing the high-density behavior of strongly interacting matter and compute the postmerger GW emission of a BNS merger remnant. We employ the frequency domain waveform model $\tt{NRPMw}$ developed by Breschi et al.~\cite{Breschi:2022xnc} to recover the spectra of the said NR waveforms assuming sensitivities of the next generation GW detectors. This paper is organized as follows: in subsection \ref{subsec:NR_Simulations}, we describe the NR simulations used in this work. In subsection \ref{subsec:Injection_Settings},  we comment upon the procedure employed to create postmerger injections from our NR dataset. Following this in subsection \ref{subsec:Parameter_Estimation}, we briefly recapitulate the methodology for Bayesian inference of parameters given an analytic BNS waveform model. We present our results in section \ref{sec:Results} where we classify our (postmerger) parameter estimation (PE) analysis in two categories with different choices of priors. Primarily in subsection \ref{subsec:Inspiral_Informed_Postmerger_PE}, we take inspiral-informed Gaussian priors on masses and tidal parameters for the postmerger. Secondarily, we present a test case in Appendix \ref{Inspiral_Agnostic_PE:results_for_all_simulations} wherein we assume broad priors for the postmerger model $\tt{NRPMw}$'s parameters and perform an inspiral-agnostic PE. In subsection \ref{subsec:Postmerger_PE_with_CE}, we repeat the postmerger analysis with the CE detectors: the broadband CE-40 and the narrowband postmerger optimized CE-20. In subsection \ref{subsec:Probing_QCD_Phase_Transitions}, we use an NR informed EOS insensitive relation to probe phase transitions at a given postmerger SNR. Finally, we conclude the paper in section \ref{sec:Conclusions}. In the appendices, we provide results for all our simulations as well as a miscellany of supplemental results. In appendices \ref{Inspiral_Agnostic_PE:results_for_all_simulations} and \ref{Inspiral_Informed_PE:results_for_all_simulations}, we provide results for the entire simulation dataset. Finally, in appendix \ref{Unconstrained_f2_f0_inference}, we provide results from a flexible configuration of the $\tt{NRPMw}$ model aimed at addressing some of the biases encountered in recovering hadronic models. 

%%%%%%%%%%%%%%%%%%%%%%%%%%%%%%%%%%%
%%%%%%% METHODS %%%%%%%%%%%%%%
%%%%%%%%%%%%%%%%%%%%%%%%%%%%%%%%%%%

\section{\label{sec:Methods}Methods}

%%%%%%%%%%%%%%%%%%%%%%%%%%%%%%%%%%%
%%%%%%% NR SIMULATIONS %%%%%%%%%%%%%%
%%%%%%%%%%%%%%%%%%%%%%%%%%%%%%%%%%%

\subsection{NR Simulations}
\label{subsec:NR_Simulations}

\begin{figure*}[t]
	\includegraphics[width=0.49\textwidth]{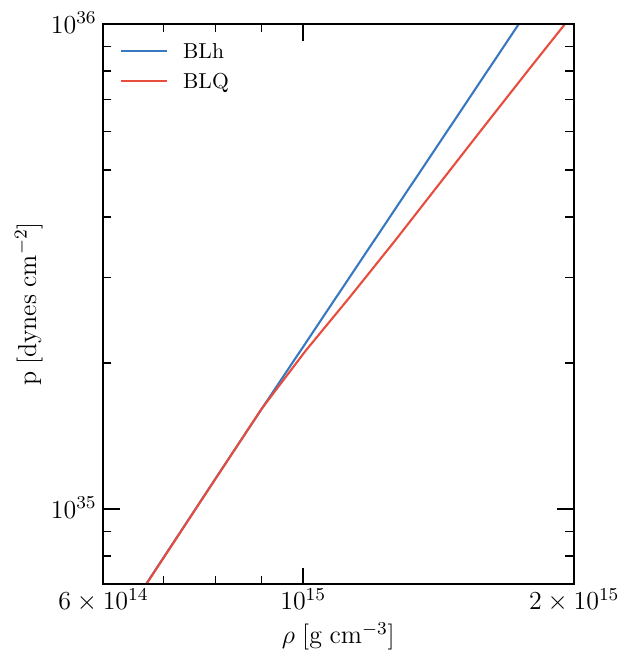}
	\includegraphics[width=0.49\textwidth]{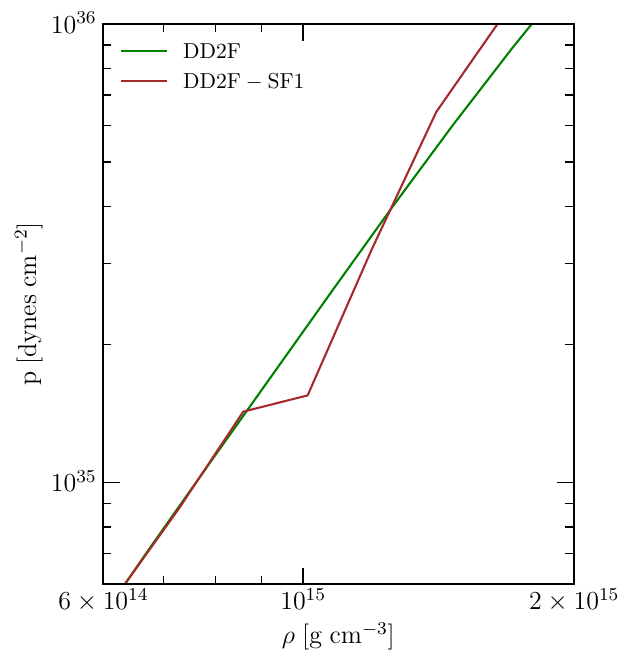}
	\caption{Pressure - density curves for the $T=0$ (zero temperature) slice of the equations of state (EOSs) used in this work. BLh and DD2F EOSs contain only nucleonic degrees of freedom whereas BLQ and DD2F-SF1 also include a prescription for a $1^{\rm{st}}$ order phase transition to deconfined quarks. Such a phase transition leads to a loss of pressure at high densities $\rho \sim 10^{15} \rm{g\;cm^{-3}}$.}
	\label{fig:TOV}
\end{figure*}

We summarize the NR simulations used in this work in Table \ref{tab:NR_Simulation_Summary}. Our dataset primarily consists of BNS merger simulations with hadronic and quark EOSs presented in ref.~\cite{Prakash:2021wpz}. We also perform merger simulations with two additional EOSs DD2F~ \cite{Typel:2009sy, Alvarez-Castillo:2016oln} and DD2F-SF1~\cite{Fischer:2017lag, Bauswein:2018bma} to include effects from different treatments of strongly-interacting matter. The mergers we consider produce remnants that do not collapse promptly and result in a finite postmerger GW signal (see Table \ref{tab:NR_Simulation_Summary}). We employ the numerical infrastructure in ref. \cite{Prakash:2021wpz} and references therein for all our NR simulations. In particular, we solve the equations of General Relativistic Hydrodynamics (GRHD) in the 3+1 Valencia Formulation \cite{Banyuls:1997zz} using the publicly available code $\tt{WhiskyTHC}$ \cite{Radice:2012cu, Radice:2013hxh, Radice:2013xpa}. We employ the $\tt{CTGamma}$ \cite{Pollney:2009yz, Reisswig:2013sqa} code available as part of the $\tt{Einstein Toolkit}$ \cite{EinsteinToolkit:2023_05} to solve for the spacetime in the Z4c formulation \cite{Bernuzzi:2009ex, Hilditch:2012fp} of the Einstein's equations. We use the $\tt{WeylScal4}$ and $\tt{Multipole}$ thorns to compute the spin $s=-2$ weighted spherical harmonics of the Newman-Penrose scalar $\Psi_4$, from which we extract the GW strain of the $\ell=2, m = 2$ mode. Additionally, we employ a zeroth moment M0 scheme \cite{Radice:2016dwd} to solve for the neutrino energies and neutrino number densities. We construct initial data assuming irrotational binaries in quasi-circular orbits using the pseudo spectral code $\tt{Lorene}$ \cite{Gourgoulhon:2000nn}. The binaries are situated at an initial separation of 45 km. Finally, we employ the $\tt{Carpet}$ \cite{Schnetter:2003rb, Reisswig:2012nc} code for providing the adaptive mesh refinement (AMR) infrastructure. 

To probe multiple possibilities in the high-density regime of QCD, we take a selection of 4 finite-temperature EOSs namely BLh \cite{Bombaci:2018ksa, Logoteta:2020yxf}, DD2F \cite{Typel:2009sy, Alvarez-Castillo:2016oln}, BLQ \cite{Prakash:2021wpz, Kashyap:2021wzs, Perego:2021mkd, Breschi:2022ens} and DD2F-SF1\cite{Fischer:2017lag, Bauswein:2018bma}. Of these, the BLh and DD2F EOSs contain only nucleonic degrees of freedom whereas the BLQ and DD2F-SF1 EOSs implement a $1^{st}$ order phase transition to deconfined quark matter while having the same low-density behavior as the BLh and DD2F EOSs, respectively. The BLQ EOS employs a Gibbs construction to combine the hadronic and quark phases resulting in a mixed phase of deconfined quarks and hadrons. There is a gradual increase in the percentage of deconfined quarks with non-zero temperatures and densities $\gtrsim 3 \rho_{\mathrm{nuc}}$ where $\rho_{\mathrm{nuc}} = 2.7 \times 10^{14} \mathrm{g\;cm^{-3}}$ is the nuclear saturation density. The DD2F-SF1 EOS on the other hand employs a Maxwell construction that allows for a less gradual transition to the deconfined quark phase as compared to the BLQ EOS.

As previously found in Bauswein et al.~\cite{Bauswein:2018bma}, the BNS models evolved with the DD2F-SF family of EOSs display large deviations from the EOS insensitive relation between the postmerger peak frequency $f_2^{\rm peak}$ and tidal deformability $\Lambda$. On the other hand, models with the BLQ EOS \cite{Prakash:2021wpz} predict postmerger peak frequencies that are within range of those spanned by hadronic EOSs and obey the $f_2^{\rm peak} - \kappa_2^T$ relation obtained in ref.~\cite{Breschi:2019srl} where $\kappa_2^T$ is the tidal polarizability defined in the same reference. It is important to emphasize that the EOS insensitive relation obtained and the simulation setup employed in ref.~\cite{Bauswein:2018bma} is not the same as the one used in ref. \cite{Breschi:2019srl}. Therefore, for consistent comparison, we performed simulations with the DD2F-SF1 EOS with our GRHD infrastructure and find that models with this EOS also display large deviations with the $f_2^{\rm peak} - \kappa_2^T$ relation. We also note that the simulations presented in our work are computed in full general relativity (GR) whereas the ones from Bauswein et al. \cite{Bauswein:2018bma} consider a conformal flatness condition to solve for the Einstein's equations.

Additionally, we consider unequal mass mergers for the BLh and BLQ EOSs to account for the impact of mass ratios. With this diversity in the choice of EOSs and the masses of BNS mergers, our study provides reasonable estimates of the GW detectability of QCD phase transitions in BNS mergers. In addition to that, we would like to remark here that even though the waveform model $\tt{NRPMw}$ is trained on a large number of NR simulations spanning 21 EOSs, simulations with DD2F and DD2F-SF1 EOSs have not been utilized for training the model and therefore validate the model's performance.

%%%%%%%%%%%%%%%%%%%%%%%%%%%%%%%%%%%
%%%%%%% INJECTION SETTINGS %%%%%%%%%%%%%%
%%%%%%%%%%%%%%%%%%%%%%%%%%%%%%%%%%%

\subsection{Injection Settings}
\label{subsec:Injection_Settings}

\begin{figure}
\includegraphics[width=\columnwidth]{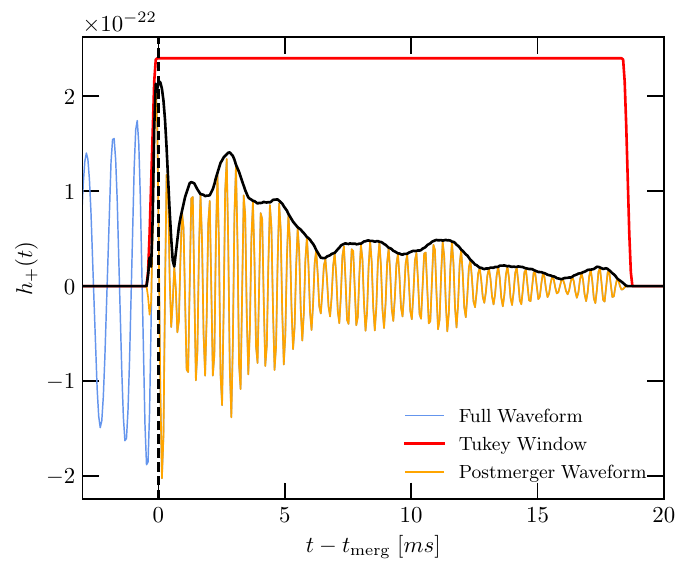}

\caption{Extraction of the postmerger waveform from an NR waveform by applying a Tukey window. This windowed waveform upon spline interpolation and zero-padding is then injected in a noise-less configuration of the ET/CE detectors for parameter estimation using \tt{NRPMw}.}
\label{fig:injection}
\end{figure}

In this section, we describe the procedure for constructing postmeregr injections from our NR simulations for a Bayesian inference study. In particular, we scale the GW strain obtained from NR simulations and introduce it in a data stream which serves to simulate the incoming GW in a detector. To compute the $\ell=2, m=2$ GW strain output from the NR simulations, we first evaluate the Newman-Penrose scalar $\Psi_4$ on coordinate spheres in a multi-polar spherical harmonic basis. This scalar (for the $\ell = 2, m = 2$ mode) is then integrated twice in time using fixed frequency integration \cite{Reisswig:2010di} to obtain the quadrupolar strain $h_+$ and $h_{\times}$. Fixed frequency integration also helps remove secular drifts in the strain amplitude that may arise because of direct integration of $\Psi_4$.

We define the time of merger $t_{\rm{merg}}$ as the time when the GW amplitude of the $\ell=2,\;m=2$ mode, i.e., $(h_{+}^2 + h_{\times}^2)^{1/2}$ is maximum. We construct the injections by considering only the postmerger portion of the NR waveform starting from $t_{\rm{merg}}$ up until the termination of the waveform. For the remnants that collapse into a black hole (BH), we define a time of BH formation $t_{\mathrm{BH}}$ (Table \ref{tab:NR_Simulation_Summary}) as the time when the minimum value of the lapse function in the computational grid drops below 0.2 which approximately corresponds to the formation of an apparent horizon for non-spinning binaries. We extract the postmerger signal ($t>t_{\rm{merg}}$) by employing a Tukey window \cite{1455106} available as part of the $\tt{scipy}$ library. In particular, we use a windowing ansatz $w$ of the form
\begin{equation}
	w(t, t_0, t_1, \delta) = 
		\begin{cases}
			0 \;\; \text{if } t < t_0 \\
			\tau(t, \delta) \;\; \text{if } t \in [t_0, t_1] \\
			0 \;\; \text{if } t > t_1
		\end{cases}
\end{equation}
where $\tau$ denotes the standard Tukey window of width $|t_1 - t_0|$ and a shape parameter $\delta$ that controls the fraction of the window inside the tapered region. Furthermore, we spline interpolate the waveforms to a sampling rate of 16,384 Hz and zero pad them to a signal segment of 1~s, as shown in Fig. \ref{fig:injection}. To systematically disentangle the effects of QCD phase transitions on the GW strain from the effects of detector noise, we construct noise-less injections. The posteriors on model parameters recovered in such a noise-less configuration approximate the average over those recovered from multiple Gaussian noise realizations. Finally, we scale the waveforms by a factor of the inverse luminosity distance $D_{L}^{-1}$ and the spin $s=-2$ weighted spherical harmonics $ _{-2} Y_{2, 2}(\iota=0, \psi=0)$ for a face-on configuration to consistently maintain a postmerger SNR of 10 in the ET detector network or in the CE-20 detector. This corresponds to placing each BNS system at different luminosity distances with respect to the detector as described in tables \ref{tab:Injection_Property_II} and \ref{tab:Injection_Property_IA}.

\begin{table*} 
\caption{ 
A summary of NR simulations employed in this work. The corresponding postmerger waveforms are used in the construction of injections for the next generation GW detectors and for the subsequent Bayesian inference. EOS represents the equation of state, $m_1$ and $m_2$ the gravitational masses of the binary ($m_1>m_2$), $q$ the mass ratio, $\Lambda_i$s the tidal deformabilities and $t_{\mathrm{BH}}$ the time of black hole formation expressed relative to the time of merger $t_{\mathrm{merg}}$.} 
\label{tab:NR_Simulation_Summary} 
 \begin{center} {
 \begin{tabular}{c c c c c c c} 
 \hline\hline 
 \\ 
$\mathrm{EOS}$ & $m_1\;[M_{\odot}]$ & $m_2\;[M_{\odot}]$  & $q$ & $\Lambda_1$ & $\Lambda_2$ & $t_{\mathrm{BH}} - t_{\mathrm{merg}}\;[\mathrm{ms}]$ \\ 
\hline
\hline
$\mathrm{BLh}$ & 1.298 & 1.298 & 1.0 & 701.901 & 701.901 & HMNS \\
$\mathrm{BLQ}$ & 1.298 & 1.298 & 1.0 & 701.901 & 701.901 & 15.95 \\ 
\hline
$\mathrm{BLh}$ & 1.481 & 1.257 & 1.178 & 295.467 & 856.064 & HMNS \\
$\mathrm{BLQ}$ & 1.481 & 1.257 & 1.178 & 295.467 & 856.064 & 3.54 \\
\hline
$\mathrm{BLh}$ & 1.398 & 1.198 & 1.167 & 435.735 & 1145.850 & HMNS \\
$\mathrm{BLQ}$ & 1.398 & 1.198 & 1.167 & 435.735 & 1145.850 & 17.2 \\
\hline
$\mathrm{BLh}$ & 1.363 & 1.363 & 1.0 & 515.379 & 515.379 & HMNS \\
$\mathrm{BLQ}$ & 1.363 & 1.363 & 1.0 & 515.379 & 515.379 & 4.1 \\
\hline
$\mathrm{DD2F}$ & 1.289 & 1.289 & 1.0 & 707.511 & 707.511 & HMNS \\
$\mathrm{DD2F-SF1}$ & 1.289 & 1.289 & 1.0 & 707.511 & 707.511 & 42.36 \\

\hline 
\hline 
\end{tabular} 
} 
\end{center} 
\end{table*}

%%%%%%%%%%%%%%%%%%%%%%%%%%%%%%%%%%%
%%%%%%% PARAMETER ESTIMATION %%%%%%%%%%%%%%
%%%%%%%%%%%%%%%%%%%%%%%%%%%%%%%%%%%

\subsection{Parameter Estimation}
\label{subsec:Parameter_Estimation}

For our postmerger PE analysis, we employ the nested sampler $\tt{UltraNest}$ \cite{johannes_buchner_2021_4636924} included as part of the \href{https://github.com/matteobreschi/bajes}{$\tt{bajes}$} code \cite{Breschi:2021wzr}. Our configuration employs $5\times10^{3}$ live points and a maximum of $5\times10^{4}$ iterations for the Monte Carlo sampler. We choose a Gaussian-noise likelihood \cite{Thrane:2018qnx} defined as
\begin{equation}
	\begin{split}
	\mathrm{log}\left(\mathcal{L}(d|\boldsymbol{\theta})\right) = -\frac{1}{2} \sum_{j} \mathrm{log} \left( 2\pi S_j \right) \\
	- \frac{1}{2} \sum_j \langle d_j - \mu(\boldsymbol{\theta}) | d_j - \mu(\boldsymbol{\theta}) \rangle
	\end{split}	
\end{equation}
where the summation index $j$ runs over the three arms in the case of the ET detector, $S_j$ denotes the power spectral density of the corresponding detector, $\mu({\boldsymbol{\theta}})$ is the $\tt{NRPMw}$ model evaluated for the parameter set $\boldsymbol{\theta}$ and $d$ represents the data stream of the injection. In the case of the CE detector, we fix $j$ to correspond to the narrow-band 20km postmerger optimized configuration. The inner product $\langle .|. \rangle$ between two signals say $a(f)$ and $b(f)$ in the frequency domain is given by
\begin{equation}
	\langle a(f) | b(f) \rangle = 4\, \Re\, {\bigintssss_{f_{\mathrm{min}}}^{f_{\mathrm{max}}}\frac{a^*(f) b(f)}{S_j(f)}} df
\end{equation}
We take $f_{min}$ and $f_{max}$ to be 1024 and 8192 $\mathrm{Hz}$, respectively to include the postmerger domain of the signal. 

We take $d_j$ to denote the data stream in each arm of the detector i.e. $d_j = s_j + n_j$, where $s_j$ and $n_j$ respectively denote the signal and noise in the detector. For noise-less injections, $d_j$ is given exclusively by the signal projected onto the individual detectors i.e.
\begin{align}
	d_j(f) = \mathrm{F}_{j; +}(\mathrm{RA., DEC.}, \psi) h_{\rm{+}}(f) + \nonumber \\
	 \mathrm{F}_{j; \times}(\mathrm{RA., DEC.}, \psi) h_{\rm{ \times}}(f)
\end{align} where $\mathrm{F}_{j; +}$ and $\mathrm{F}_{j; \times}$ denote the antenna pattern
functions of the $j^{th}$ arm of the ET detector (or a CE-20 detector) and RA., DEC. and $\psi$ denote the right ascension, declination and the polarization angle of the binary respectively. The injected signal corresponds to the strain from NR simulations.

The joint posterior distribution function(PDF) of the posterior samples corresponding to the parameters of the $\tt{NRPMw}$ model is given by the Bayes' theorem as 
\begin{align}
	p(\boldsymbol{\theta}|d) = \frac{\mathcal{L}(d | \boldsymbol{\theta}) \pi(\boldsymbol{\theta})}{\mathcal{Z}}
\end{align} 
where $\mathcal{Z}$ denotes the marginalized likelihood or the evidence for the data stream and $\pi(\boldsymbol{\theta})$ denotes the prior PDFs for the model parameters. Finally, to compute the individual posteriors ($\theta_i$) of the model parameters, we marginalize the joint PDF over the corresponding parameters to obtain 
\begin{align}
	p(\theta_i | d) = \bigintssss \left( \prod_{k \neq i} d\theta_k\right) p(\boldsymbol{\theta}|d)
	\label{compute_posteriors}
\end{align}

In the $\tt{NRPMw}$ model presented in Breschi et al. \cite{Breschi:2022xnc}, the postmerger frequency parameter $f_2$ is decided by a fit to an EOS insensitive relation (see Table I of ref. \cite{Breschi:2022xnc}) with $\kappa_2^T$ and accounted for deviations by using the re-calibration parameter $\delta f_2$. In this work, we will assume $f_2$ to be an unconstrained parameter over which we can sample in a Bayesian framework. In other words, this means migrating $f_2$ from the set of $\boldsymbol{\theta}_{\mathrm{fit}}$ to $\boldsymbol{\theta}_{\mathrm{free}}$, where $\boldsymbol{\theta}_{\mathrm{fit}}$ and $\boldsymbol{\theta}_{\mathrm{free}}$ are respectively the sets of fitted parameters and free parameters for $\tt{NRPMw}$, as defined in ref. \cite{Breschi:2022xnc}. The motivation behind making $f_2$ unconstrained lies in the fact that we do not want our results to be informed in any way by the $f_2-\kappa_2^T$ relation. Throughout this work, we will refer to the global maxima in the reconstructed postmerger spectra as $f_2^{\mathrm{peak}}$ to avoid confusion with the $f_2$ parameter of the $\tt{NRPMw}$ model which is a carrier frequency evolving linearly with time. We would like to stress that even though $f_2^{\rm peak}$ and $f_2$ are close numerically, they are not the same quantity. $f_2^{\rm peak}$
is a property of the reconstructed spectra whereas $f_2$ is a parameter of the $\tt{NRPMw}$ model. Posteriors on $f_2^{\rm peak}$ are computed from the global postmerger maxima of the reconstructed signal which in turn depends on $f_2$ and other parameters. In a nutshell, $f_2^{\rm peak}$ is influenced by the choice of $f_2$ but not the other way around. Throughout this work, we will refer to this updated model with the unconstrained $f_2$ parameter as $\tt{NRPMw}$. For comparison, we have also presented calculations in subsection \ref{subsec:Probing_QCD_Phase_Transitions} with the original model of Breschi et al.~\cite{Breschi:2022xnc} where $f_2$ is constrained by $\kappa_2^T$ and we call this model as $\tt{NRPMw\_v1}$. Finally, to explore a more flexible configuration of the model, we unconstrain not only $f_2$ but also $f_0$ which is the parameter for radial oscillation modes. We refer to this version of the model as $\tt{NRPMw\_v2}$ and describe it in appendix \ref{Unconstrained_f2_f0_inference}.

%%%%%%%%%%%%%%%%%%%%%%%%%%%%%%%%%%%
%%%%%%% CHOICE OF PRIORS %%%%%%%%%%%%%%
%%%%%%%%%%%%%%%%%%%%%%%%%%%%%%%%%%%

\subsection{Choice of Priors}
\label{subsubsec:Choice_of_Priors}

\begin{table} 
\caption{ 
Prior ranges for the parameters of the $\tt{NRPMw}$ model as well as the extrinsic parameters in an inspiral agnostic setting. In particular, the priors on $M$ and $q$ have been set in accordance to ref. \cite{Callister:2021gxf} so as to maintain a uniform distribution in $m_1$ and $m_2$. 
} 
\label{tab:inspiral_agnostic_priors_table} 
 \begin{center} {
 \begin{tabular}{c c c c} 
 \hline\hline 
 \\ 
$\mathrm{parameter}$ & $\mathrm{min}$ & $\mathrm{max}$ & $\mathrm{Type}$  \\ 
 \hline 
 \hline 
$\mathrm{M\;[M_{\odot}]}$ & 1 & 6 & ref. \cite{Callister:2021gxf} \\
$q$ & 1 & 2 & ref. \cite{Callister:2021gxf} \\
$\chi_1$ & -0.2 & 0.2 & aligned spin \\
$\chi_2$ & -0.2 & 0.2 &  aligned spin \\
$\Lambda_1$ & 0 & 4000 & Uniform \\
$\Lambda_2$ & 0 & 4000 & Uniform \\
$\mathrm{R.A.}$ & 0 & 2$\pi$ & Uniform \\
$\mathrm{DEC.}$ & $-\pi/2$ & $\pi/2$ & Cosinusoidal \\
$\mathrm{cos} \;\iota$ & -1 & 1 & Uniform \\
$\psi$ & 0 & $\pi$ & Uniform \\
$D_L\;\mathrm{[Mpc]}$ & 5 & 500 & Volumetric \\
$t_{\mathrm{coll}}/M$ & 1 & 3000 & Uniform \\
$M^2 \alpha_{\mathrm{peak}}$ & -$10^{-4}$& $10^{-4}$& Uniform \\
$\phi_{\mathrm{PM}}$ & 0 & 2$\pi$ & Uniform \\
$f_2\;[\mathrm{kHz}]$ & 1.5 & 5 & Uniform \\
$\delta (Mf_0)$ & -1 & 2 & $\rm{Gaussian}^{\mu=0}_{\sigma=0.449}$\\
$\delta (Mf_{\mathrm{mrg}}/\nu)$ & -0.2 & 0.2 & $\rm{Gaussian}^{\mu=0}_{\sigma=0.026}$ \\
$\delta (A_{\mathrm{mrg}}/M)$ & -0.2 & 0.2 & $\rm{Gaussian}^{\mu=0}_{\sigma=0.018}$ \\
$\delta (M/t_0)$ & -0.5 & 0.5 & $\rm{Gaussian}^{\mu=0}_{\sigma=0.092}$ \\
$\delta (A_0/M)$ & -1 & 4 & $\rm{Gaussian}^{\mu=0}_{\sigma=0.663}$ \\
$\delta (A_1/M)$ & -1 & 2 & $\rm{Gaussian}^{\mu=0}_{\sigma=0.152}$ \\
$\delta (A_2/M)$ & -1 & 2 & $\rm{Gaussian}^{\mu=0}_{\sigma=0.385}$ \\
$\delta (A_3/M)$ & -1 & 2 & $\rm{Gaussian}^{\mu=0}_{\sigma=0.269}$ \\
$\delta (M^2\mathcal{I}m(\alpha_{\mathrm{fus}})/\nu)$ & -4 & 4 & $\rm{Gaussian}^{\mu=0}_{\sigma=0.751}$ \\
$\delta (M\mathcal{R}e(\beta_{\mathrm{peak}}))$ & -1 & 2 & $\rm{Gaussian}^{\mu=0}_{\sigma=0.27}$ \\
$\delta (M\Delta_{\mathrm{fm}})$& -1 & 4 & $\rm{Gaussian}^{\mu=0}_{\sigma=0.744}$ \\
$\delta (M\Gamma_{\mathrm{fm}})$ & -1 & 4 & $\rm{Gaussian}^{\mu=0}_{\sigma=0.977}$ \\

\hline 
\hline 
\end{tabular} 
} 
\end{center} 
\end{table}

\begin{table} 
\caption{ 
Prior ranges for the parameters of the $\tt{NRPMw}$ model and the extrinsic parameters in an inspiral informed setting. We constrain priors on $M$, $q$, $\Lambda_1$ and $\Lambda_2$ from the inspiral signal. In this table, we show details for the prior distribution employed for the $1.398 M_{\odot} - 1.198 M_{\odot}$ binary with the BLh EOS. The type of priors remains the same for all models in our work.
} 
\label{tab:inspiral_informed_priors_table} 
 \begin{center} {
 \begin{tabular}{c c c c} 
 \hline\hline 
 \\ 
$\mathrm{parameter}$ & $\mathrm{min}$ & $\mathrm{max}$ & $\mathrm{Type}$  \\ 
 \hline 
 \hline 
$\mathrm{M\;[M_{\odot}]}$ & 2.641 & 2.652 & $\rm{Gaussian}^{\mu=2.646}_{\sigma=0.001}$ \\
$q$ & 1.11 & 1.22 & $\rm{Gaussian}^{\mu=1.17}_{\sigma=0.01}$ \\
$\chi_1$ & -0.2 & 0.2 & aligned spin \\
$\chi_2$ & -0.2 & 0.2 &  aligned spin \\
$\Lambda_1$ & 363.94 & 559.70 & $\rm{Gaussian}^{\mu=448.24}_{\sigma=18.79}$ \\
$\Lambda_2$ & 1030.17 & 1203.71 & $\rm{Gaussian}^{\mu=1123.61}_{\sigma=21.28}$ \\
$\mathrm{R.A.}$ & 0 & 2$\pi$ & Uniform \\
$\mathrm{DEC.}$ & $-\pi/2$ & $\pi/2$ & Cosinusoidal \\
$\mathrm{cos} \;\iota$ & -1 & 1 & Uniform \\
$\psi$ & 0 & $\pi$ & Uniform \\
$D_L\;\mathrm{[Mpc]}$ & 5 & 500 & Volumetric \\
$t_{\mathrm{coll}}/M$ & 1 & 3000 & Uniform \\
$M^2 \alpha_{\mathrm{peak}}$ & -$10^{-4}$& $10^{-4}$& Uniform \\
$\phi_{\mathrm{PM}}$ & 0 & 2$\pi$ & Uniform \\
$f_2\;[\mathrm{kHz}]$ & 1.5 & 5 & Uniform \\
$\delta (Mf_0)$ & -1 & 2 & $\rm{Gaussian}^{\mu=0}_{\sigma=0.449}$\\
$\delta (Mf_{\mathrm{mrg}}/\nu)$ & -0.2 & 0.2 & $\rm{Gaussian}^{\mu=0}_{\sigma=0.026}$ \\
$\delta (A_{\mathrm{mrg}}/M)$ & -0.2 & 0.2 & $\rm{Gaussian}^{\mu=0}_{\sigma=0.018}$ \\
$\delta (M/t_0)$ & -0.5 & 0.5 & $\rm{Gaussian}^{\mu=0}_{\sigma=0.092}$ \\
$\delta (A_0/M)$ & -1 & 4 & $\rm{Gaussian}^{\mu=0}_{\sigma=0.663}$ \\
$\delta (A_1/M)$ & -1 & 2 & $\rm{Gaussian}^{\mu=0}_{\sigma=0.152}$ \\
$\delta (A_2/M)$ & -1 & 2 & $\rm{Gaussian}^{\mu=0}_{\sigma=0.385}$ \\
$\delta (A_3/M)$ & -1 & 2 & $\rm{Gaussian}^{\mu=0}_{\sigma=0.269}$ \\
$\delta (M^2\mathcal{I}m(\alpha_{\mathrm{fus}})/\nu)$ & -4 & 4 & $\rm{Gaussian}^{\mu=0}_{\sigma=0.751}$ \\
$\delta (M\mathcal{R}e(\beta_{\mathrm{peak}}))$ & -1 & 2 & $\rm{Gaussian}^{\mu=0}_{\sigma=0.27}$ \\
$\delta (M\Delta_{\mathrm{fm}})$& -1 & 4 & $\rm{Gaussian}^{\mu=0}_{\sigma=0.744}$ \\
$\delta (M\Gamma_{\mathrm{fm}})$ & -1 & 4 & $\rm{Gaussian}^{\mu=0}_{\sigma=0.977}$ \\

\hline 
\hline 
\end{tabular} 
} 
\end{center} 
\end{table}

With the advent of the next generation of GW detectors, it is expected for binaries that are loud enough that their postmergers can be detected, masses and tidal deformabilities will be measured accurately from the inspiral \cite{Branchesi:2023mws, Gupta:2023lga}. Therefore, the most accurate PE would result from an analysis of the full signal, i.e., inspiral and postmerger. However, performing Bayesian inference on the full signal is computationally expensive. In this work, we adopt a two-fold approach in the sense that we analyze the inspiral and postmerger signals using separate inference codes. From the inspiral inference, we compute posteriors on total gravitational mass $M$, mass ratio $q$ and the tidal deformabilities $\Lambda_i$s, all of which for loud signals are Gaussians to a good approximation. In a separate inference for the postmerger, we constrain the prior bounds of the postmerger model by supplying the Gaussian priors thus obtained. We refer the reader to subsection \ref{subsec:Inspiral_Informed_Postmerger_PE} for the detailed procedure to compute these priors. 

On the other hand in appendix \ref{Inspiral_Agnostic_PE:results_for_all_simulations}, we only use the postmerger inference with a set of priors that are independent of the inspiral signal. These priors are detailed in Table \ref{tab:inspiral_agnostic_priors_table}. For both the choices of priors, we do not include the $f_2-\kappa_2^T$ relation in the model.

Providing a comparison between results obtained from an inspiral-informed and inspiral-agnostic choice of priors is essential. As we will make explicit in this work, the choice of priors has minimal influence on the recovery of $f_2^{\rm peak}$ which is solely estimated from the postmerger. However, estimating $f_2^{\rm peak}$ is not the only pre-requisite for detecting phase transitions. Phase transitions are detected by quantifying violations of EOS insensitive relations between the postmerger $f_2$ and the inspiral $\kappa_2^T$. With the inspiral-agnostic priors, $f_2^{\rm peak}$ is well measured but there are large uncertainties in the measurement of $\kappa_2^T$ (see appendix \ref{Inspiral_Agnostic_PE:results_for_all_simulations}). This can be mitigated by supplying priors that are informed about the tidal properties and masses from the inspiral signal. This is precisely what we observe with the choice of inspiral-informed priors where our sampler essentially recovers the Gaussian priors set from the inspiral. Another motivation behind such a comparative study with different choices of priors is to demonstrate the $\tt{NRPMw}$ model's performance when subjected to different degrees of independence in the sampling of the prior parameter space. 

When a parameter of the $\tt{NRPMw}$ model is constrained by fits to EOS insensitive relations, we employ corresponding recalibration parameters to account for the uncertainties in these relations. The priors on all the recalibration parameters $\delta \boldsymbol{\theta}_{\rm fit}$ are distributed normally around a mean value of zero with a variance decided by the relative standard deviation between the scatter of NR simulations and the EOS insensitive relation. When we make a parameter independent of these fits as in the case of $f_2$, we ignore the corresponding recalibration parameter.

%%%%%%%%%%%%%%%%%%%%%%%%%%%%%%%%%%%
%%%%%%% RESULTS %%%%%%%%%%%%%%
%%%%%%%%%%%%%%%%%%%%%%%%%%%%%%%%%%%

\section{\label{sec:Results}Results}

%%%%%%%%%%%%%%%%%%%%%%%%%%%%%%%%%%%
%%%%%%% INSPIRAL-INFORMED POSTMERGER PE  %%%%%%%%%%%%%%
%%%%%%%%%%%%%%%%%%%%%%%%%%%%%%%%%%%
\subsection{Inspiral-Informed Postmerger PE}
\label{subsec:Inspiral_Informed_Postmerger_PE}

\begin{figure*}[t]
	\includegraphics[width= \textwidth]{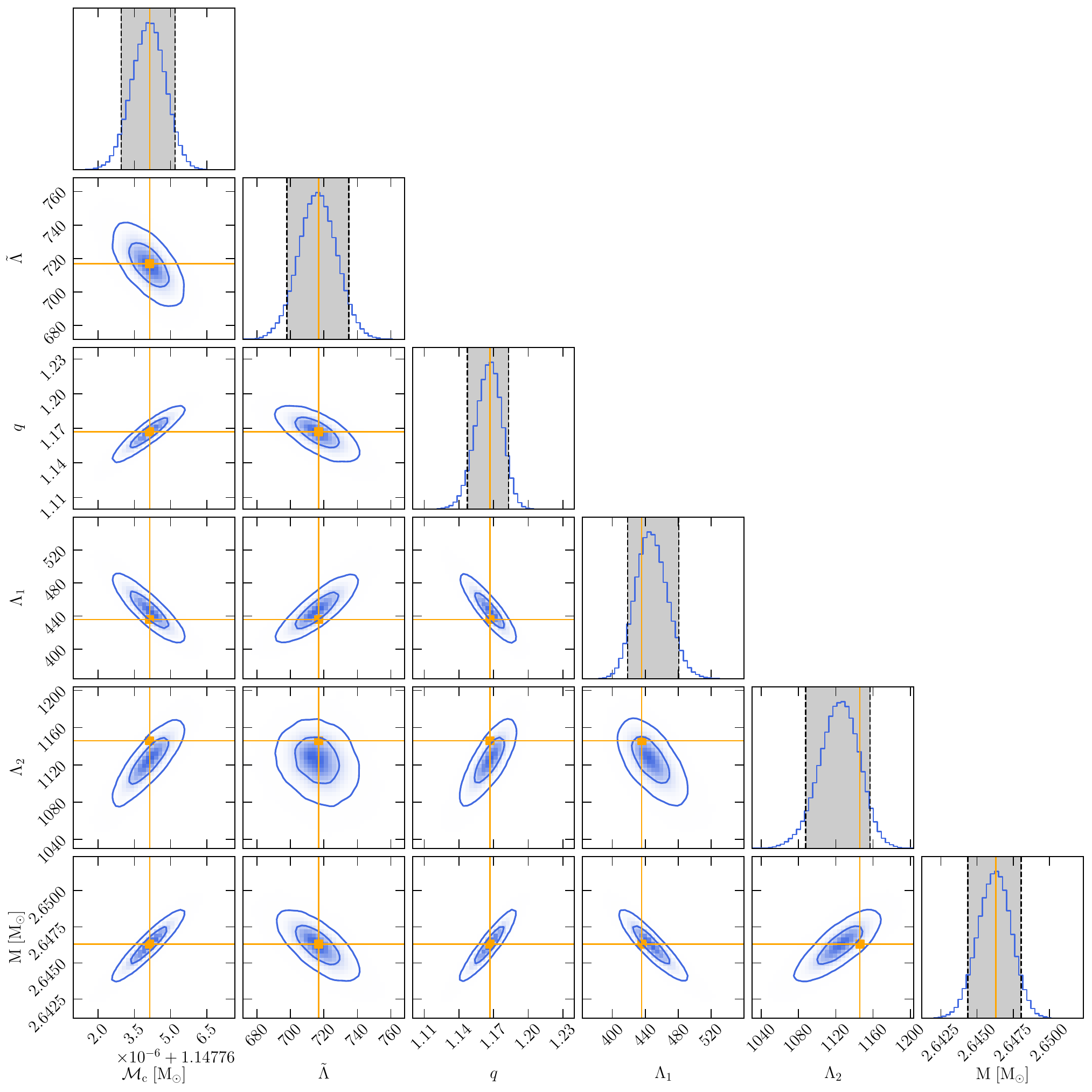}
	\caption{Corner plot depicting the posteriors on chirp mass $\mathcal{M}_c$, tidal deformability $\tilde{\Lambda}$, mass ratio $q$, individual tidal parameters $\Lambda_i$s and the total mass $M$ for the binary 1.398 $M_{\odot}$ - 1.198 $M_{\odot}$ with the BLh EOS. These posterior PDFs are computed from a self-consistent injection recovery of the $\tt{TaylorF2}$ waveform model corresponding to the binary parameters presented in Table \ref{tab:NR_Simulation_Summary} for the ET detector configuration.}
	\label{fig:inspiral_PE_corner_5}
\end{figure*}

\begin{table*} 
\caption{ 
A summary of the properties of postmerger injections corresponding to the NR simulations reported in table \ref{tab:NR_Simulation_Summary}. In this table we present these properties for the choice of priors that is informed by the inspiral signal as described in subsection \ref{subsec:Inspiral_Informed_Postmerger_PE}. In particular, GW model represents the specific configuration of the NRPMw model utilized for the recovery. We us the NRPMw model in three configurations namely, $\tt{NRPMw}$, where the $f_2$ parameter is \textbf{unconstrained} by the $f_2 - \kappa_2^T$ relation, $\tt{NRPMw\_v1}$ where the $f_2$ parameter \textbf{is constrained} by the quasi universal relation, and the most flexible $\tt{NRPMw\_v2}$ configuration where both $f_2$ and $f_0$ are unconstrained from their respective quasi universal relations. Detector is the GW datector used for the recovery of postmerger injections, $f_{2; \mathrm{Injected}}^{\rm peak}$ and $f_{2;\mathrm{Recovered}}^{\rm peak}$ are respectively the injected and recovered postmerger peak frequencies and $D_{\ell}$ is the luminosity distance of the binary from the detector. In the last two columns, we report the postmerger signal to noise ratios of the injected and recovered signals.}
\label{tab:Injection_Property_II} 
 \begin{center} {
 \begin{tabular}{c c c c c c c c c c c} 
 \hline\hline 
 \\ 
$\mathrm{Index}$ & $\mathrm{EOS}$ & $m_1\;[M_{\odot}]$ & $m_2\;[M_{\odot}]$  & $\mathrm{GW\;Model}$ & $\mathrm{Detector}$ & $f_{2;\mathrm{Injected}}^{\mathrm{peak}}\;[\mathrm{kHz}]$ & $f_{2;\;\mathrm{Recovered}}^{\mathrm{peak}}\;[\mathrm{kHz}]$ & $D_\ell\;[\mathrm{Mpc}]$ & $\rho_{\mathrm{injected}}$& $\rho_{\mathrm{recovered}}$   \\ 
\hline
\hline

1 & $\mathrm{BLh}$ & 1.298 & 1.298 & $\tt{NRPMw}$ & ET & 2.804 & $2.842^{+0.041}_{-0.025}$ & 89.049 & 10 & $8.73^{+1.08}_{-1.62} $\\
2 & $\mathrm{BLQ}$ & 1.298 & 1.298 & $\tt{NRPMw}$ & ET & 2.927 & $2.924^{+0.016}_{-0.025}$ & 93.474 & 10 & $8.61^{+1.10}_{-1.71} $\\
\hline
3 & $\mathrm{BLh}$ & 1.481 & 1.257 & $\tt{NRPMw}$ & ET & 2.962 & $2.948^{+0.016}_{-0.025}$ & 97.503 & 10 & $8.66^{+1.17}_{-1.49} $\\
4 & $\mathrm{BLQ}$ & 1.481 & 1.257 & $\tt{NRPMw}$ & ET & 3.143 & $3.145^{+0.303}_{-0.106}$ & 83.434 & 10 & $8.75^{+1.20}_{-1.19} $\\
\hline
5 & $\mathrm{BLh}$ & 1.398 & 1.198 & $\tt{NRPMw}$ & ET & 2.758 & $2.825^{+0.082}_{-0.066}$ & 87.027 & 10 & $7.82^{+0.95}_{-1.22} $\\
6 & $\mathrm{BLQ}$ & 1.398 & 1.198 & $\tt{NRPMw}$ & ET & 2.955 & $2.957^{+0.025}_{-0.033}$ & 87.500 & 10 & $8.75^{+1.17}_{-1.70} $\\
\hline
7 & $\mathrm{BLh}$ & 1.363 & 1.363 & $\tt{NRPMw}$ & ET & 3.074 & $3.055^{+0.033}_{-0.025}$ & 97.282 & 10 & $7.97^{+1.16}_{-1.66} $\\
8 & $\mathrm{BLQ}$ & 1.363 & 1.363 & $\tt{NRPMw}$ & ET & 3.197 & $3.268^{+0.09}_{-0.066}$ & 78.449 & 10 & $8.60^{+1.33}_{-1.35} $\\
\hline
9 & $\mathrm{DD2F}$ & 1.289 & 1.289 & $\tt{NRPMw}$ & ET & 2.889 & $2.916^{+0.025}_{-0.025}$ & 93.284 & 10 & $8.56^{+1.13}_{-1.70} $\\
10 & $\mathrm{DD2F-SF1}$ & 1.289 & 1.289 & $\tt{NRPMw}$ & ET & 3.354 & $3.284^{+0.213}_{-0.172}$ & 78.247 & 10 & $8.53^{+1.25}_{-1.25} $\\
11 & $\mathrm{DD2F-SF1}$ & 1.289 & 1.289 & $\tt{NRPMw}$ & ET & 3.354 & $3.325^{+0.106}_{-0.041}$ & 52.165 & 15 & $12.34^{+2.17}_{-1.87} $\\
\hline
12 & $\mathrm{DD2F}$ & 1.289 & 1.289 & $\tt{NRPMw}$ & CE-20 & 2.888 & $2.916^{+0.025}_{-0.025}$ & 118.467 & 10 & $8.59^{+1.12}_{-2.14} $\\
13 & $\mathrm{DD2F-SF1}$ & 1.289 & 1.289 & $\tt{NRPMw}$ & CE-20 & 3.375 & $3.276^{+0.221}_{-0.09}$ & 89.078 & 10 & $ 8.48^{+1.15}_{-1.78} $\\
\hline
14 & $\mathrm{DD2F-SF1}$ & 1.289 & 1.289 & $\tt{NRPMw\_v1}$ & ET & 3.354 & $3.243^{+0.066}_{-0.106}$ & 78.247 & 10 & $8.40^{+1.23}_{-1.21} $\\
\hline
15 & $\mathrm{BLQ}$ & 1.298 & 1.298 & $\tt{NRPMw\_v1}$ & ET & 2.927 & $2.924^{+0.016}_{-0.025}$ & 93.474 & 10 & $8.55^{+1.12}_{-1.65} $\\
\hline
16 & $\mathrm{BLh}$ & 1.298 & 1.298 & $\tt{NRPMw\_v2}$ & ET & 2.804 & $2.834^{+0.033}_{-0.025}$ & 89.049 & 10 & $8.60^{+1.12}_{-1.62} $\\
\hline 
\hline 
\end{tabular} 
} 
\end{center} 
\end{table*}

In this subsection, we present results for the postmerger PE analysis by taking the priors on $M$, $q$, $\Lambda_1$, and $\Lambda_2$ as Gaussian (normal) priors that are informed from the inspiral signal. Since the NR waveforms simulate only the last few orbits before merger and for a reliable estimate of masses and tidal parameters we require a longer inspiral data stream, we employ the $\tt{TaylorF2}$ waveform model \cite{Sathyaprakash:1991mt, Bohe:2013cla, Arun:2008kb, Mikoczi:2005dn, Bohe:2015ana, Mishra:2016whh, Poisson:1997ha} to simulate the inspiral signal targeted at the parameters of the binaries listed in table \ref{tab:Injection_Property_II}. The inspiraling binaries are assumed to be non-spinning and situated at the most optimal sky location corresponding to the detectors (either ET or CE).

We perform a self-consistent injection recovery with the $\tt{TaylorF2}$ waveform model in the ET (or CE-40) noise configuration and compute posteriors on the chirp mass $\mathcal{M}_c$, tidal deformability $\tilde{\Lambda}$ as defined in ref.~\cite{Wade:2014vqa}, mass ratio $q$, individual tidal parameters $\Lambda_i$s and the total mass $M$ for all the hadronic models listed in Table \ref{tab:Injection_Property_II}. For this purpose, we employ the publicly available $\tt{Bilby}$ framework \cite{Ashton:2018jfp, Romero-Shaw:2020owr, Ashton:2021anp} that utilizes relative binning \cite{Zackay:2018qdy} for the computation of posteriors. 

In Fig. \ref{fig:inspiral_PE_corner_5}, we show the posterior PDFs from the self-consistent injection recovery of the $\tt{TaylorF2}$ model targeted to simulate a long inspiral of the $1.398 M_{\odot} - 1.198 M_{\odot}$ binary with the BLh EOS. We see that the chirp mass $\mathcal{M}_c$ is extremely well measured with a standard deviation of $6.77 \times 10^{-7}M_{\odot}$. The posterior PDFs for the tidal parameters $\Lambda_i$s are refined by recomputing them via the universal relations presented in \cite{Chatziioannou:2018vzf} by taking the $M$, $q$, $\Lambda_1$ and $\Lambda_2$ inspiral posteriors as inputs. This could be a potential source of systematic errors which we have underestimated given the uncertainties in these relations as pointed out in ref.~\cite{Kastaun:2019bxo}. We note that this being an asymmetric merger ($q\neq$1), we have an accurate determination of $q$ and consequently $\Lambda_1$ and $\Lambda_2$ . For an equal-mass merger, the injected value of $q=1$, lies on the edge of the priors for the sampler and results in a biased measurement. This biased measurement of $q$ also lends bias to the measurement of $\Lambda_1$ and $\Lambda_2$ as is shown in Fig. \ref{fig:inspiral_PE_corner_1}. Nevertheless, even for $q=1$ mergers, symmetric tidal combinations such as $\kappa_2^T$ which is used as a probe for phase transitions by the $\tt{NRPMw}$ model are well estimated to within 90\%. 

We approximate the inspiral posteriors on $M$, $q$, $\Lambda_1$ and $\Lambda_2$ with Gaussian functions that have the same average as that of the inspiral posterior and a standard deviation equal to a quarter of the full 2 $\sigma$ width of the posterior. This choice allows us to be sufficiently conservative and establish a lower bound for measurement accuracy, which will only be improved if one chooses more restrictive priors and/or consider correlations between the priors. We remark that, at the moment, the $\tt{bajes}$ infrastructure does not support a specification of correlated priors. We extract these Gaussian profiles from the simulations run with hadronic EOSs and use them as priors for a postmerger PE for both hadronic and quark simulations. This is because, as discussed in \ref{subsec:NR_Simulations}, the hadronic and quark EOSs have the same low-density EOS and therefore the same tidal deformabilities. We would like to emphasize that while this approach is not a replacement for a full inspiral-postmerger inference, it provides reliable estimates for masses and tidal polarizabilities from the inspiral signal. The signal in the inspiral corresponding to a postmerger SNR of 10, has an SNR of $\sim 600$. The standard deviation in the inspiral estimates of the total mass range between $10^{-4} M_{\odot} - 10^{-3} M_{\odot}$. In addition to that, the percentage error in $\kappa_2^T$, i.e., deviation between the injected $\kappa_2^T$ and the $50^{\rm th}$ percentile of the recovered posterior ranges from 0.1\%-0.5\%. The advantages of employing independent pipelines for inspiral and postmerger is two-fold. First, the usage of relative binning for the inspiral signal significantly reduces computational cost. Second, the posteriors on masses and tidal parameters obtained in this analysis help constrain the priors for the postmerger inference thereby providing estimates on the performance of $\tt{NRPMw}$ in a hitherto untested domain of restricted sampling freedom.

\begin{figure*}[t]
	\includegraphics[width=0.49\textwidth]{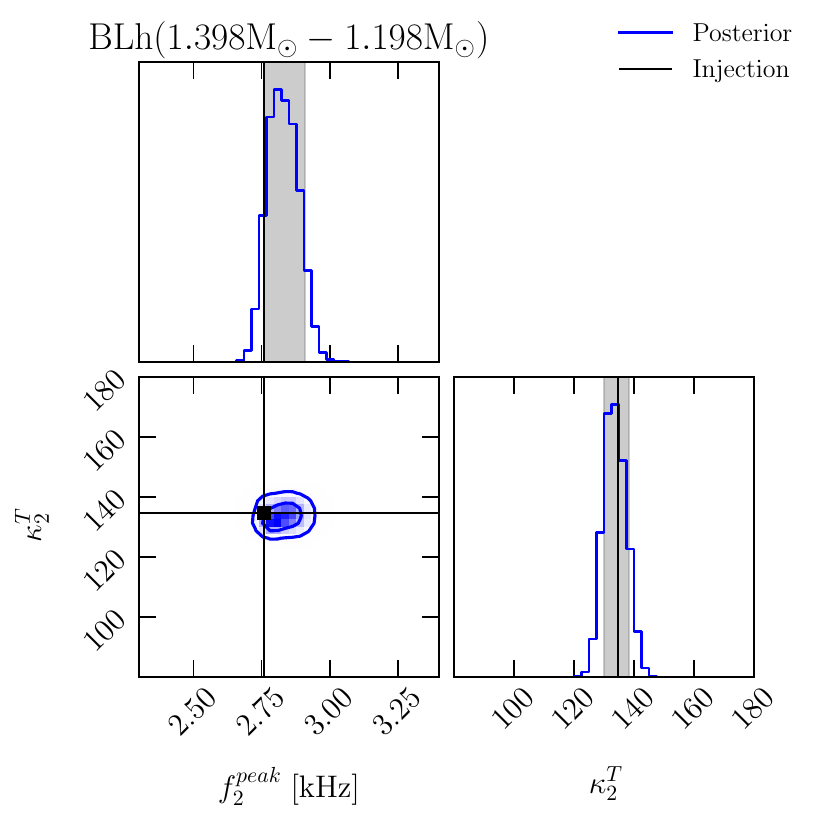}
	\includegraphics[width=0.49\textwidth]{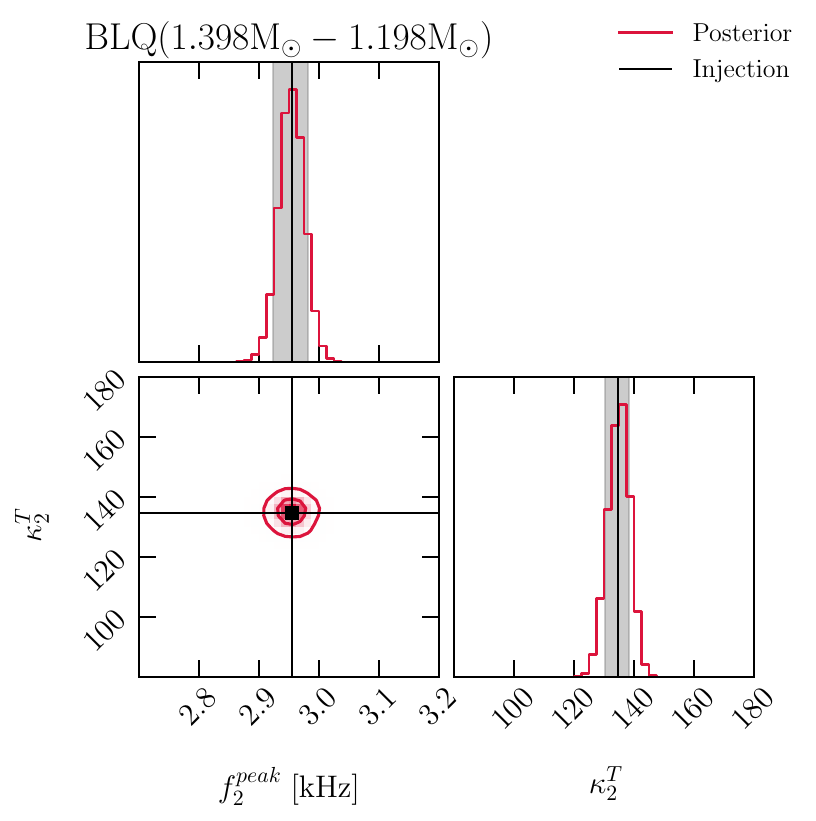}
	\caption{\textit{Left Panel:} The posterior distribution on the postmerger peak frequency $f_2^{\mathrm{peak}}$ and the tidal polarizability $\kappa_2^T$ for the binary $1.398 M_{\odot} - 1.198 M_{\odot}$ with the BLh EOS at a postmerger SNR of 10. We also show the 90\% and 50 \% contour levels for the joint PDF. \textit{Right panel:} The same calculation for the corresponding quark model. Shown in grey shaded regions are the 90\% CIs for the respective posteriors. We note that for the quark EOS, $\tt{NRPMw}$  is able to recover the injected $f_2^{\mathrm{peak}}$ to within 90\% CIs however, for the BLh (hadronic) case, the injected value lies at the boundary of the $5^{\rm{th}}$ percentile. Nevertheless, for both the cases the injection lies within the 90\% contour of the joint PDF.}
	\label{fig:Inspiral_Informed_PE_f2_k2T_5_6}
\end{figure*}

\begin{figure}
	\includegraphics[width=\columnwidth]{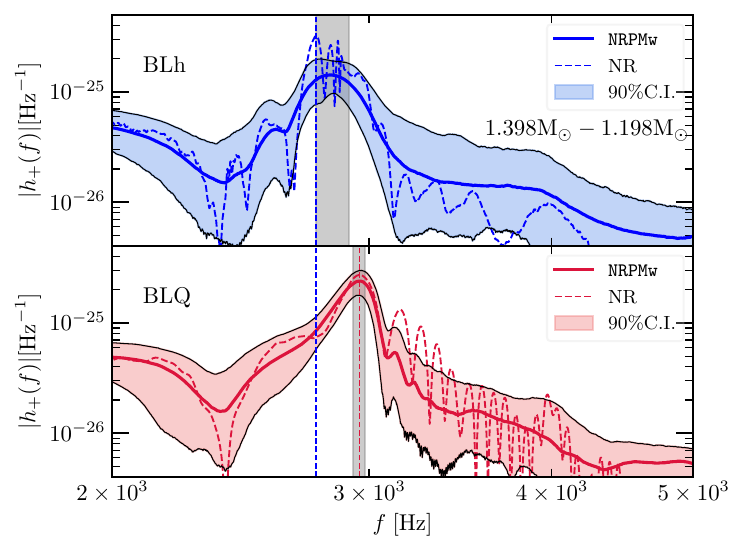}
	\caption{Reconstructed spectra for the binary $1.398 M_{\odot} - 1.198 M_{\odot}$ with the BLh and BLQ EOSs at a postmerger SNR of 10. The dotted curves represnt the injected spectra and the solid curves represent the median reconstructed signal by $\tt{NRPMw}$. We also show 90\% CIs on the reconstructed signal in the shaded regions. Vertical dotted lines correspond to the injected $f_2^{\mathrm{peak}}$ and the grey shaded regions represent the 90\% CIs on the $f_2^{\mathrm{peak}}$ posteriors.}
	\label{fig:reconstructed_spectra_5_6}
\end{figure}

In Fig. \ref{fig:Inspiral_Informed_PE_f2_k2T_5_6}, we show our results for an inspiral informed PE for the representative case of the $1.398 M_{\odot} - 1.198 M_{\odot}$ merger with the BLh and BLQ EOSs. We note that for the model with the BLQ EOS, the 90\% credible interval (CI) estimated by $\tt{NRPMw}$ for the posterior of $f_2^{\rm peak}$ contains the injected value whereas for the model with the BLh EOS, the 95\% CI of the $f_2^{\rm peak}$ posterior contains the injection. Additionally, in Fig. \ref{fig:reconstructed_spectra_5_6}, we present the reconstructed frequency spectra for the same pair of simulations using $\tt{NRPMw}$. We emphasize that the CIs for the hadronic and quark case do not overlap, implying that at a postmerger SNR of 10, the two models can be distinguished.

We summarize our results for the detectability of $f_2^{\mathrm{peak}}$ for all other simulations in our work in Figs. \ref{fig:Inspiral_Informed_PE_f2_k2T_1234}, \ref{fig:Inspiral_Informed_PE_f2_k2T_78910} and \ref{fig:reconstructed_spectra_II_123478910}. We report that for the binaries $1.481 M_{\odot} - 1.257 M_{\odot}$ and $1.363 M_{\odot} - 1.363 M_{\odot}$ with the BLh EOS, the 90\% CIs estimated for the $f_2^{\rm peak}$ posterior contain the injection. For the rest of the hadronic models, i.e., $1.398 M_{\odot} - 1.198 M_{\odot}$ with BLh, $1.289 M_{\odot} - 1.289 M_{\odot}$ with DD2F and $1.298 M_{\odot} - 1.298 M_{\odot}$ with BLh, 95 \%, 98 \% and 99.5 \% CIs of $f_2^{\rm peak}$ contain the injection respectively. For the hadronic models, we identify a systematic bias that leads to an overestimation of $f_2^{\rm peak}$ by $\tt{NRPMw}$. This bias primarily arises because of the presence of multiple (>2) amplitude modulations in the postmerger signal. As mentioned in Fig. 2 of ref. ~\cite{Breschi:2022xnc}, $\tt{NRPMw}$ is designed to capture the peaks of only the first two amplitude modulations, following which it models a damped sinusoidal decay of the postmerger amplitude. In the subsection \ref{subsubsec:multi_amplitude_modulations}, we discuss this bias in detail.
On the other hand, for the quark EOSs, we observe that the 90\% CIs for $f_2^{\rm peak}$ contain the injected $f_2^{\mathrm{peak}}$ except for the $1.363 M_{\odot} - 1.363 M_{\odot}$ merger where the 95\% CIs contain the injection.

%%%%%%%%%%%%%%%%%%%%%%%%%%%%%%%%%%%%%%%%%%%%%%%%%%%%%%%%%%%%%%%%%%%%%%%%
%%%%%%%%%%%%%%%%%% Biases due to multiple amp modulations %%%%%%%%%%%%%%%%
%%%%%%%%%%%%%%%%%%%%%%%%%%%%%%%%%%%%%%%%%%%%%%%%%%%%%%%%%%%%%%%%%%%%%%%%%

\subsubsection{Biases due to multiple amplitude modulations}
\label{subsubsec:multi_amplitude_modulations}

\begin{figure*}[t]
	\includegraphics[width=0.49\textwidth]{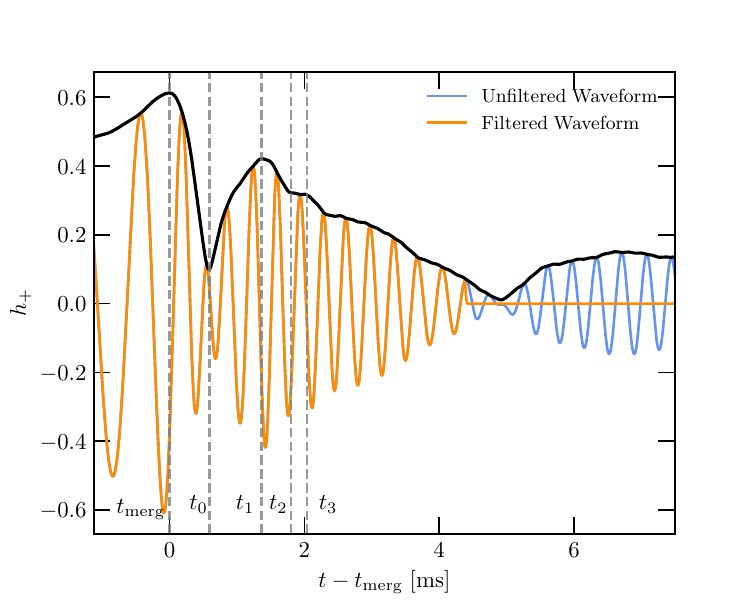}
	\includegraphics[width=0.49\textwidth]{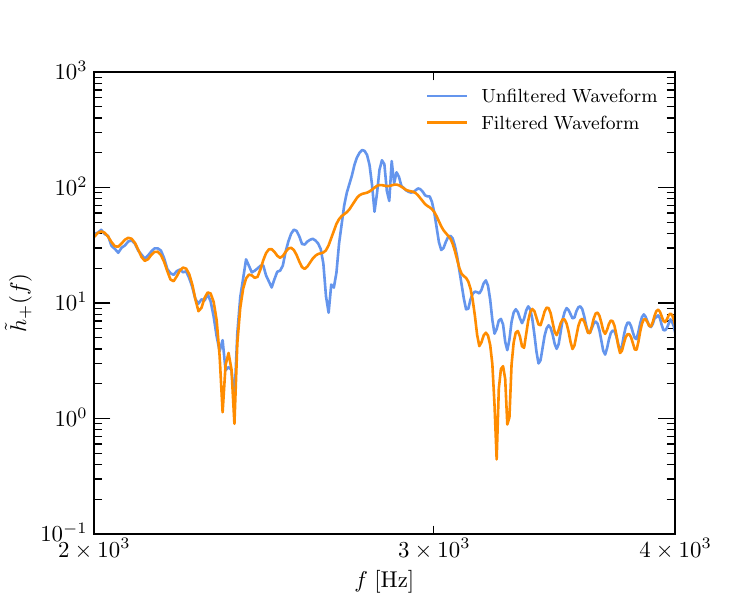}
	\includegraphics[width=0.49\textwidth]{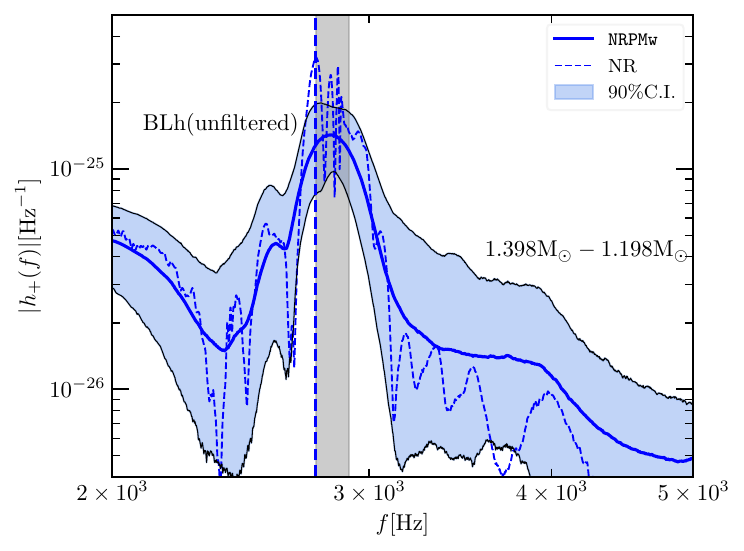}
	\includegraphics[width=0.49\textwidth]{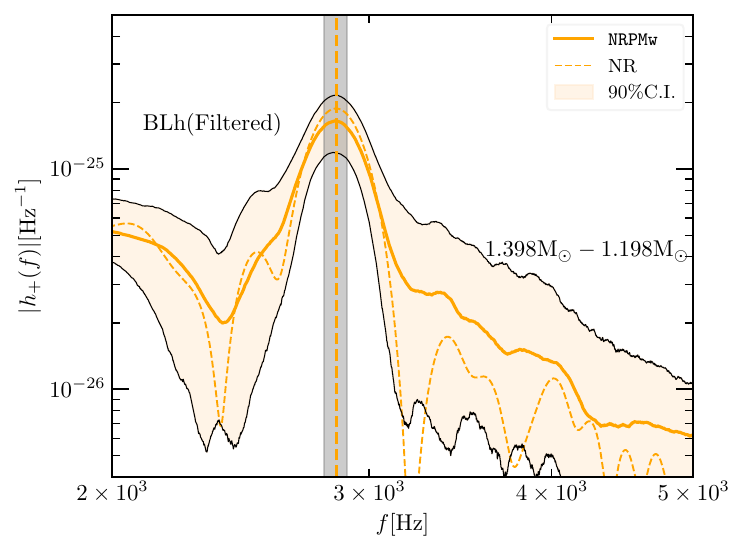}
	\caption{\textit{Upper Left panel:} The time domain postmerger waveform for the BLh - $1.398 M_{\odot} - 1.198 M_{\odot}$ binary with and without an exponential filter. $\tt{NRPMw}$ works best for the early postmerger where it can capture the $1^{\rm{st}}$ two amplitude modulations that peak at $t_1$ and $t_3$ respectively. The model, as of now, cannot capture subsequent amplitude modulations. \textit{Upper Right Panel:} The frequency spectra of the corresponding unfiltered and filtered waveforms that show a shift in $f_2^{\mathrm{peak}}$ upon exclusion of amplitude modulations at $t-t_{\mathrm{merg}}\gtrsim 5 \;\mathrm{ms}$. \textit{Bottom Panel:} The reconstructed spectra for the unfiltered waveform (left) and the filtered waveform (right) that show the bias in $f_2^{\mathrm{peak}}$'s measurement because of the $3^{rd}$ and subsequent amplitude modulations. We see that upon filtering these modulations, the model is able to capture the $f_2^{\mathrm{peak}}$ to within 90\% CIs.}
	\label{fig:multiple_amplitude modulations}
\end{figure*}

A characteristic feature of our hadronic simulations, in particular the binaries $1.398 M_{\odot} - 1.198 M_{\odot}$ (BLh), $1.289 M_{\odot} - 1.289 M_{\odot}$ (DD2F) and $1.298 M_{\odot} - 1.298 M_{\odot}$ (BLh), is the existence of multiple amplitude modulations in the $\ell=2, m=2$ mode of the GW strain. From our NR simulations of hadronic EOSs, we note that these modulations are typically observed in the early postmerger signal, i.e., when the remnant has just formed and undergoes large dynamical deformations resulting in amplitude-modulated GW emissions. $\tt{NRPMw}$, as of now, is unable to capture multiple modulations in the postmerger amplitude and attempts to reconstruct amplitude modulations beyond 2 via damped sinusoids. This leads to a biased overestimation of $f_2^{\mathrm{peak}}$ as is evidenced in Figs. \ref{fig:Inspiral_Informed_PE_f2_k2T_5_6}, \ref{fig:Inspiral_Informed_PE_f2_k2T_1234} and \ref{fig:Inspiral_Informed_PE_f2_k2T_78910}. 

In this section, we explore in detail the major source of this systematic bias, i.e., the multiple amplitude modulations. In Fig. \ref{fig:multiple_amplitude modulations}, we show the time domain waveform for the $1.398 M_{\odot} - 1.198 M_{\odot}$ binary with the BLh EOS that exhibits multiple amplitude modulations. In line with the convention for nodal points presented in ref.~\cite{Breschi:2022xnc}, we denote positions of the merger as $t_{\mathrm{merg}}$, the $1^{st}$ two postmerger maxima as $t_1$ and $t_3$ and their corresponding intermediate minima as $t_0$ and $t_2$. $\tt{NRPMw}$ only includes amplitude modulations until $t_3$ beyond which the amplitude is described by a damped sinusoid. We introduce an exponential filtering function 
\begin{equation}
	F(t) = 1/(1+\mathrm{exp}(t-t_{\mathrm{cutoff}}))
	\label{eq:filter}
\end{equation} where $t_{\mathrm{cutoff}}$ denotes the point where the filter cuts off the strain. 
We take $t_{\mathrm{cutoff}}$ to be near the position of the third amplitude modulation and filter off the subsequent signal to disentangle the effects of subsequent modulations. In the right panel of Fig.~\ref{fig:multiple_amplitude modulations}, we show the frequency spectra of this filtered waveform against that of the unfiltered waveform. We note that the subsequent amplitude modulations for $t-t_{\mathrm{merg}}\gtrsim 5\;\mathrm{ms}$ lead to multiple oscillations near $f_2^{\mathrm{peak}}$. Such closely-spaced oscillations in the frequency domain are not a morphology that can be captured by $\tt{NRPMw}$ and the model tends to construct an average over these peaks leading to a bias. On the other hand, when such modulations have been filtered out, the model captures the peak of the filtered spectra much better.

We also refer the reader to appendix \ref{Unconstrained_f2_f0_inference} for a brief discussion on how one can start to mitigate this bias by modifying the $\tt{NRPMw}$ model and making it more flexible to capture multiple modulations.  

We would like to emphasize that even though we have shown for our hadronic systems that removing multiple (>2) amplitude oscillations can help remove biases in the measurement of $f_2^{\rm peak}$, we only report results from the unfiltered, i.e., complete waveforms for the purposes of detecting phase transitions. This is because, in a realistic detection scenario, it is rather artificial to engineer the waveforms to support a recovery of $f_2^{\rm peak}$ to within some confidence level. Additionally, we would like to emphasize that there is still scope for improvement in the contemporary BNS waveform models to capture the above-mentioned morphologies and that additional avenues apart from shifts in $f_2^{peak}$ need to be explored for a holistic examination of QCD phase transition effects.

%%%%%%%%%%%%%%%%%%%%%%%%%%%%%%%%%%%
%%%%%%% POSTMERGER PE WITH CE %%%%%%%%%%%%%%
%%%%%%%%%%%%%%%%%%%%%%%%%%%%%%%%%%%

\subsection{Postmereger PE with Cosmic Explorer}
\label{subsec:Postmerger_PE_with_CE}

In this subsection, we repeat the inspiral informed PE analysis described in \ref{subsec:Inspiral_Informed_Postmerger_PE} on the $1.289 M_{\odot} - 1.289 M_{\odot}$ binaries with the DD2F and DD2F-SF1 EOSs but with a difference that now we employ the Cosmic Explorer sensitivities for recovering our models. The configurations we employ are a broad-band 40~km detector and a narrow-band 20~km detector which has been optimized for postmerger and has increased sensitivity in the 2-4 $\mathrm{kHz}$ regime. The advantage of the enhanced sensitivities of the CE-20 detector is that for the same postmerger SNR of 10, we will observe more distant and therefore more frequent mergers. We inject the $\tt{TaylorF2}$ predicted inspiral for the hadronic model $1.289 M_{\odot} - 1.289 M_{\odot}$ with the DD2F EOS in the broad-band CE-40 configuration. This binary is now placed at a distance of 118.467 Mpc so as to produce a postmerger SNR of 10 in CE-20 configuration. This is because we would like to harness the sensitivities of the CE detectors most optimally. CE-40 has higher sensitivity at low frequencies corresponding to the inspiral signal and therefore it is utilized for estimating the mass and tidal parameters from an inspiral signal (as described in \ref{subsec:Inspiral_Informed_Postmerger_PE}). On the other hand, CE-20 has increased sensitivities in the kilohertz regime corresponding to the emission frequencies of the BNS remnant and hence is utilized for the postmerger PE. 

\begin{figure}
	\includegraphics[width=\columnwidth]{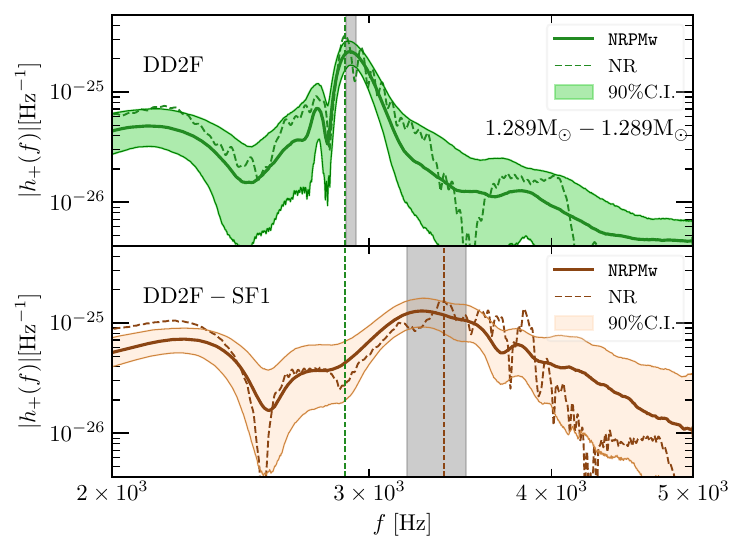}
	\caption{The reconstructed spectra corresponding to an inspiral informed postmerger PE for the binary $1.289 M_{\odot} - 1.289 M_{\odot}$ with the DD2F and DD2F-SF1 EOSs, computed with the postmerger optimized CE-20 detector. Like in the case of recovery from Einstein Telescope (Fig. \ref{fig:reconstructed_spectra_II_123478910}), here also we see that multiple amplitude modulations can bias the recovery of $f_2^{\mathrm{peak}}$ for DD2F.}
	\label{fig:reconstructed_spectra_CE_910}
\end{figure}

\begin{figure*}[t]
	\includegraphics[width=0.49\textwidth]{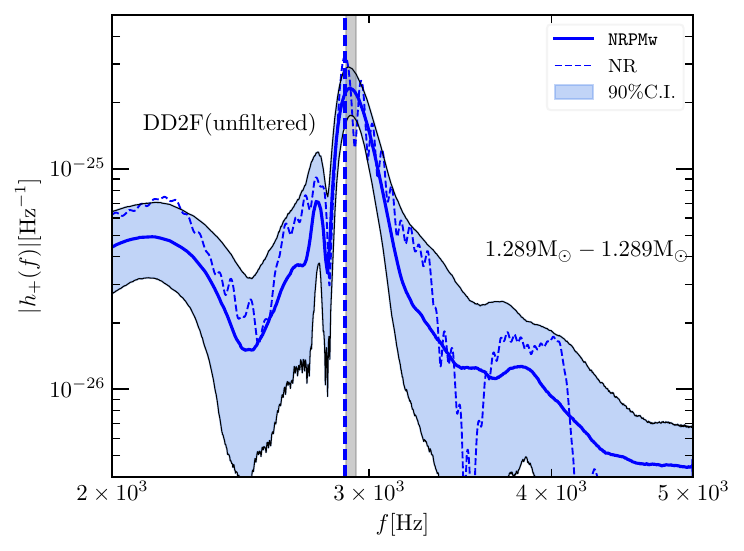}
	\includegraphics[width=0.49\textwidth]{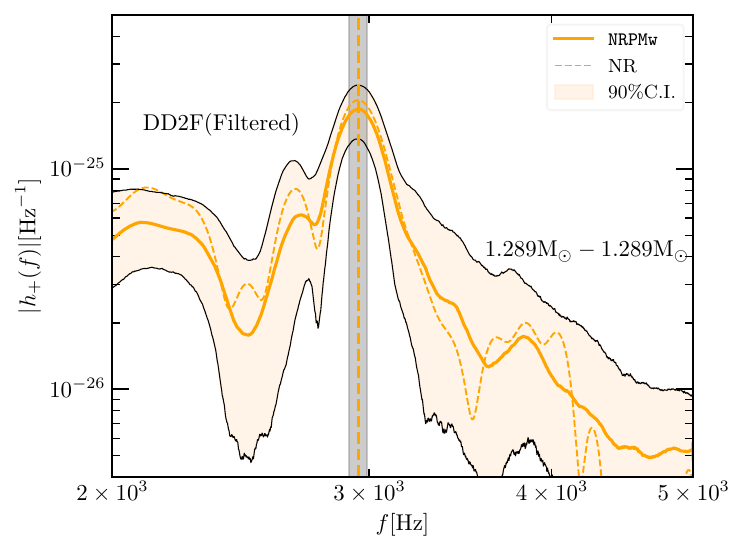}
	\caption{Same calculation as in fig. \ref{fig:multiple_amplitude modulations} to show the impact of multiple amplitude modulations on the recovery of $f_2^{\mathrm{peak}}$ for the hadronic DD2F simulation computed with the CE-20 detector. Here also we observe that filtering out the $3^{\rm{rd}}$ and subsequent modulation in the postmerger amplitude can result in an accurate recovery of $f_2^{\mathrm{peak}}$ to within 90\% CIs.}
	\label{fig:reconstructed_spectra_CE_filtered_910}
\end{figure*}

In Fig.~\ref{fig:reconstructed_spectra_CE_910}, we present a reconstruction of the postmerger amplitude spectrum recovered using the CE-20 detector by the $\tt{NRPMw}$ model. We see yet again that the measurement of $f_2^{\mathrm{peak}}$ for the hadronic model is overestimated due to the multiple amplitude modulations in the time-domain GWs from the hadronic model which we show explicitly in Fig. \ref{fig:reconstructed_spectra_CE_filtered_910}. The quality of the reconstruction of spectra and the accuracy of recovery of $f_2^{\rm peak}$ is similar in both the ET and CE-20 detectors with the only advantage being the increased rates of observation of BNS mergers with CE-20. 

So far we have demonstrated that the $\tt{NRPMw}$ model along with the sensitivities of the ET detector and the CE detector can reliably detect and reconstruct the postmerger signal which is evidenced by the fact that we recover most ($\approx$8) of the injected SNR (Table \ref{tab:Injection_Property_II}). In addition, we have shown that our model is capable of recovering (albeit with some bias) the $f_2^{\rm peak}$ frequency and distinguishing the same between the hadronic and quark models at a postmerger SNR of 10. We would like to emphasize that detecting and distinguishing the $f_2^{\rm peak}$ frequency is not sufficient for inferring the occurrence of a phase transition in a realistic observational setting. The latter requires quantifying violations from EOS insensitive relations (see subsection \ref{subsec:Probing_QCD_Phase_Transitions}). Since such relations involve inspiral tidal parameters in addition to the postmerger $f_2^{\rm peak}$, it is imperative that we have reliable estimates of the tidal parameters. In this regard, the utility of inspiral informed priors becomes clear. We can see that for all our models be it hadronic or quark, the 90\% CIs for $\kappa_2^T$ posterior by the $\tt{NRPMw}$ model contain the injected value. There exists no information about masses or the tidal properties from the postmerger signal alone (at least at a postmerger SNR of 10) and our postmerger model essentially recovers these priors. In contrast, with the priors that are agnostic of the inspiral signal, as in appendix \ref{Inspiral_Agnostic_PE:results_for_all_simulations}, the estimates of the tidal $\kappa_2^T$ are dominated by large errors which in turn will make an inference of QCD phase transitions difficult from the EOS insensitive quasi universal relations. 

%%%%%%%%%%%%%%%%%%%%%%%%%%%%%%%%%%%%%%%%%%%%%%%%%%%%
%%%%%%% Probing QCD Phase Transitions %%%%%%%%%%%%%%
%%%%%%%%%%%%%%%%%%%%%%%%%%%%%%%%%%%%%%%%%%%%%%%%%%%%

\subsection{Probing QCD Phase Transitions}
\label{subsec:Probing_QCD_Phase_Transitions}

\begin{figure*}[t]
\includegraphics[width=0.49\textwidth]{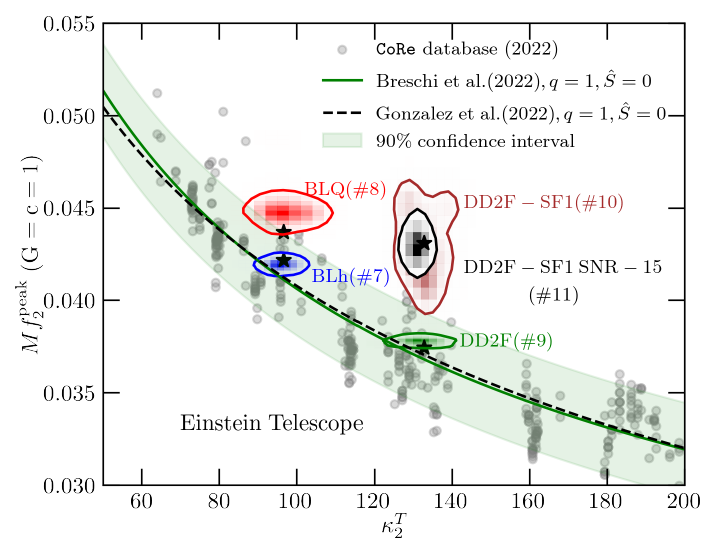}
\includegraphics[width=0.49\textwidth]{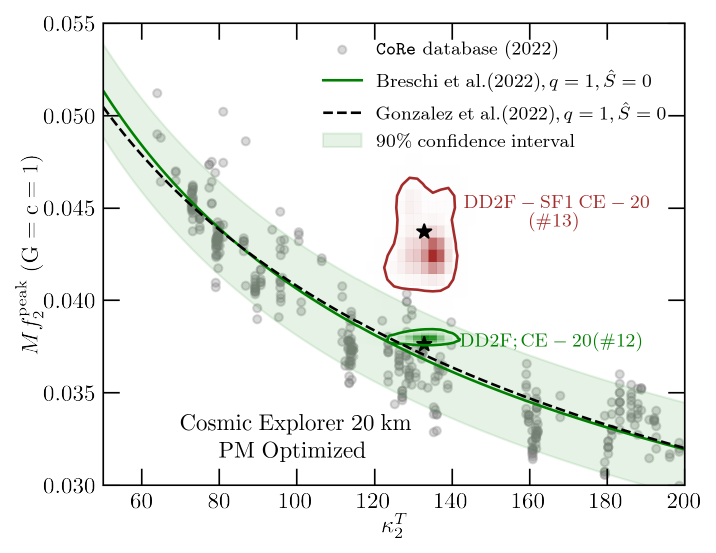}
\includegraphics[width=0.49\textwidth]{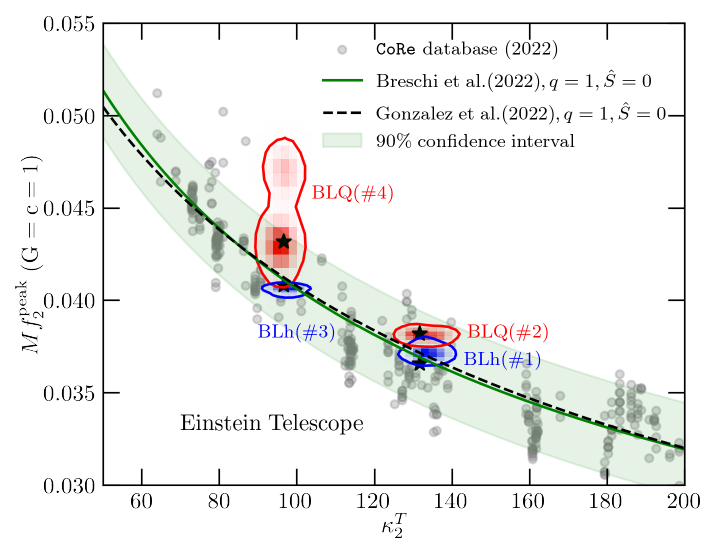}
\includegraphics[width=0.49\textwidth]{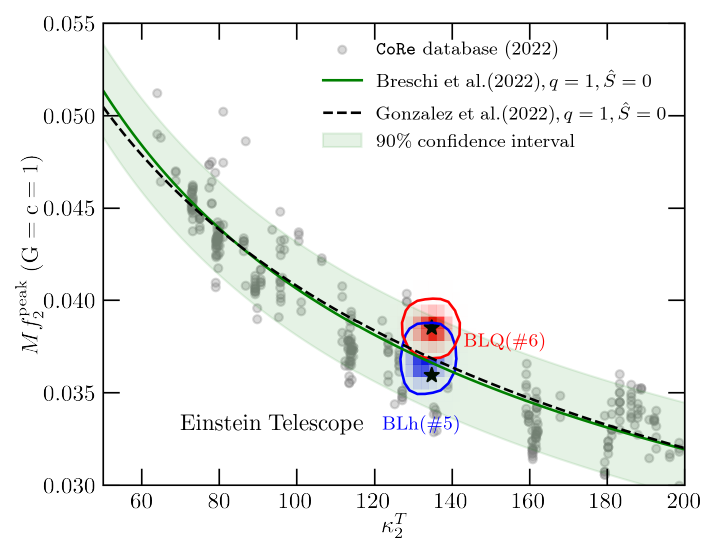}

\caption{Quasi Universal relation from Breschi et al.~\cite{Breschi:2022xnc} shown along with this work's Bayesian inference error estimates. Shown in grey scatter points, are the hadronic simulations from the $\tt{CoRe}$ database along with the fitting functions from \cite{Breschi:2022xnc} and \cite{Gonzalez:2022mgo} for non-spinning and symmetric binaries. The light green shaded region represents a 90\% confidence interval corresponding to the fit function from \cite{Breschi:2022xnc} which is also implemented in the $\tt{NRPMw}$ model. Even though \cite{Gonzalez:2022mgo} updates upon the fit coefficients in \cite{Breschi:2022xnc}, the two are within 90\% confidence of each other. Black stars denote the injected values in a 2D parameter space of $Mf_2$ and $\kappa_2^T$. The colored shaded regions represent the 90\% contours of the 2D joint posteriors on $Mf_2$ and $\kappa_2^T$ obtained in this study. In parenthesis we depict the simulation index of the binaries as defined in Table \ref{tab:Injection_Property_II}. \textit{Top Panels:} We show binaries which are non-degenerate with respect to each other upto 90\% CIs and with the universal relation. \textit{Bottom Panels:} We show models which are not mutually distinguishable to 90 \%.}
\label{fig:QUR}
\end{figure*}

As we have previously remarked, detection of a postmerger signal and a reliable recovery and distinguishability of $f_2^{\rm peak}$ is necessary but not sufficient for probing QCD phase transitions. Previous works \cite{Bauswein:2018bma, Bauswein:2020ggy, Breschi:2023mdj} suggest the utility of EOS insensitive relations, in particular between $f_2^{\rm peak}$ and a tidal parameter be it $\tilde{\Lambda}$ or $\kappa_2^T$, in probing such phase transitions. Specifically, if EOS softening effects by such phase transitions produce deviations from the aforementioned relations that are non-degenerate with other hadronic models, one can ascertain the occurrence of a phase transition with some confidence. This requires that we have reliable estimates of not just $f_2^{\rm peak}$ but also of tidal properties. Comparing NR simulations of the postmerger signal at different resolutions can only provide the former as a one-dimensional error estimate because tidal properties are fixed upon assuming a specific equation of state. The only way we can compute a joint uncertainty of $f_2^{\rm peak}$ and $\kappa_2^T$ is by Bayesian inference of the postemrger signal which is informed of the tidal properties from the inspiral (and of course a Bayesian inference on the full signal). 

We employ the fitting function obtained in ref.~\cite{Breschi:2022xnc} with reference to the \href{http://www.computational-relativity.org/gwdb/}{$\tt{CoRe}$} database. This fitting function improves upon the QUR obtained in refs.~\cite{Breschi:2019srl} by explicitly including the effects of inspiral spins and taking into account additional GRHD simulations performed with $\tt{WhiskyTHC}$ and $\tt{BAM}$ infrastructures. In Fig. \ref{fig:QUR}, we plot the QUR fitting function for symmetric binaries that are non-spinning, along with an ensemble of simulations that form a part of the $\tt{CoRe}$ database. We also show the 90\% confidence levels for the fit describing symmetric binaries. To this collection, we add the injections presented in this work with their error estimates that are essentially the 90\% contour levels of the 2-dimensional joint posteriors for mass-rescaled $Mf_2^{\rm peak}$ and $\kappa_2^T$ obtained with the choice of inspiral informed priors. 

In subsection \ref{subsec:Inspiral_Informed_Postmerger_PE} and appendix \ref{Inspiral_Informed_PE:results_for_all_simulations}, we have provided evidence for mutual distinguishability between hadronic and quark models based on the non-degeneracy of the 90\% CIs of the $f_2^{\rm peak}$ posteriors. In this section, we present a discussion on detecting phase transitions based on non-degeneracies between the \textbf{joint} $f_2^{\rm peak} - \kappa_2^T$ posteriors and comparing them with the EOS insensitive relation of Breschi et al.~\cite{Breschi:2022xnc}. In the first (upper-left) panel of Fig.\ref{fig:QUR}, we present hadronic and quark models that, at a postmerger SNR of 10, are mutually distinguishable as is seen by the absence of any overlap between the corresponding joint $Mf_2^{\rm peak} - \kappa_2^T$ posteriors. These models are the $1.363 M_{\odot} - 1.363 M_{\odot}$ binary for the BLh and BLQ EOSs and the $1.289 M_{\odot} - 1.289 M_{\odot}$ binary for the DD2F and DD2F-SF1 EOSs. For the $1.363 M_{\odot} - 1.363 M_{\odot}$ binary, we notice that even though the hadronic and quark models are distinguishable (up to 90\% confidence), the quark model's joint posterior is degenerate with other hadronic EOSs, implying that a postmerger SNR of 10, we cannot conclusively confirm a phase transition for this binary. On the other hand, for the $1.289 M_{\odot} - 1.289 M_{\odot}$ binary with the DD2F-SF1 EOSs, we notice that the injection and the corresponding joint posteriors do not overlap with the universal relation, implying that at a postmerger SNR of 10, we can confirm the presence of a phase transition. We do however caution the reader about a possible caveat. The conclusion that whether we can confirm a phase transition to some confidence is sensitive to the particular universal relation used. The 90\% contours of the joint posterior with DD2F-SF1 EOS, even though not overlapping with the universal relation's error margin, are very close to them and systematics in the universal relation may change our conclusions. Such systematics may result from updating the coefficients of the fit upon adding more simulations. At higher postmerger SNRs, detectability avenues will improve. This is made concrete by an additional model recovered at a higher postmerger SNR of 15, where we find that the DD2F-SF1 model's joint posteriors shrink and are even more removed from the universal relation than the same model at postmerger SNR 10. 

Similarly, in the second panel (top right) of the figure \ref{fig:QUR}, we repeat the calculations for the case of $1.289 M_{\odot} - 1.289 M_{\odot}$ binary with the DD2F and DD2F-SF1 EOS, assuming the Cosmic Explorer (CE-20) detector sensitivity. We notice here that the joint posteriors corresponding to the quark EOS are "more" non-degenerate with the universal relation as compared to the same model recovered from the Einstein Telescope sensitivity. Consequently, at a postmerger SNR of 10, we can confirm the presence of a phase transition. The better performance of the CE-20 detector as compared to the Einstein Telescope's recovery, is not entirely unexpected. Indeed we note that for the injected $f_2^{\rm peak}$ frequencies close to 3 kHz, the CE-20 detector is more sensitive than the ET detector. 

On the other hand, in the second (bottom left) and third (bottom right) panels of figure \ref{fig:QUR}, we show binaries for which the 90\% contours of the joint posterior overlap between the hadronic and quark models. These models include the $1.298 M_{\odot} - 1.298 M_{\odot}$, $1.481 M_{\odot} - 1.257 M_{\odot}$ and $1.398 M_{\odot} - 1.198 M_{\odot}$ binaries with the BLh and BLQ EOSs. For the quark models of these binaries, at a postmerger SNR of 10, the presence of a QCD phase transition cannot be ascertained given the degeneracy with other hadronic EOSs. 

\begin{figure*}[t]
	\includegraphics[width=0.49\textwidth]{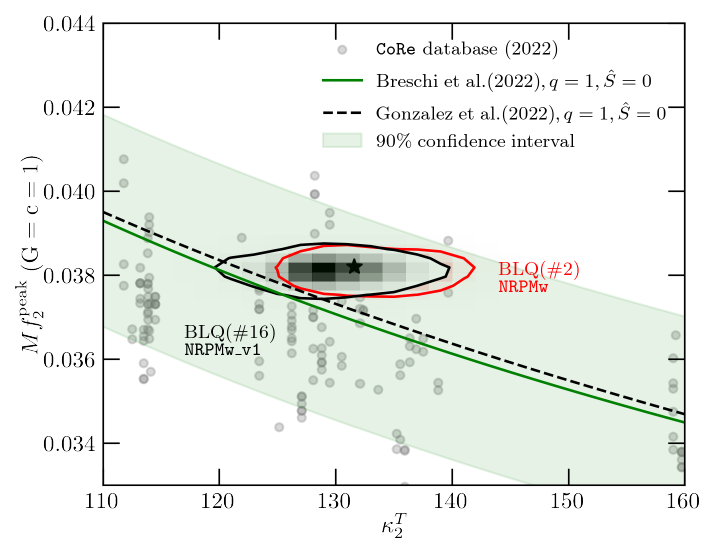}
	\includegraphics[width=0.49\textwidth]{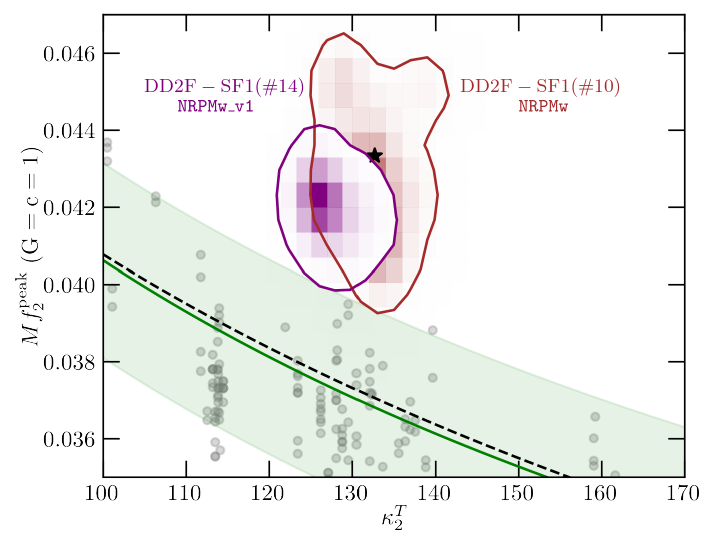}
	\caption{A comparison of the joint inference of $f_2^{\rm peak}$ and $\kappa_2^T$ between two configurations of $\tt{NRPMw}$, i.e., excluding the $f_2^{\rm peak} - \kappa_2^T$ universal relation ($\tt{NRPMw}$) and including the $f_2^{\rm peak}$ universal relation ($\tt{NRPMw\_v1}$). \textit{Left Panel:} Results for the $1.298 M_{\odot}- 1.298 M_{\odot}$ binary with the BLQ EOS whose injection follows the universal relation. Both the model configurations can recover the injection with the data  (injection) slightly preferring the QUR informed model $\tt{NRPMw\_v1}$. \textit{Right panel}: Results for the $1.289 M_{\odot} - 1.289 M_{\odot}$ binary with the DD2F-SF1 EOS whose injection strongly violates the universal relation. The data (injection) slightly prefers the more flexible QUR uninformed model ($\tt{NRPMw}$) as it has larger flexibility than $\tt{NRPMw\_v1}$ in reference to recovering injections that violate the universal relation.}
	\label{fig:QUR_f2_free}
	\end{figure*}

Therefore, in a nutshell, even though postmerger waveforms from the hadronic models may be distinguishable from the corresponding quark models by virtue of non-degeneracy of $f_2^{\rm peak}$ posteriors, they may still be degenerate with each other in a two-dimensional space of $f_2^{\rm peak} - \kappa_2^T$ uncertainties. Furthermore, when there is no degeneracy in a joint measurement of $f_2^{\rm peak}$ and $\kappa_2^T$, a postmerger SNR of 10 can confirm the presence of a phase transition only if the model violates the universal relation strongly, i.e., $\Delta f_2^{\rm peak} \approx 455$ Hz ($\gtrsim 1.6 \;\sigma$). At a postmerger SNR of 10, systematics in the universal relation may also play a role in influencing conclusions about the detectability of phase transitions. However, for louder binaries with SNR $\sim 15$, phase transitions of the type predicted by the DD2F-SF1 model can be confirmed with a higher confidence. 

At this stage we also present a test of the EOS insensitive relations with reference to detecting QCD phase transitions in Fig.~\ref{fig:QUR_f2_free}. To this aim, we test two configurations of our model. Firstly, we use the $\tt{NRPMw}$ configuration employed throughout this work where the $f_2$ parameter is independent of the $f_2 - \kappa_2^T$ universal relation from Breschi et al.~\cite{Breschi:2022xnc}, i.e., the universal relation is ignored. Secondly, we employ the original model configuration of Breschi et al. (called $\tt{NRPMw\_v1}$ in this work) where $f_2$ is decided by the universal relation. In particular, we are posing the question that given a signal, whether the inclusion of the $f_2 - \kappa_2^T$ universal relation in the model can play a role in detecting a "strong" phase transition. In the left panel of Fig.\ref{fig:QUR_f2_free}, we present results for the $1.298 M_{\odot}- 1.298 M_{\odot}$ binary with the BLQ EOS whose injection is consistent with the universal relation. We note that for both model configurations, the 90\% contour of the joint PDF contains the injection at a postmerger SNR of 10. To quantify this comparison we compute the Bayes' factor for the two hypotheses, i.e., inference with and without the universal relation respectively. We find that $\rm{log}\;\mathcal{B}^{\rm{with\;QUR}}_{without\;QUR} = 2.53^{+0.27}_{-0.27}$ indicating a weak preference towards the QUR informed $\tt{NRPMw\_v1}$ model. On the other hand, in the right panel of Fig.~\ref{fig:QUR_f2_free}, we present the same calculation for the $1.289 M_{\odot} - 1.289 M_{\odot}$ binary with the DD2F-SF1 EOS. Since in this case, the injection is inconsistent with the universal relation, including the same in the model tends to drive the joint posterior toward the universal relation and away from the injection. This is evidenced by the fact that the 90\% contour of the $\tt{NRPMw\_v1}$ model does not contain the injection whereas the injection is well captured within the joint posterior of the more flexible $\tt{NRPMw}$ model. To quantify the same statement, the $\rm{log}\;\mathcal{B}^{\rm{without\;QUR}}_{with\;QUR} = 2.24^{+0.27}_{-0.27}$ at a postmerger SNR of 10 indicating a weak preference towards the more flexible $\tt{NRPMw}$ model with respect to detecting phase transitions that strongly violate the universal relations.

%%%%%%%%%%%%%%%%%%%%%%%%%%%%%%%%%%%
%%%%%%% CONCLUSIONS %%%%%%%%%%%%%%
%%%%%%%%%%%%%%%%%%%%%%%%%%%%%%%%%%%

\section{Conclusions}
\label{sec:Conclusions}

In this work we have demonstrated with the help of Bayesian inference, an avenue to probe QCD phase transitions using the postmerger GW emission from BNS merger remnants. We have considered NR simulations with substantial diversity in masses and treatments of quark matter. Equipped with the enhanced kilohertz sensitivities of the next-generation GW detectors along with a frequency domain waveform model $\tt{NRPMw}$, we have shown that it is possible to reliably detect strong phase transiitons at postmerger SNRs as low as 10. 

To model the influence of deconfined quarks on the dynamics of BNS merger remnants, we employ the BLQ and DD2F-SF1 EOSs which model the deconfined quark phase by Gibbs construction and Maxwell's construction respectively. In the case of a merger, these treatments lead to remnants with very different properties, most notably differences in the postmerger peak frequencies. We construct the postmerger signals by windowing out the inspiral signal from our NR waveforms and injecting the signal thus obtained in a noise-less configuration of ET or CE detectors. 

We perform independent Bayesian inference calculations on the inspiral and the postmerger signals using the $\tt{Bilby}$ (via the $\tt{TaylorF2}$ model) and $\tt{bajes}$ (via the $\tt{NRPMw}$ model) codes, respectively. We compute the posteriors of total mass, mass ratio and tidal deformabilities which are expectedly Gaussian to a good approximation (except for $q=1$ biases). These posteriors help inform the priors for the postmerger PE analysis which provides the posteriors on $f_2^{\mathrm{peak}}$. We find that $\tt{NRPMw}$ model can reliably recover the postmerger signal as is evidenced by the recovered SNRs (Table \ref{tab:Injection_Property_II} and Table \ref{tab:Injection_Property_IA}). Additionally, at a postmerger SNR of 10, the model can also recover the $f_2^{\rm peak}$ frequency and distinguish the same between a hadronic and quark model to upto 90\% confidence. 

Our work also serves to present new test cases to which our waveform model has been applied as a means to evaluate its validity. We have presented for the first time, the behavior of the model in an inspiral-informed PE setting and tested its performance on morphologically complex NR waveforms. It is noteworthy that simulations from DD2F, DD2F-SF1 EOSs are also the ones that the model has not been trained on. For these cases too we get reliable signal reconstruction and recover most of the SNR.

We have provided a complimentary analysis by employing the CE-40 and CE-20 detectors. The advantage of utilizing the CE detectors for this purpose is two fold. First, with enhanced postmerger sensitivities, BNS mergers can be probed at larger luminosity distances and hence more frequently. Second, a combination of broad-band CE-40 detector and a narrow band postmerger optimized CE-20 detector is optimal for a holistic detection because of increased sensitivities in the inspiral (by CE-40)  and the postmerger (by CE-20). For sources with postmerger SNR of 10 in CE-20, we have used the CE-40 detector to compute posteriors on masses and tidal parameters which serve as priors on the postmerger PE analysis via the CE-20 detector. We find no major differences in the inference of $f_2^{\rm peak}$ or the quality of signal reconstruction as compared to inference with the ET detcetor.

We emphasize that even though $\tt{NRPMw}$ coupled with the enhanced sensitivities of the upcoming generation of GW detectors, can reliably detect and distinguish the $f_2^{\rm peak}$ frequencies at a postmerger SNR of 10, it is not sufficient to probe QCD phase transitions. We compare the joint posterior estimates on $f_2^{\rm peak}$ and $\kappa_2^T$ in reference to the $f_2 - \kappa_2^T$ universal relation from Breschi et al.~\cite{Breschi:2022xnc} and find that starting at postmerger SNRs of 10, we can claim a detection of a first order phase transition but only for models that violate the universal relations by more than 1.6 $\sigma$. We also demonstrate a slight preference towards the model configuration which is independent of the universal relation in detecting "strong" phase transitions by a $\rm{log}\;\mathcal{B}^{\rm{without\;QUR}}_{with\;QUR} = 2.24^{+0.27}_{-0.27}$.

For final remarks, Bayesian inference is done on waveforms that have a rich morphological structure and therefore we speculate that indicators of QCD phase transitions may not be exclusively encoded in $f_2^{\mathrm{peak}}$. This warrants exploration of alternative signatures of phase transitions, e.g., imprints in the postmerger amplitude or the lifetimes of remnants. Our work calls for efforts in several directions. First, as we have shown, the current waveform models need to be improved to take into account additional waveform morphologies like multiple amplitude modulations which can be a significant source of bias at high enough SNRs. Additionally, even though $\tt{NRPMw}$ does include a prescription for modelling the black hole ringdown, we have omitted the same in favor of ease of computation. The ringdown spectrum and the ensuing quasi-normal modes can be important for constraining QCD phase transitions from short lived remnants or promptly collapsing binaries where such phase transitions can play a role \cite{Perego:2021mkd, Kashyap:2021wzs, Bauswein:2020aag}. Second, the universal relations can themselves involve systematic biases which can be sourced from uncertainties in the physics modelled in the simulations. Such biases may shift the universal relations in the $f_2^{\rm peak} - \kappa_2^T$ plane affecting conclusions about the occurrence of phase transitions. Lastly, improvements are required in improving the postmerger convergence of contemporary NR codes \cite{Espino:2022mtb} as will be required by large SNR detections from the next generation detectors. Overall, the prospects of detecting a QCD phase transition with the enhanced sensitivities of the upcoming detectors, seem not too pessimistic. A single GW170817 like event, provided a postmerger is also observed, can in theory constrain QCD phase transitions.

%%%%%%%%%%%%%%%%%%%%%%%%%%%%%%%%%%%
%%%%%%% ACKNOWLEDGMENTS %%%%%%%%%%%%%%
%%%%%%%%%%%%%%%%%%%%%%%%%%%%%%%%%%%

\section{Acknowledgments}
AP would like to thank Alejandra Gonzales for providing the postmerger data from the second release of the $\tt{CoRe}$ database for the updated fits in Fig. \ref{fig:QUR}. AP would also like to thank Prof K.~G.~Arun and Dr.~Arnab Dhani for many useful discussions, a careful reading of the manuscript and their comments.

DR acknowledges funding from the U.S. Department of Energy, Office of
Science, Division of Nuclear Physics under Award Number(s) DE-SC0021177
and from the National Science Foundation under Grants No. PHY-2011725,
PHY-2020275, PHY-2116686, and AST-2108467.
Simulations were performed on PSC Bridges2, SDSC Expanse, TACC's Stampede2
(NSF XSEDE allocation TG-PHY160025).
This research used resources of the National Energy Research Scientific
Computing Center, a DOE Office of Science User Facility supported by the
Office of Science of the U.S.~Department of Energy under Contract
No.~DE-AC02-05CH11231. Computations for this research were performed on the Pennsylvania State University’s Institute for Computational and Data Sciences’ Roar supercomputer.

SB acknowledge funding from from the EU Horizon under ERC Consolidator Grant, no. InspiReM-101043372.

\appendix

%%%%%%%%%%%%%%%%%%%%%%%%%%%%%%%%%%%%%%%%%%%%%%%%%%%%%%%%%%%%%%%%%%%%%%%%%%%%%%%%%%%%%%%%%%%
%%%%%%% Appendix: Inspiral Agnostic PE: results for all simulations %%%%%%%%%%%%%%
%%%%%%%%%%%%%%%%%%%%%%%%%%%%%%%%%%%%%%%%%%%%%%%%%%%%%%%%%%%%%%%%%%%%%%%%%%%%%%%%%%%%%%%%%%%

\section{Inspiral Agnostic PE: results for all simulations}
\label{Inspiral_Agnostic_PE:results_for_all_simulations}

\begin{table*} 
\caption{Same properties as presented in table \ref{tab:Injection_Property_II} but now for the choice of inspiral agnostic priors.} 
\label{tab:Injection_Property_IA} 
 \begin{center} {
 \begin{tabular}{c c c c c c c c c c c} 
 \hline\hline 
 \\ 
$\mathrm{Index}$ & $\mathrm{EOS}$ & $m_1\;[M_{\odot}]$ & $m_2\;[M_{\odot}]$  & $\mathrm{GW\;Model}$ & $\mathrm{Detector}$ & $f_{2;\mathrm{Injected}}^{\mathrm{peak}}\;[\mathrm{kHz}]$ & $f_{2;\;\mathrm{Recovered}}^{\mathrm{peak}}\;[\mathrm{kHz}]$ & $D_\ell\;[\mathrm{Mpc}]$ & $\rho_{\mathrm{injected}}$& $\rho_{\mathrm{recovered}}$   \\ 
\hline
\hline

1 & $\mathrm{BLh}$ & 1.298 & 1.298 & $\tt{NRPMw}$ & ET & 2.804 & $2.825^{+0.033}_{-0.025}$ & 89.049 & 10 & $8.88^{+1.07}_{-1.59} $\\
2 & $\mathrm{BLQ}$ & 1.298 & 1.298 & $\tt{NRPMw}$ & ET & 2.927 & $2.924^{+0.025}_{-0.025}$ & 93.474 & 10 & $8.64^{+1.08}_{-1.61} $\\
\hline
3 & $\mathrm{BLh}$ & 1.481 & 1.257 & $\tt{NRPMw}$ & ET & 2.962 & $2.957^{+0.016}_{-0.025}$ & 97.503 & 10 & $8.79^{+1.12}_{-1.47} $\\
4 & $\mathrm{BLQ}$ & 1.481 & 1.257 & $\tt{NRPMw}$ & ET & 3.143 & $3.284^{+0.139}_{-0.246}$ & 83.434 & 10 & $8.72^{+1.15}_{-1.11} $\\
\hline
5 & $\mathrm{BLh}$ & 1.398 & 1.198 & $\tt{NRPMw}$ & ET & 2.758 & $2.842^{+0.098}_{-0.09}$ & 87.027 & 10 & $7.89^{+0.88}_{-1.15} $\\
6 & $\mathrm{BLQ}$ & 1.398 & 1.198 & $\tt{NRPMw}$ & ET & 2.955 & $2.973^{+0.033}_{-0.033}$ & 87.500 & 10 & $8.66^{+1.22}_{-1.43} $\\
\hline
7 & $\mathrm{BLh}$ & 1.363 & 1.363 & $\tt{NRPMw}$ & ET & 3.073 & $3.055^{+0.025}_{-0.033}$ & 97.282 & 10 & $7.97^{+1.19}_{-1.52} $\\
8 & $\mathrm{BLQ}$ & 1.363 & 1.363 & $\tt{NRPMw}$ & ET & 3.197 & $3.284^{+0.279}_{-0.147}$ & 78.449 & 10 & $8.13^{+1.21}_{-1.29} $\\
\hline
9 & $\mathrm{DD2F}$ & 1.289 & 1.289 & $\tt{NRPMw}$ & ET & 2.889 & $2.907^{+0.025}_{-0.025}$ & 93.284 & 10 & $8.51^{+1.12}_{-1.48} $\\
10 & $\mathrm{DD2F-SF1}$ & 1.289 & 1.289 & $\tt{NRPMw}$ & ET & 3.354 & $3.432^{+0.066}_{-0.139}$ & 78.247 & 10 & $8.60^{+1.38}_{-1.20} $\\
\hline 
\hline 
\end{tabular} 
} 
\end{center} 
\end{table*}

\begin{figure*}[t]
	\includegraphics[width=0.49\textwidth]{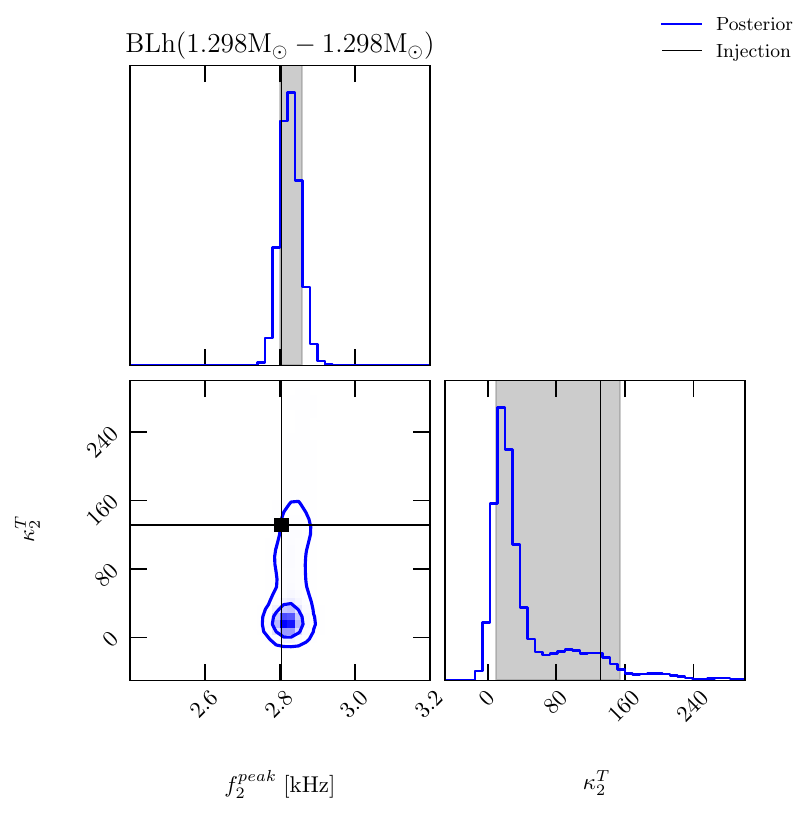}
	\includegraphics[width=0.49\textwidth]{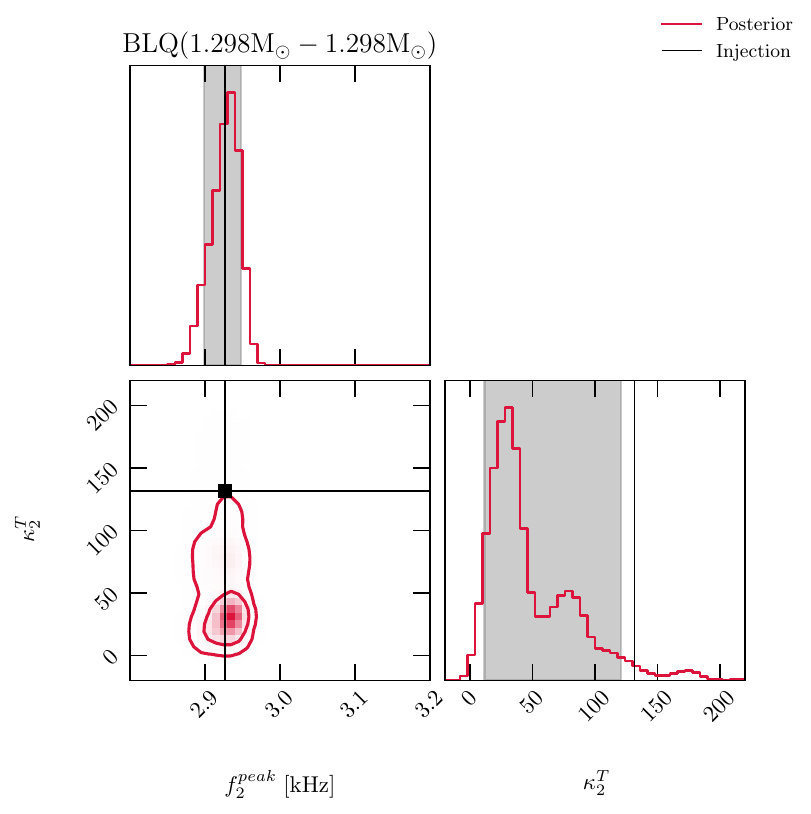}
	\includegraphics[width=0.49\textwidth]{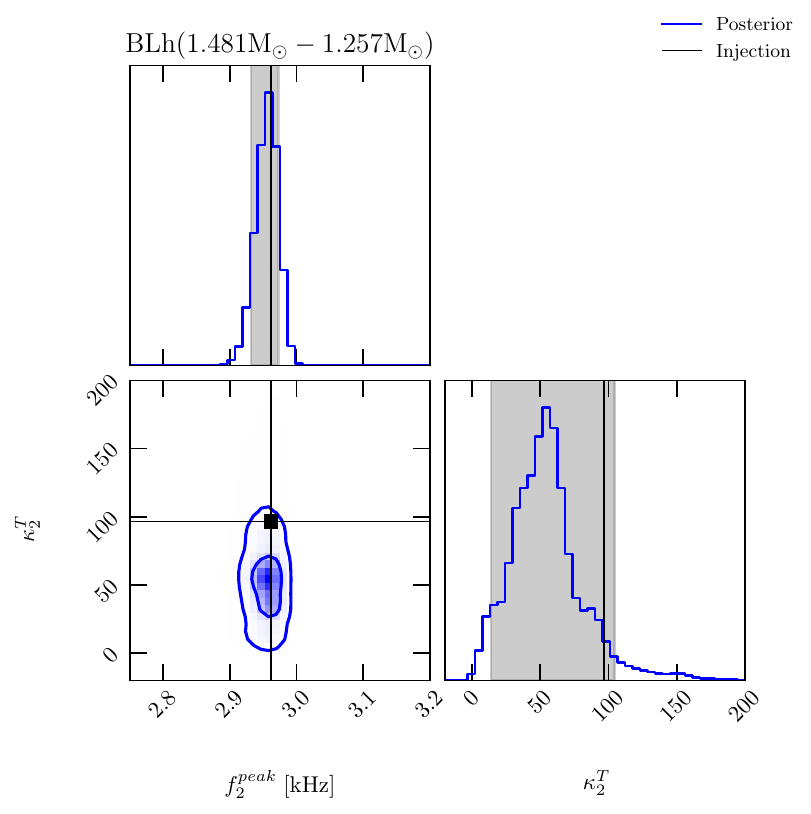}
	\includegraphics[width=0.49\textwidth]{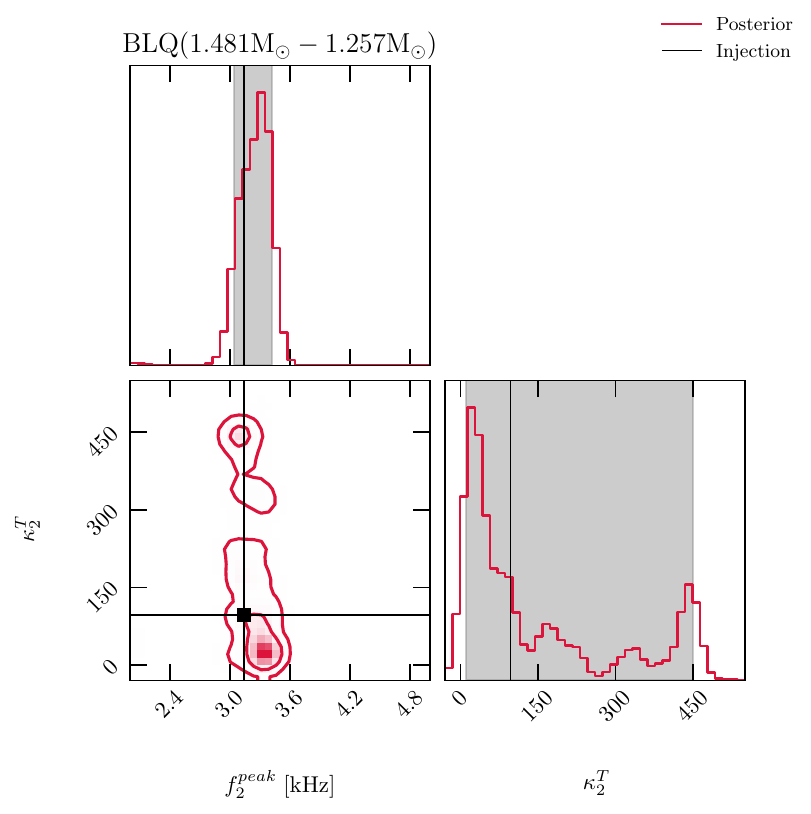}
	\caption{Same calculations as in Fig. \ref{fig:Inspiral_Informed_PE_f2_k2T_5_6}, i.e., a measurement of $f_2^{\rm peak}$ and $\kappa_2^T$ for the $1.298 M_{\odot} - 1.298 M_{\odot}$ and $1.481 M_{\odot} - 1.257 M_{\odot}$ binaries with the BLh and BLQ EOSs. In contrast to Fig. \ref{fig:Inspiral_Informed_PE_f2_k2T_5_6}, here we use a different choice of priors that are uninformed of the inspiral signal and set to wide ranges as described in Table \ref{tab:inspiral_agnostic_priors_table}. We note that the $\tt{NRPMw}$ model captures to within 90\% CIs the $f_2^{\mathrm{peak}}$ frequency for the quark and hadronic models however, the tidal polarizability $\kappa_2^{T}$ is poorly determined owing to the fact that no tidal information is present in the postmerger signal.}
	\label{fig:Inspiral_Agnostic_PE_f2_k2T_1234}
\end{figure*}

\begin{figure*}[t]
	\includegraphics[width=0.49\textwidth]{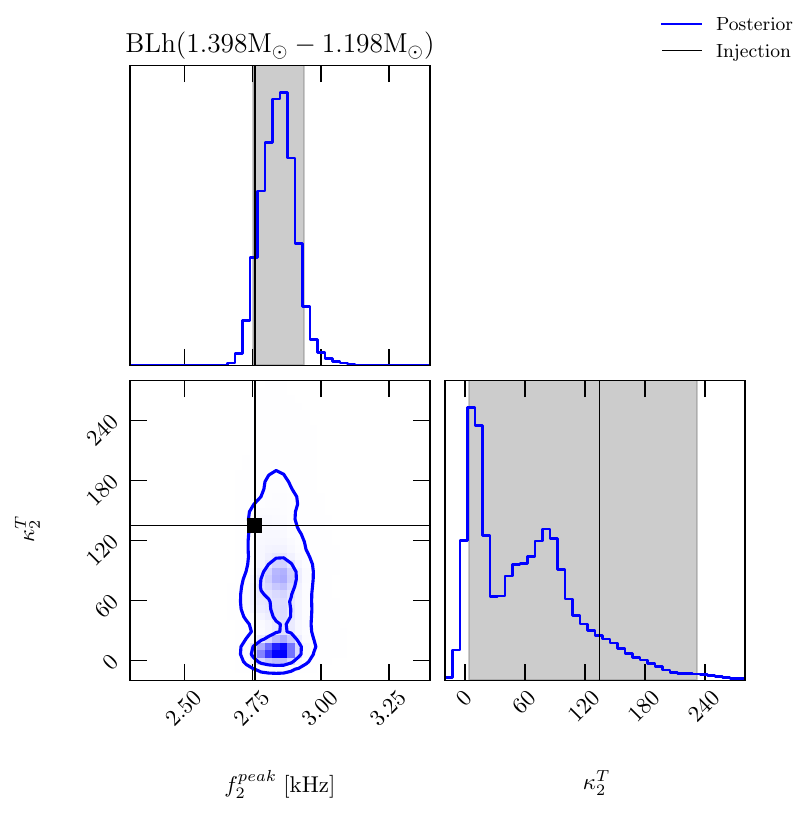}
	\includegraphics[width=0.49\textwidth]{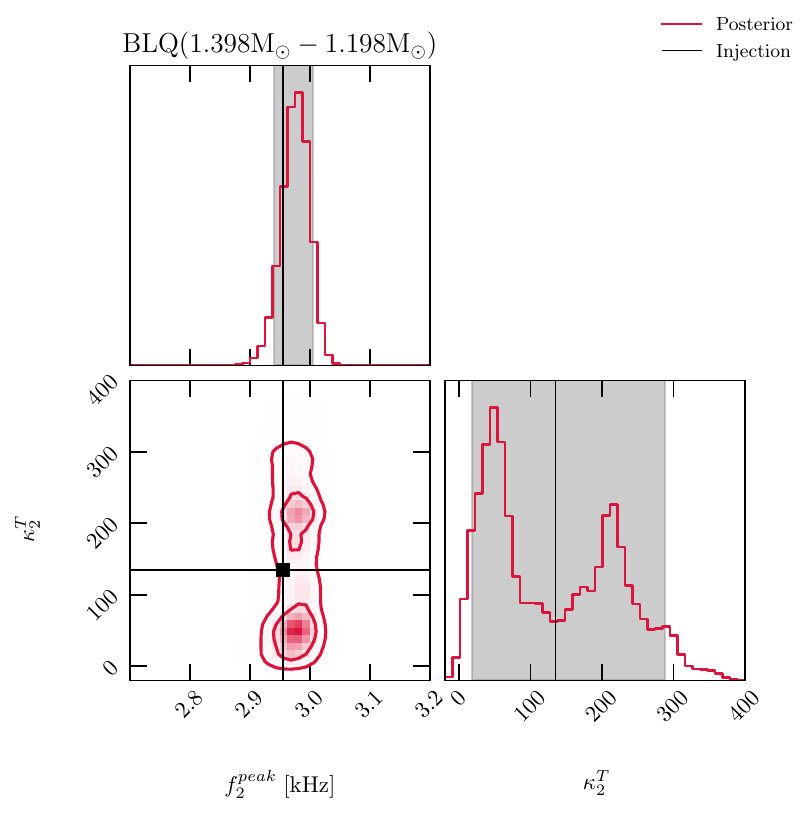}
	\caption{Same calculations as in Figure \ref{fig:Inspiral_Agnostic_PE_f2_k2T_1234} for the binary $1.398 M_{\odot} - 1.198 M_{\odot}$ simulated with the BLh and BLQ EOSs.}
	\label{fig:Inspiral_Agnostic_PE_f2_k2T_56}
\end{figure*}

\begin{figure*}[t]
	\includegraphics[width=0.49\textwidth]{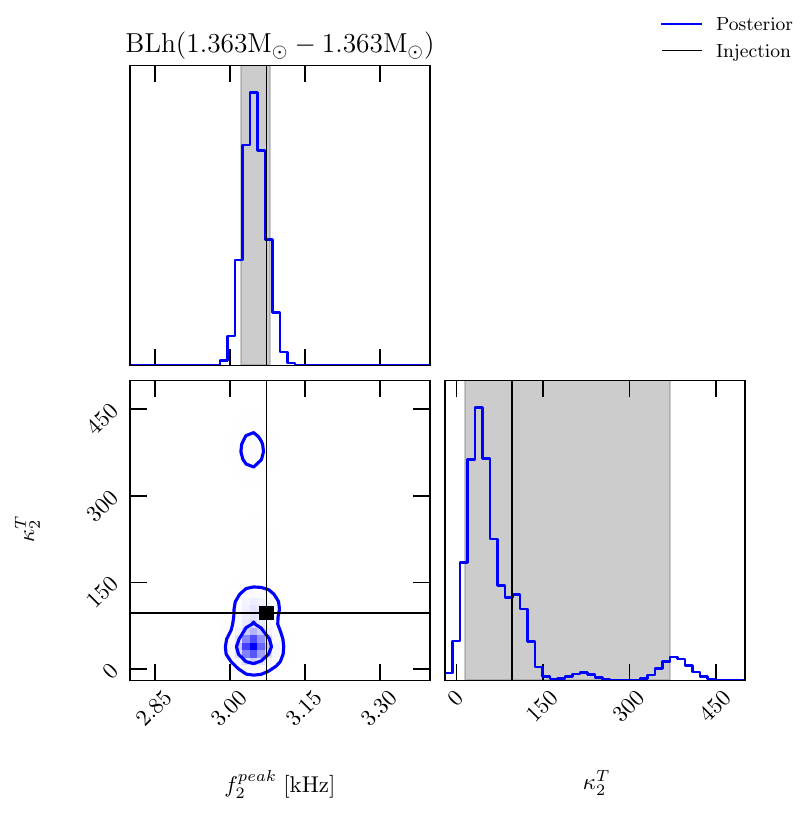}
	\includegraphics[width=0.49\textwidth]{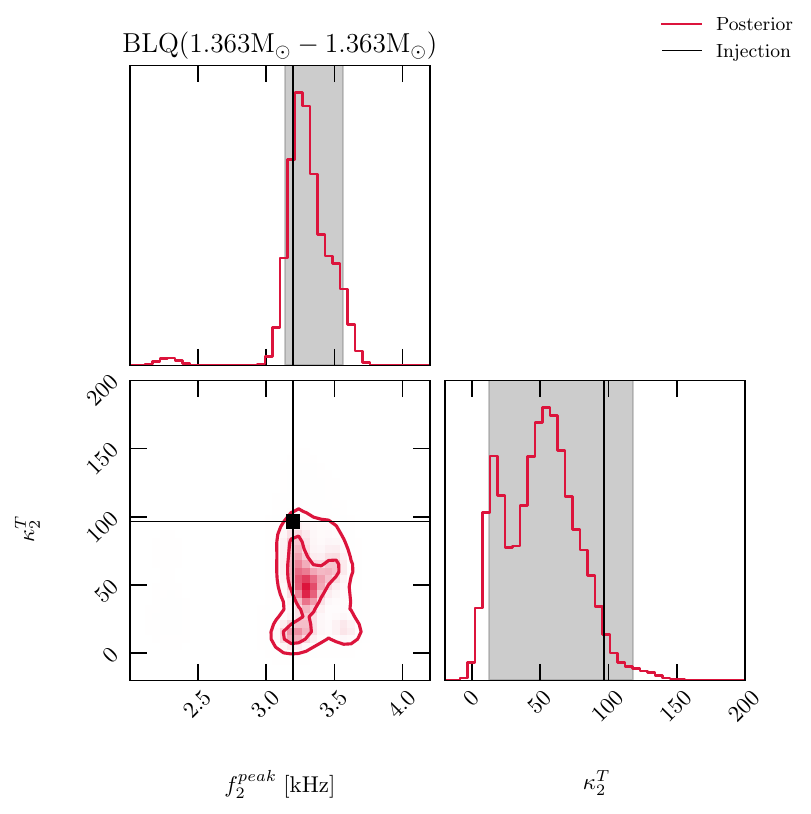}
	\includegraphics[width=0.49\textwidth]{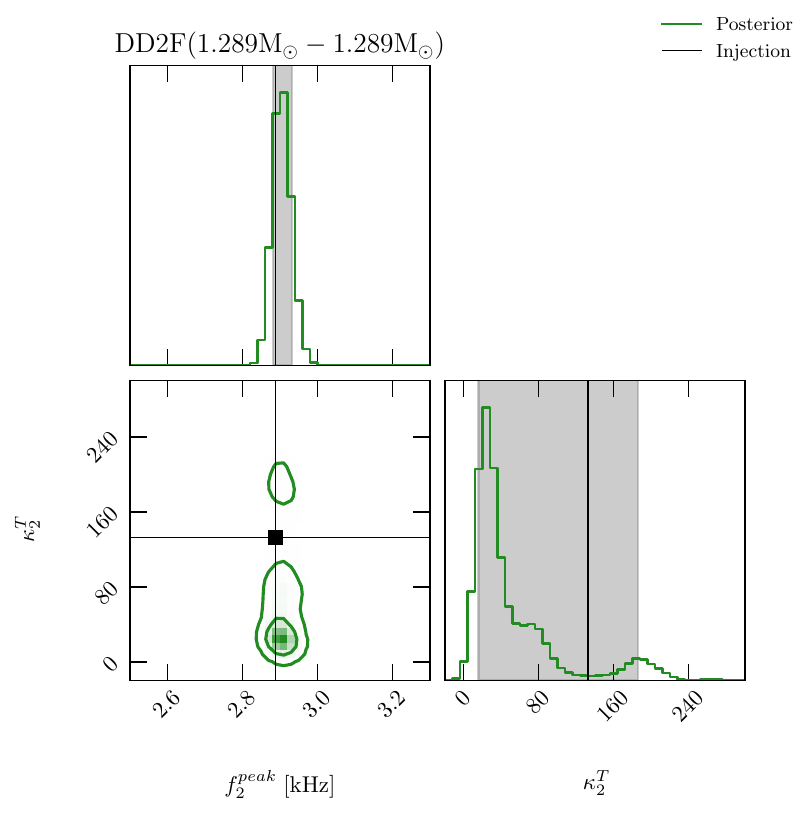}
	\includegraphics[width=0.49\textwidth]{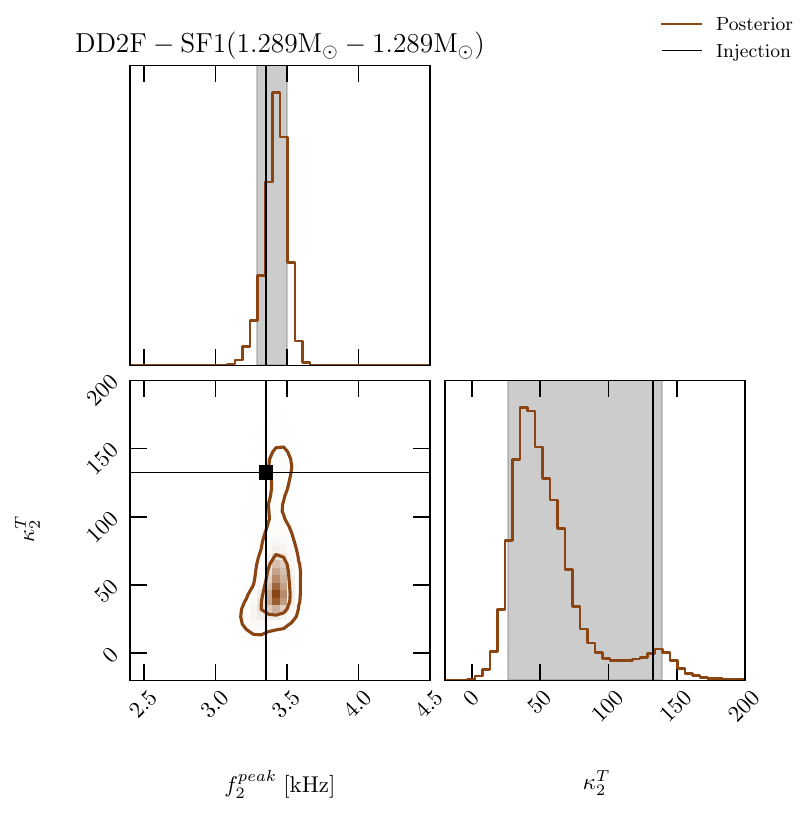}		
	\caption{Same calculations as in Figure \ref{fig:Inspiral_Agnostic_PE_f2_k2T_1234} for the binary $1.363 M_{\odot} - 1.363 M_{\odot}$ simulated with the BLh and BLQ EOSs and the binary $1.289 M_{\odot} - 1.289 M_{\odot}$ simulated with the DD2F and DD2F-SF1 EOS.}
	\label{fig:Inspiral_Agnostic_PE_f2_k2T_78910}
\end{figure*}

In this appendix, we present results for a postmerger PE of the $\tt{NRPMw}$ model's parameters wherein we set a wide range of values to the priors as described in Table \ref{tab:inspiral_agnostic_priors_table}. The choice of priors follows that in \cite{Breschi:2022ens} and is targeted at a wide range of possibilities for the GW event. To this aim, we present results for the postmerger PE of all the simulations listed in Table \ref{tab:Injection_Property_IA} performed with this choice of priors. In Figs. \ref{fig:Inspiral_Agnostic_PE_f2_k2T_1234}, \ref{fig:Inspiral_Agnostic_PE_f2_k2T_56} and \ref{fig:Inspiral_Agnostic_PE_f2_k2T_78910} we present the posterior PDFs for $f_2^{\mathrm{peak}}$ and $\kappa_2^T$. We note that the $f_2^{\rm peak}$ frequencies are recovered accurately and the injection is contained in the 90\% CIs. There appear to be no significant differences as compared to the estimation of $f_2^{\rm peak}$ from the inspiral informed choice of priors. At the same time, $\kappa_2^T$ is very poorly determined, serving to verify the fact that once the $f_2 - \kappa_2^T$ universal relation has been omitted from the model, there exists no tidal information solely from the postmerger signal. 

\begin{figure*}[t]
	\includegraphics[width=0.49\textwidth]{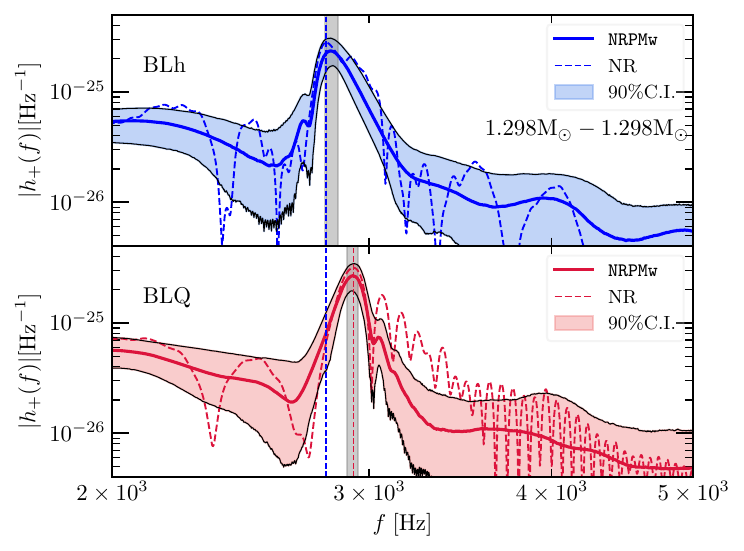}
	\includegraphics[width=0.49\textwidth]{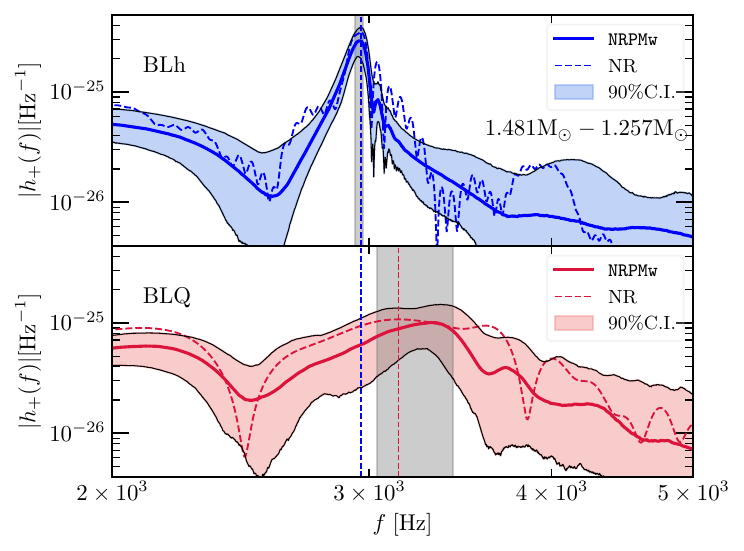}
	\includegraphics[width=0.49\textwidth]{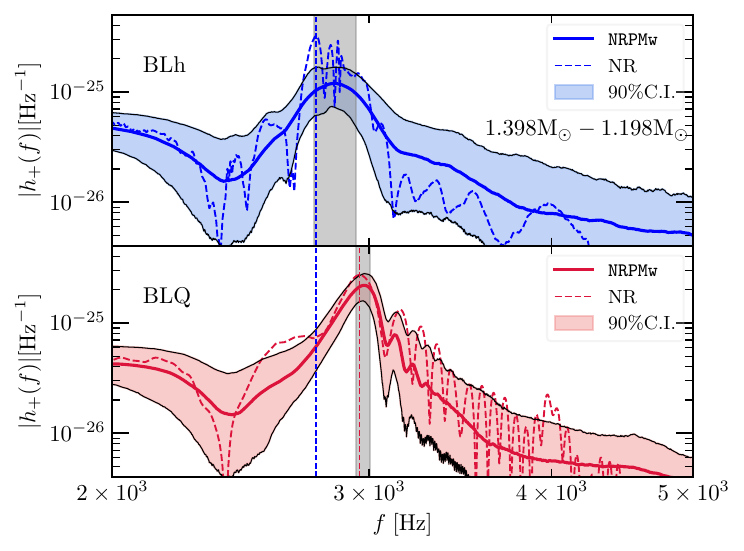}
	\includegraphics[width=0.49\textwidth]{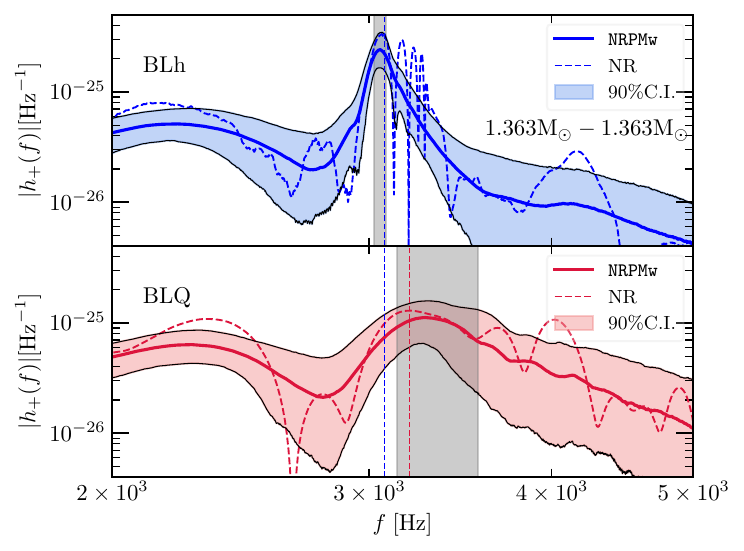}
	\includegraphics[width=0.49\textwidth]{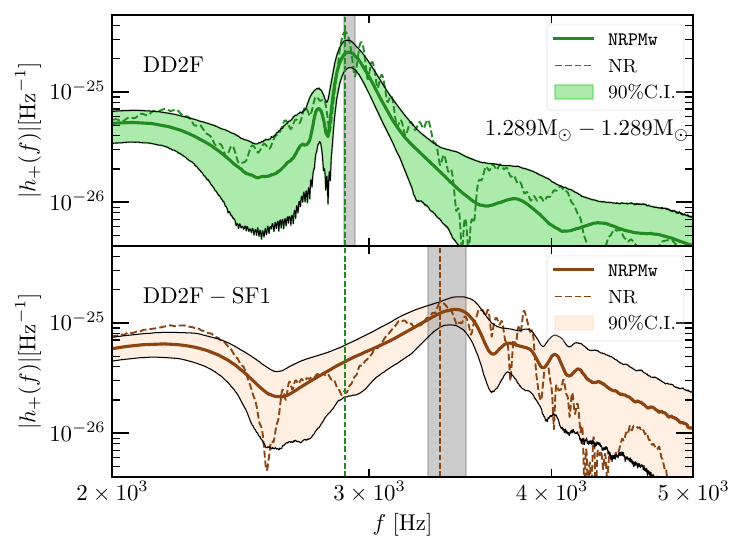}
	\caption{Same calculation as in Fig. \ref{fig:reconstructed_spectra_5_6}, i.e. reconstructed spectra for all the binaries in our work but computed with priors that are uninformed of the inspiral signal. We observe an accurate recovery of $f_2^{\mathrm{peak}}$ and distinguishability between hadronic and quark models to 90\% at a postmerger SNR of 10.}
	\label{fig:reconstructed_spectra_IA_12345678910}
\end{figure*}

Finally, in Fig.~\ref{fig:reconstructed_spectra_IA_12345678910}, we present the waveform reconstruction for the case of inspiral agnostic priors. Like in the case of the inspiral informed priors, the postmerger estimation of $f_2^{\rm peak}$ is accurate and distinguishable between hadronic and quark models. Additionally, the signal is reliably reconstructed as shown by the fact that most of the postmerger SNR is recovered (see Table \ref{tab:Injection_Property_IA}).
 
%%%%%%%%%%%%%%%%%%%%%%%%%%%%%%%%%%%%%%%%%%%%%%%%%%%%%%%%%%%%%%%%%%%%%%%%%%%%%%%%%%%%%%%%%%%
%%%%%%% Appendix: Inspiral Informed PE: results for all simulations %%%%%%%%%%%%%%
%%%%%%%%%%%%%%%%%%%%%%%%%%%%%%%%%%%%%%%%%%%%%%%%%%%%%%%%%%%%%%%%%%%%%%%%%%%%%%%%%%%%%%%%%%%

\section{Inspiral Informed PE: results for all simulations}
\label{Inspiral_Informed_PE:results_for_all_simulations}

In this appendix, we present results analogous to Figs. \ref{fig:Inspiral_Informed_PE_f2_k2T_5_6} and \ref{fig:reconstructed_spectra_5_6} for all the systems as listed in Table \ref{tab:Injection_Property_II} with the ET detector and the $\tt{NRPMw}$ model at a postmerger SNR of 10. In particular, in Figs. \ref{fig:Inspiral_Informed_PE_f2_k2T_1234} and \ref{fig:Inspiral_Informed_PE_f2_k2T_78910} we show the posterior PDFs for $f_2^{\mathrm{peak}}$ and $\kappa_2^{T}$. As mentioned previously in subsection \ref{subsec:Inspiral_Informed_Postmerger_PE}, the $\tt{NRPMw}$ model performs very well with the quark EOSs, in that the 90\% CI of $f_2^{\rm peak}$ posteriors contain the injection. However, for the hadronic simulations $1.398 M_{\odot} - 1.198 M_{\odot}$ (BLh), $1.289 M_{\odot} - 1.289 M_{\odot}$ (DD2F) and $1.298 M_{\odot} - 1.298 M_{\odot}$ (BLh), the estimation of $f_2^{\rm peak}$ is biased due to the presence of multiple amplitude modulations as explained in Fig. \ref{fig:reconstructed_spectra_II_filtered_unfiltered_19}. We also show postmerger spectra for the waveform reconstructions in Fig. \ref{fig:reconstructed_spectra_II_123478910} that serve to re-affirm the detectability and distinguishability of the $f_2^{\rm peak}$ frequencies between the hadronic and quark models.

In Figs. \ref{fig:inspiral_Informed_PE_histograms3} and \ref{fig:inspiral_Informed_PE_histograms4}, we show a comparison between posterior PDFs of the total mass $M$, mass ratio $q$ and tidal parameters $\Lambda_i$'s between the cases of inspiral informed and inspiral agnostic priors. We note the significant improvement in the estimation of masses and tidal parameters upon including inspiral information which is essentially a recovery of the priors that are informed of the inspiral signal. As we have stressed in the main text, we require reliable estimates of the inspiral signal to consistently probe QCD phase transitions from the EOS insensitive relations.

\begin{figure*}[t]
	\includegraphics[width= \textwidth]{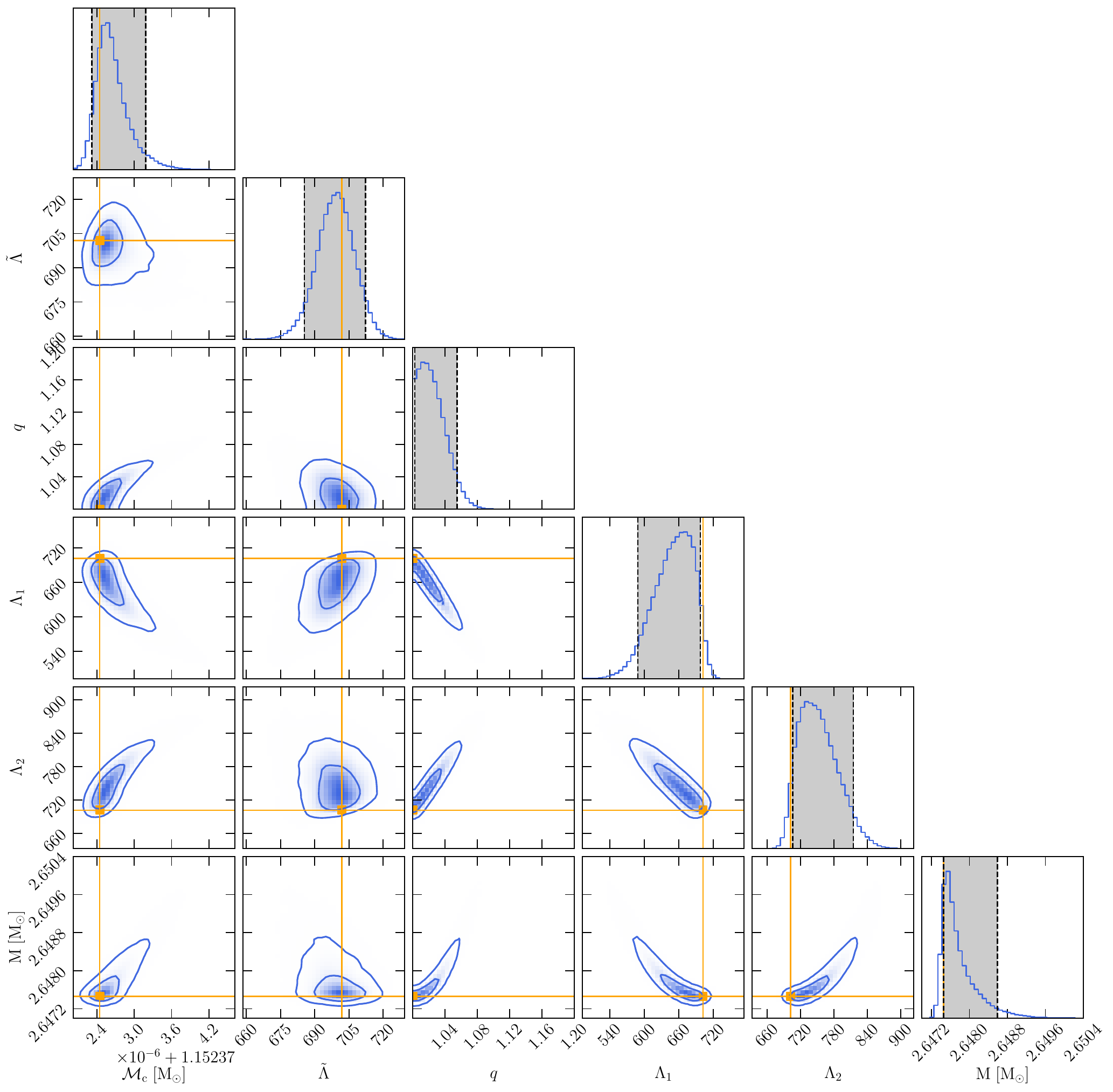}
	\caption{Corner plot depicting the posteriors on chirp mass $\mathcal{M}_c$, tidal deformability $\tilde{\Lambda}$, mass ratio $q$, individual tidal parameters $\Lambda_i$s and the total mass $M$ for the binary 1.298 $M_{\odot}$ - 1.298 $M_{\odot}$ with the BLh EOS. Owing to the fact that the mass ratio is at the edge of the priors, we have biases in the measurements of $q$, $\Lambda_1$, $\Lambda_2$ and $M$.}
	\label{fig:inspiral_PE_corner_1}
\end{figure*}

\begin{figure*}[t]
	\includegraphics[width=0.49\textwidth]{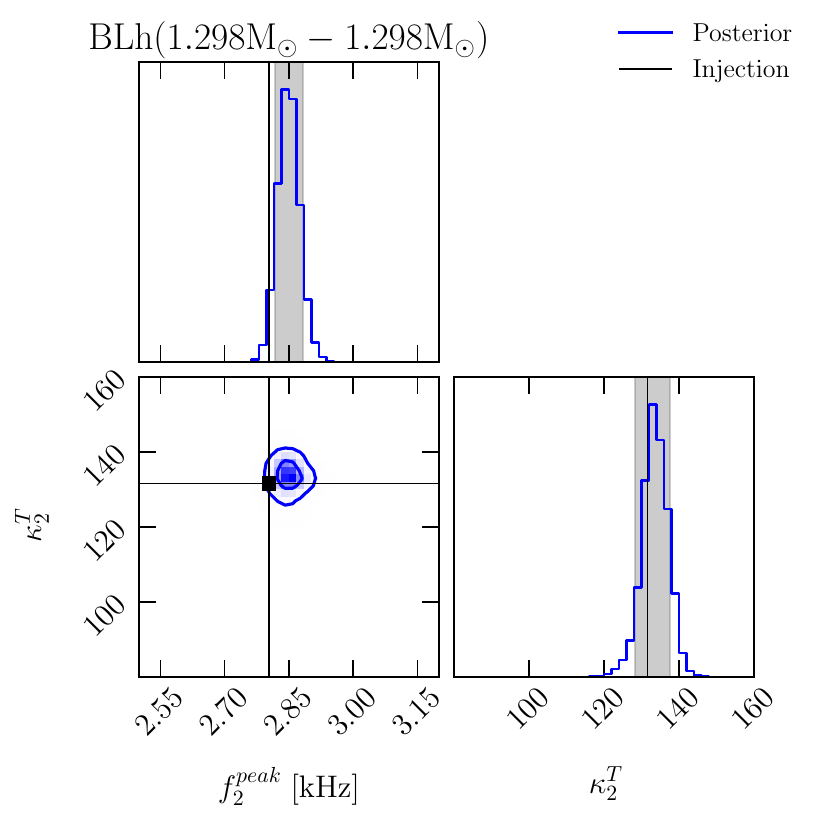}
	\includegraphics[width=0.49\textwidth]{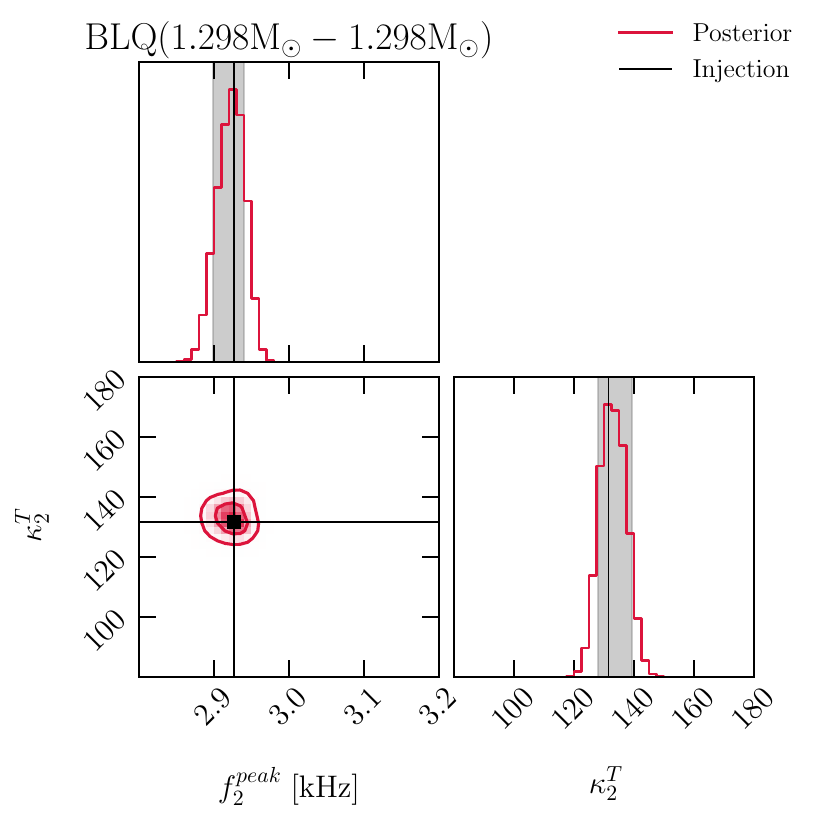}
	\includegraphics[width=0.49\textwidth]{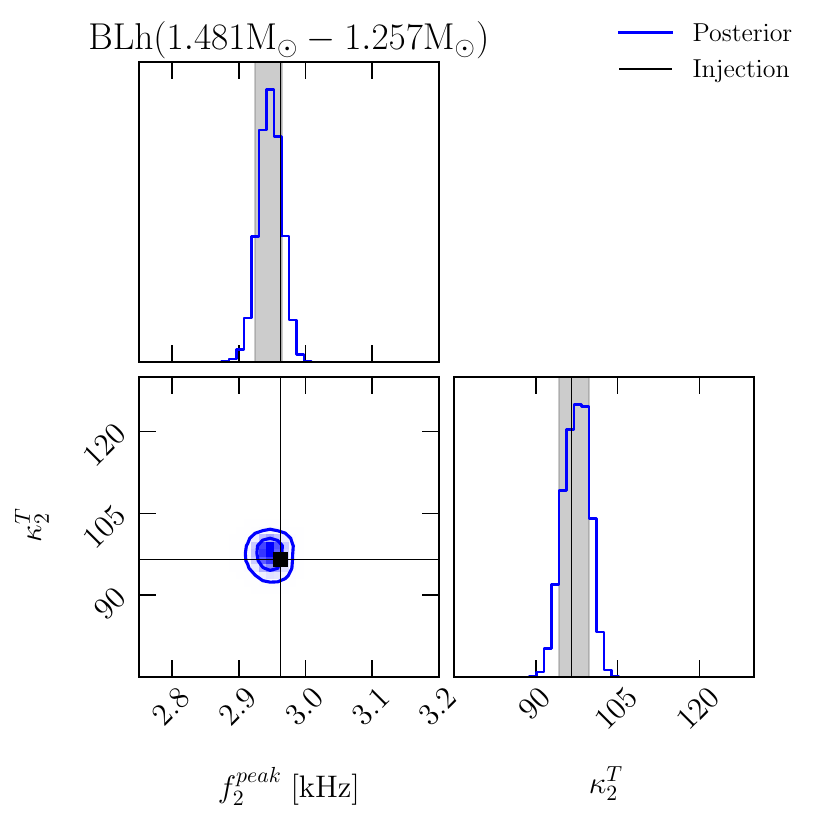}
	\includegraphics[width=0.49\textwidth]{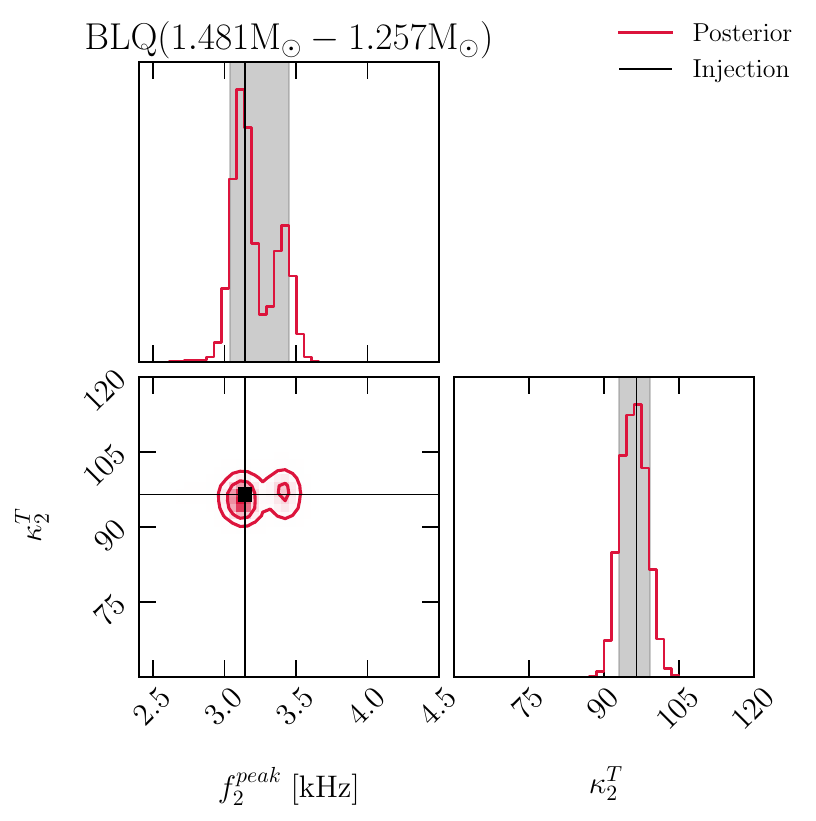}
	\caption{Same calculations as in Fig. \ref{fig:Inspiral_Informed_PE_f2_k2T_5_6} for the binaries $1.298 M_{\odot} - 1.298 M_{\odot}$ and $1.481 M_{\odot} - 1.257 M_{\odot}$ with the BLh and BLQ EOSs. We note that the $\tt{NRPMw}$ model captures to within 90\% CIs the $f_2^{\mathrm{peak}}$ frequency for the quark EOSs however the measurement of the same for hadronic model $1.298 M_{\odot} - 1.298 M_{\odot}$ suffers from a systematic bias that of multiple amplitude modulations. The double-peaked feature in the 1.481 $M_{\odot}$ - 1.257 $M_{\odot}$ binary is because this system is the shortest-lived of all our simulations due to which the uncertainties in the measurement of postmerger frequency are the highest.}
	\label{fig:Inspiral_Informed_PE_f2_k2T_1234}
\end{figure*}

\begin{figure*}[t]
	\includegraphics[width=0.49\textwidth]{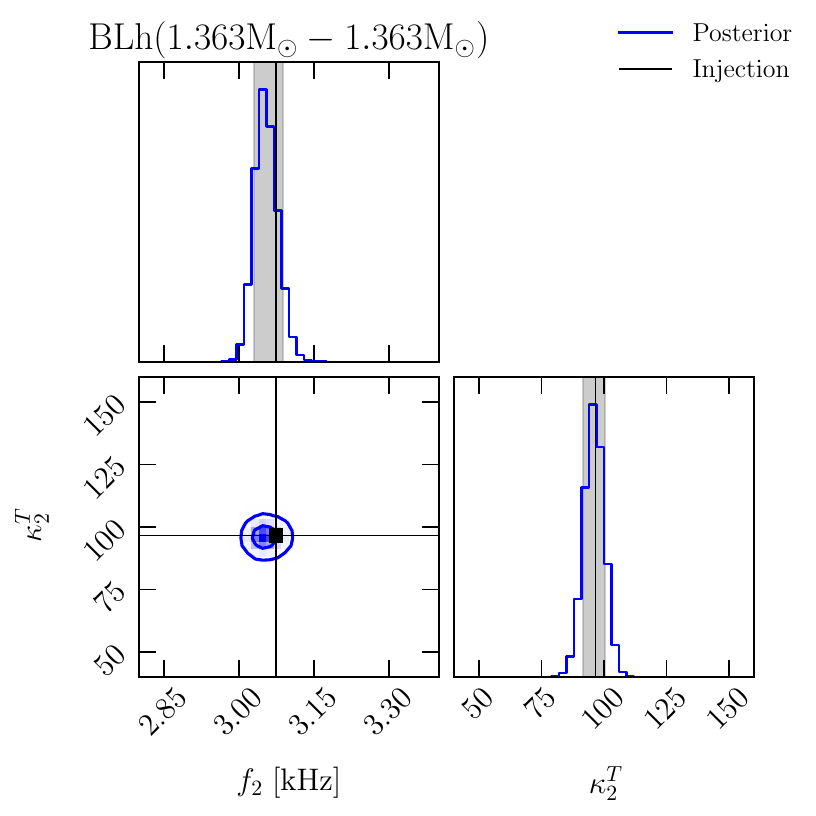}
	\includegraphics[width=0.49\textwidth]{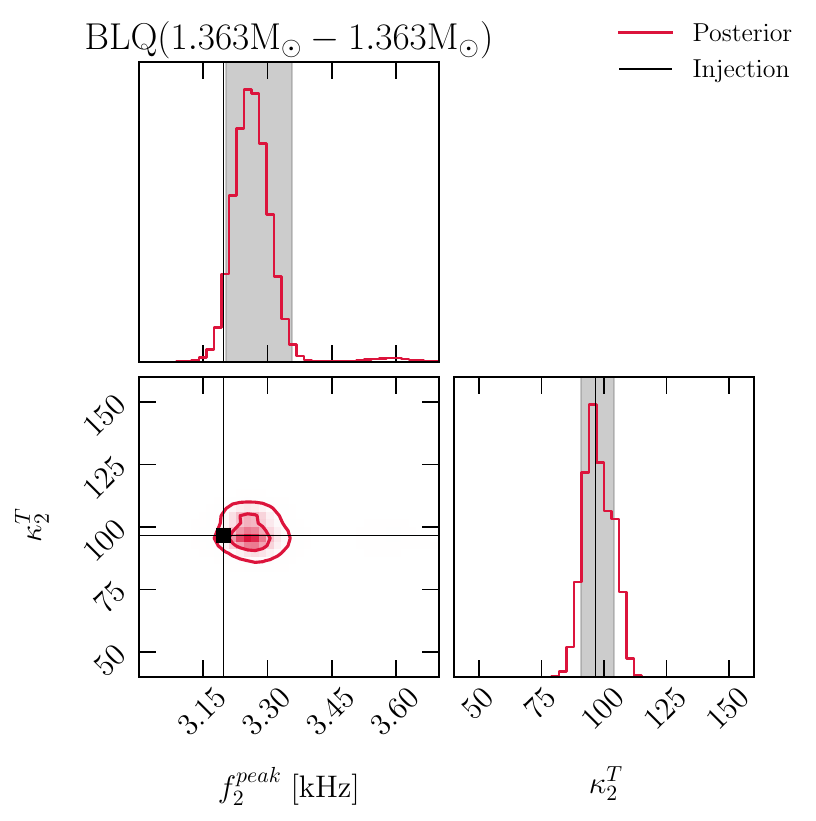}
	\includegraphics[width=0.49\textwidth]{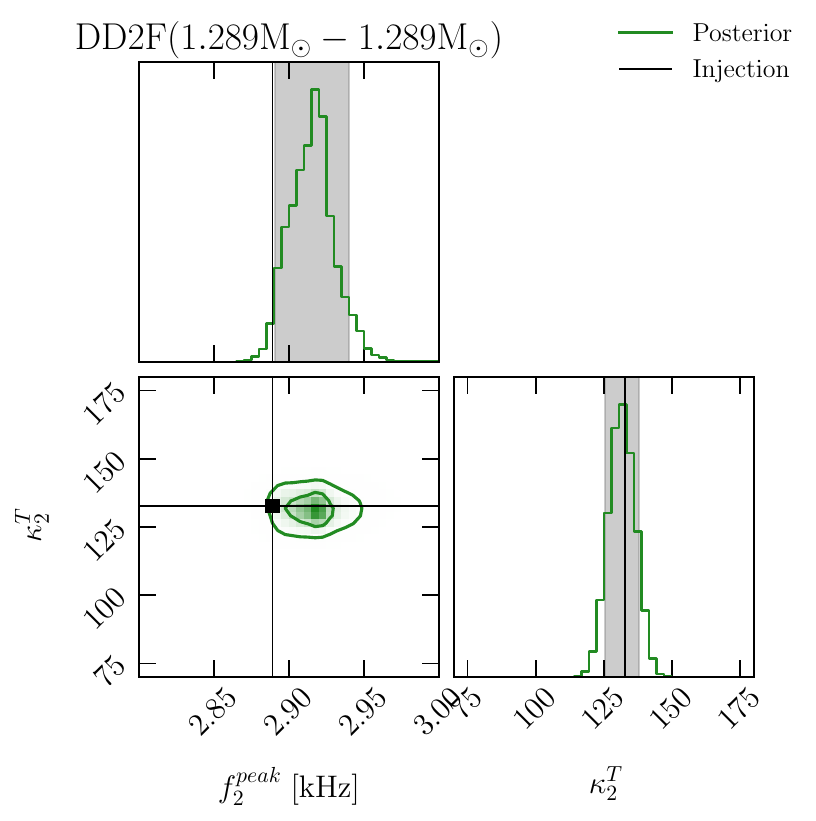}
	\includegraphics[width=0.49\textwidth]{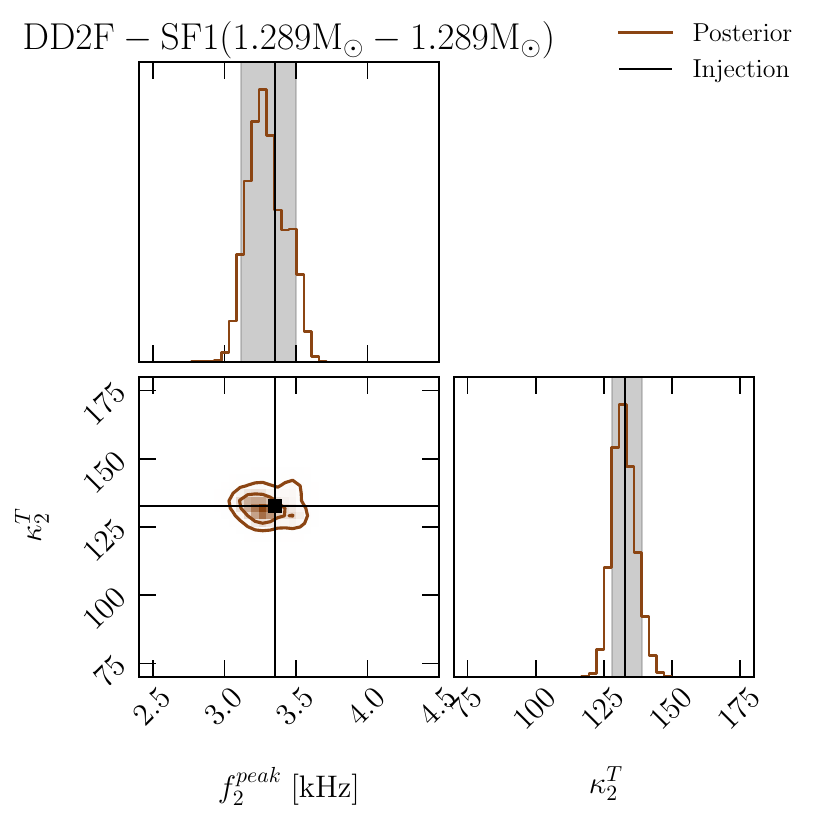}		
	\caption{Same calculations as in Fig. \ref{fig:Inspiral_Informed_PE_f2_k2T_5_6} for the binary $1.363 M_{\odot} - 1.363 M_{\odot}$ with the BLh and BLQ EOS and the binary $1.289 M_{\odot} - 1.289 M_{\odot}$ with the DD2F and DD2F-SF1 EOS.}
	\label{fig:Inspiral_Informed_PE_f2_k2T_78910}
\end{figure*}

\begin{figure*}[t]
	\includegraphics[width=0.49\textwidth]{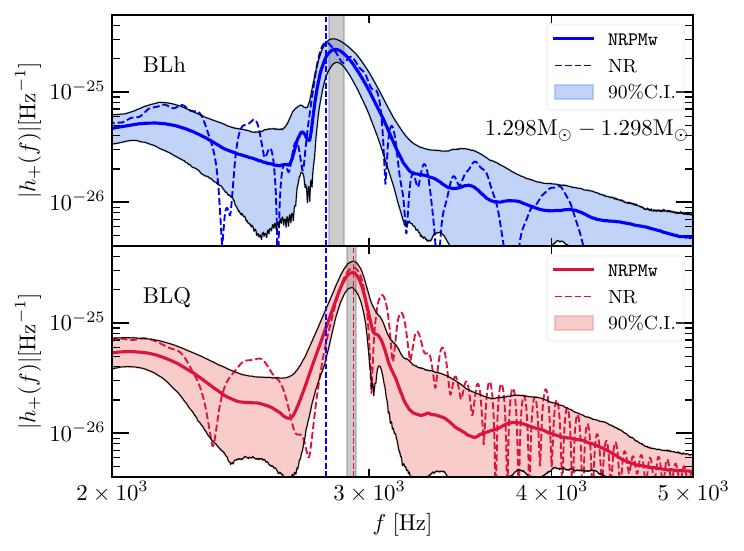}
	\includegraphics[width=0.49\textwidth]{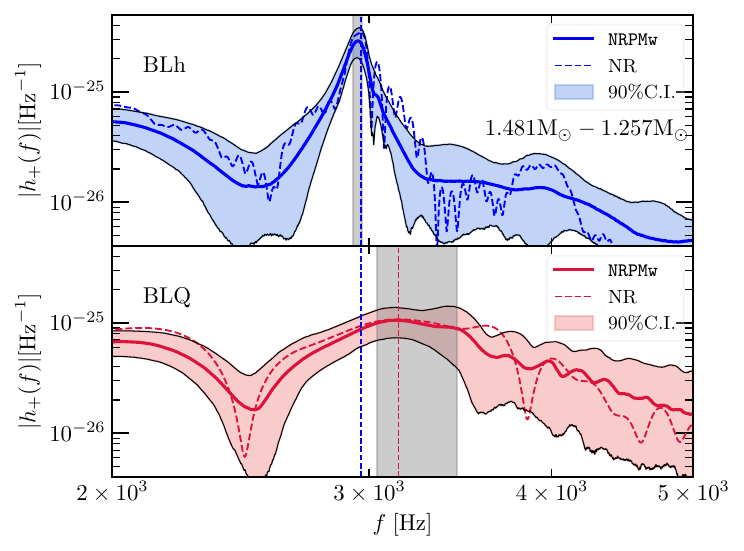}
	\includegraphics[width=0.49\textwidth]{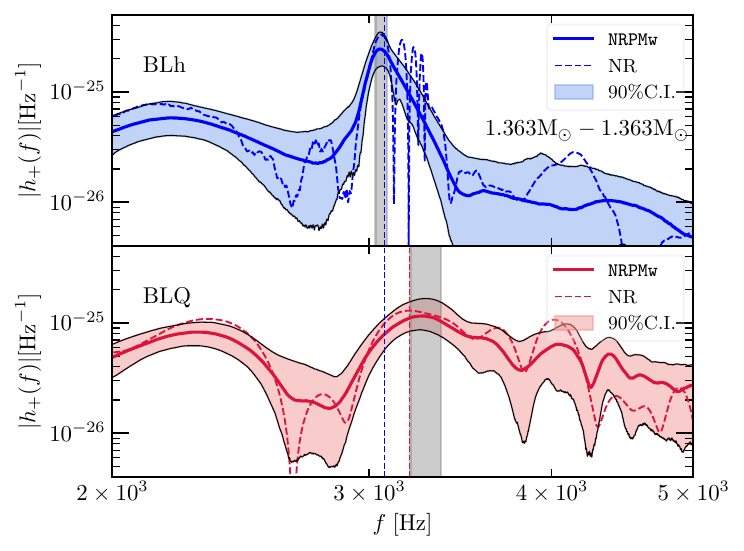}
	\includegraphics[width=0.49\textwidth]{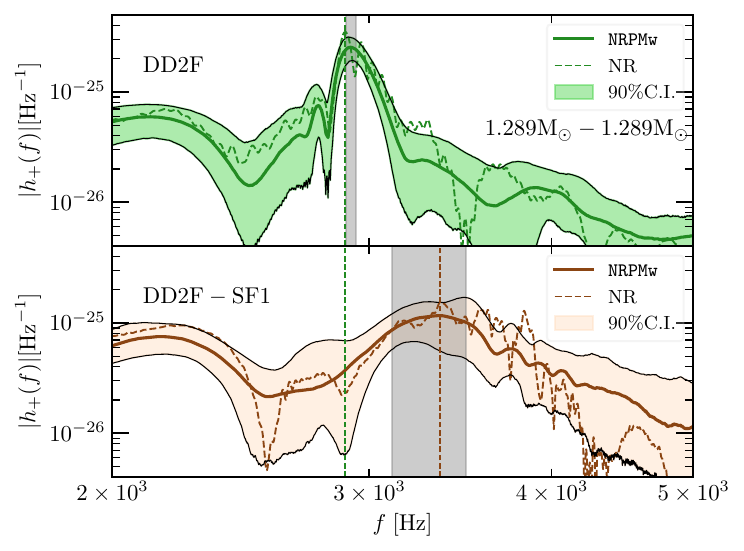}
	\caption{Same calculation as in Fig. \ref{fig:reconstructed_spectra_5_6}, i.e., reconstructed spectra for the binaries $1.298 M_{\odot} - 1.298 M_{\odot}$, $1.481 M_{\odot} - 1.257 M_{\odot}$ and $1.363 M_{\odot} - 1.363 M_{\odot}$ with the BLh and BLQ EOS as well as for the binary $1.289 M_{\odot} - 1.289 M_{\odot}$ with the DD2F and DD2F-SF1 EOS.}
	\label{fig:reconstructed_spectra_II_123478910}
\end{figure*}

\begin{figure*}[t]
	\includegraphics[width=0.49\textwidth]{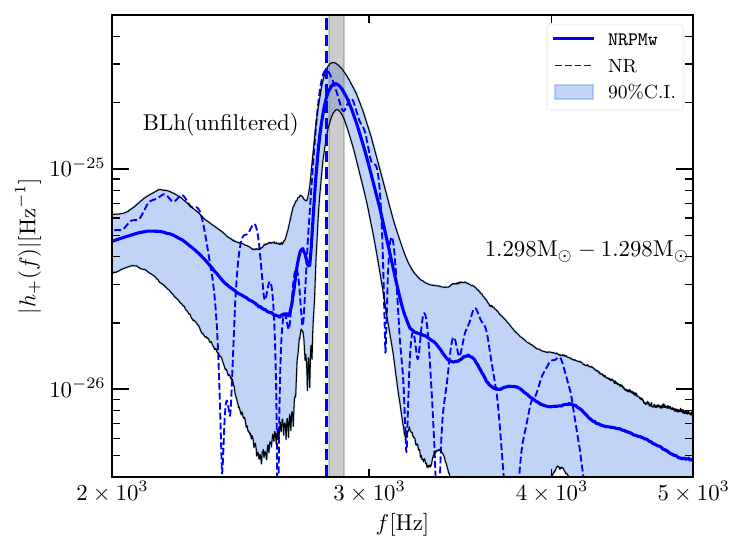}
	\includegraphics[width=0.49\textwidth]{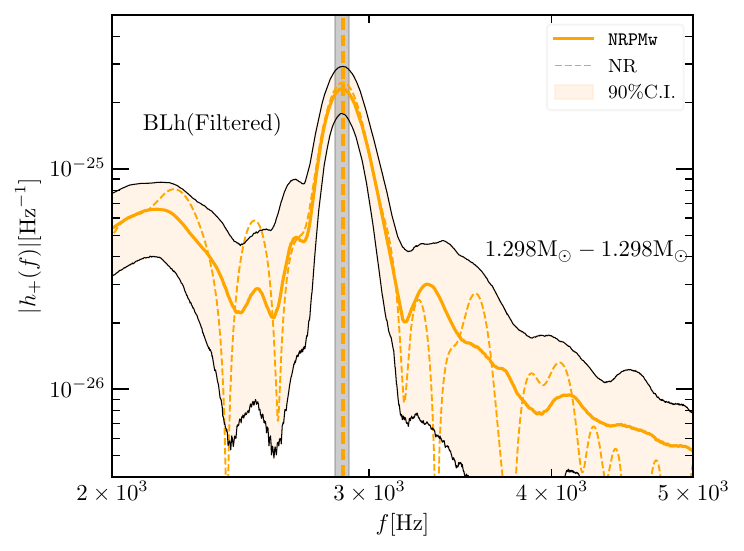}
	\includegraphics[width=0.49\textwidth]{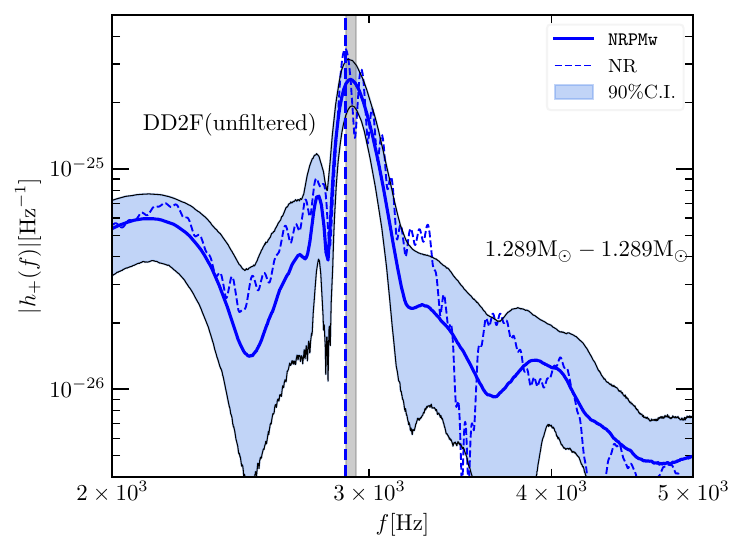}
	\includegraphics[width=0.49\textwidth]{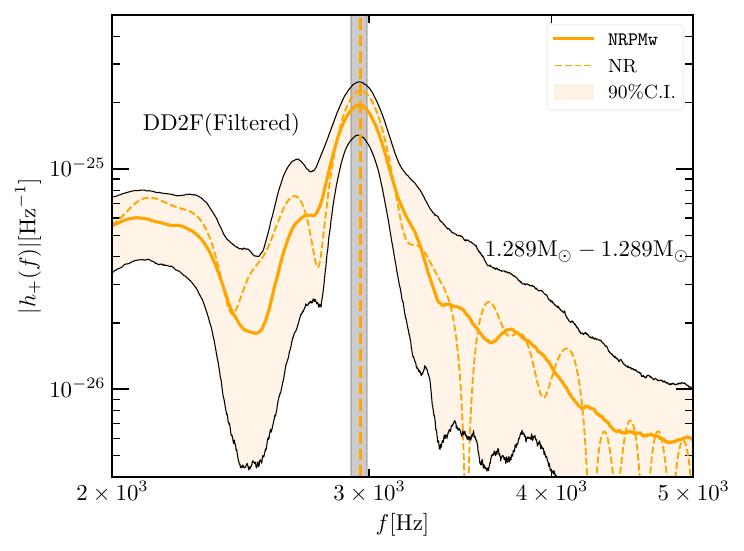}
	\caption{Same calculations as in Figure \ref{fig:multiple_amplitude modulations} for the binary $1.298 M_{\odot} - 1.298 M_{\odot}$ with the BLh EOS and the binary $1.289 M_{\odot} - 1.289 M_{\odot}$ with the DD2F EOS. Here, we show that the exclusion of more than 2 amplitude modulations in the strain can lead to recovery of the $f_2^{\mathrm{peak}}$ to within 90\% CIs at a postmerger SNR of 10.}
	\label{fig:reconstructed_spectra_II_filtered_unfiltered_19}
\end{figure*}

\begin{figure*}[t]
	\includegraphics[width=0.49\textwidth]{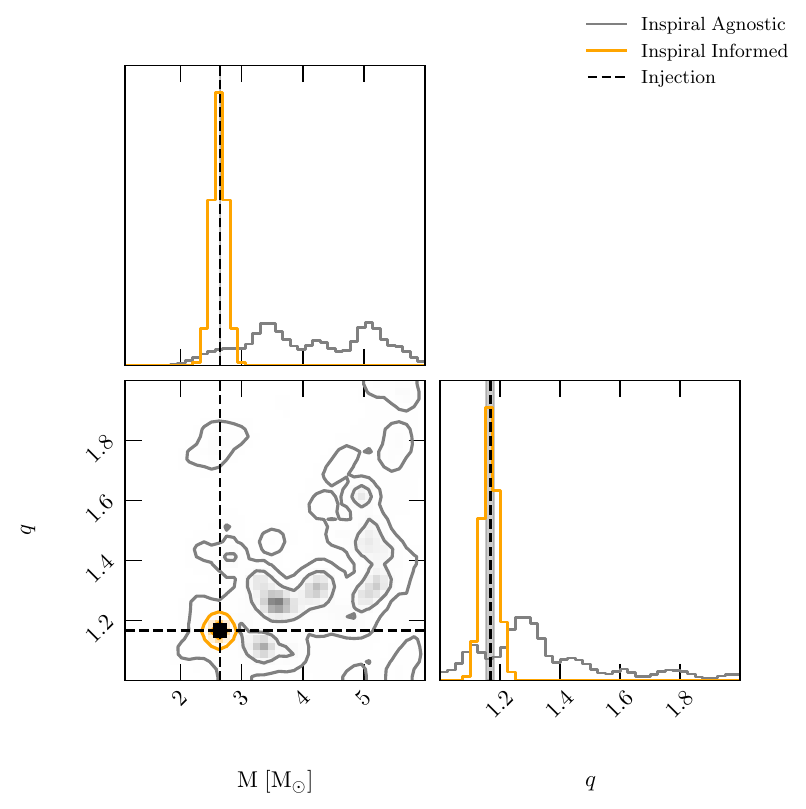}
	\includegraphics[width=0.49\textwidth]{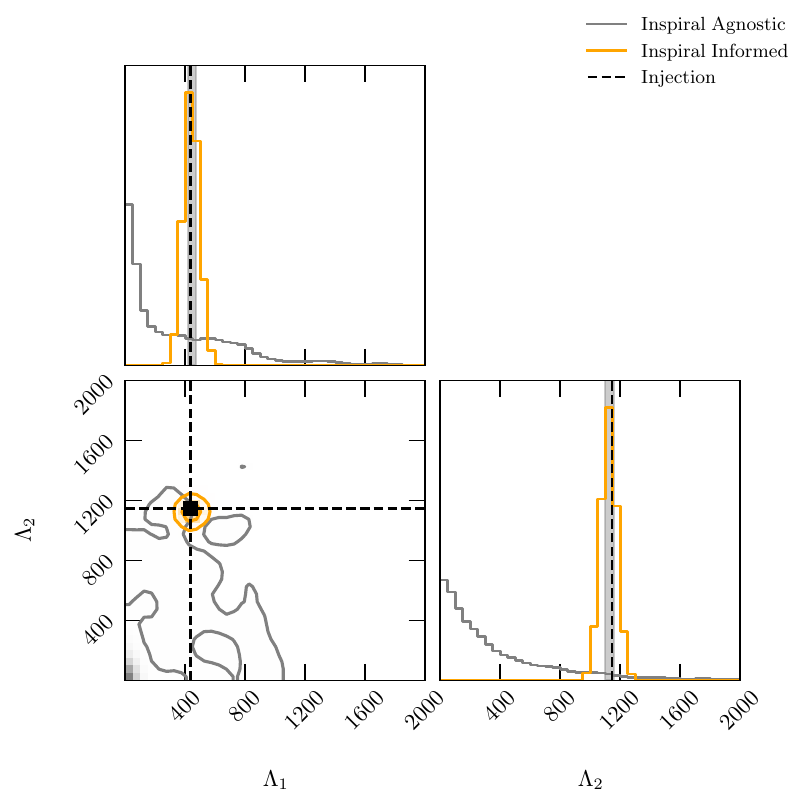}
	
	\caption{\textit{Left Panel}: The posterior distributions of the total mass and mass ratio from a postmerger PE of the binary $1.398 M_{\odot} - 1.198 M_{\odot}$ with the BLh EOS compared between the two choice of priors used in this work. \textit{Right Panel}: the posterior PDFs for the component tidal deformabilities. In both cases we notice a clear improvement in accuracy for the measurement of $M, q, \Lambda_1$ and $\Lambda_2$.}
	\label{fig:inspiral_Informed_PE_histograms3}
	\end{figure*}

	\begin{figure*}[t]
	\includegraphics[width=0.49\textwidth]{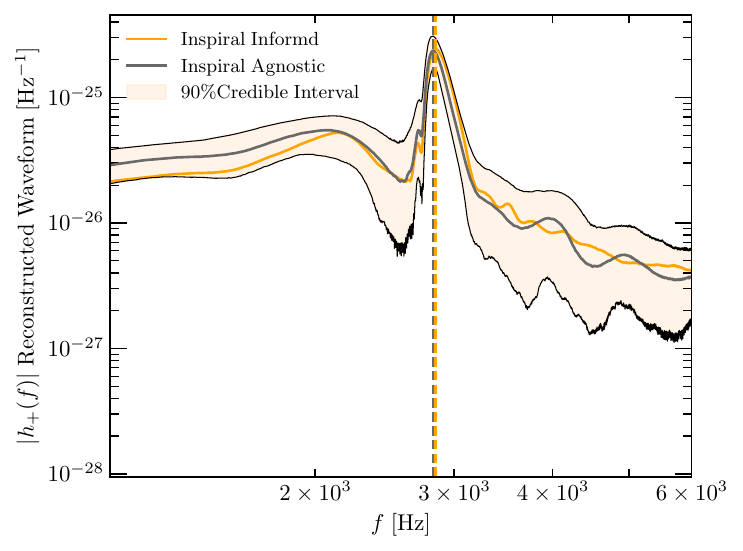}
	\caption{The reconstructed $\tt{NRPMw}$ waveforms for a postmerger PE of the binary $1.398 M_{\odot} - 1.198 M_{\odot}$ with the BLh EOS corresponding to both the choices of priors. Both the reconstructions lie within the 90\% CIs of each injection and the $f_2^{\mathrm{peak}}$ frequency shows only a miniscule deviation of $\approx$0.3\%.}
	\label{fig:inspiral_Informed_PE_histograms4}
	\end{figure*}

%%%%%%%%%%%%%%%%%%%%%%%%%%%%%%%%%%%%%%%%%%%%%%%%%%%%%%%%%%%%%%%%%%%%%%%%%%%%%%%%%%%%%%%%%%%
%%%%%%% Appendix: Unconstrained f2 f0 inference %%%%%%%%%%%%%%
%%%%%%%%%%%%%%%%%%%%%%%%%%%%%%%%%%%%%%%%%%%%%%%%%%%%%%%%%%%%%%%%%%%%%%%%%%%%%%%%%%%%%%%%%%%

\section{Inference with unconstrained $f_2$ and $f_0$ parameters}
\label{Unconstrained_f2_f0_inference}

\begin{figure*}[t]
	\includegraphics[width=0.49\textwidth]{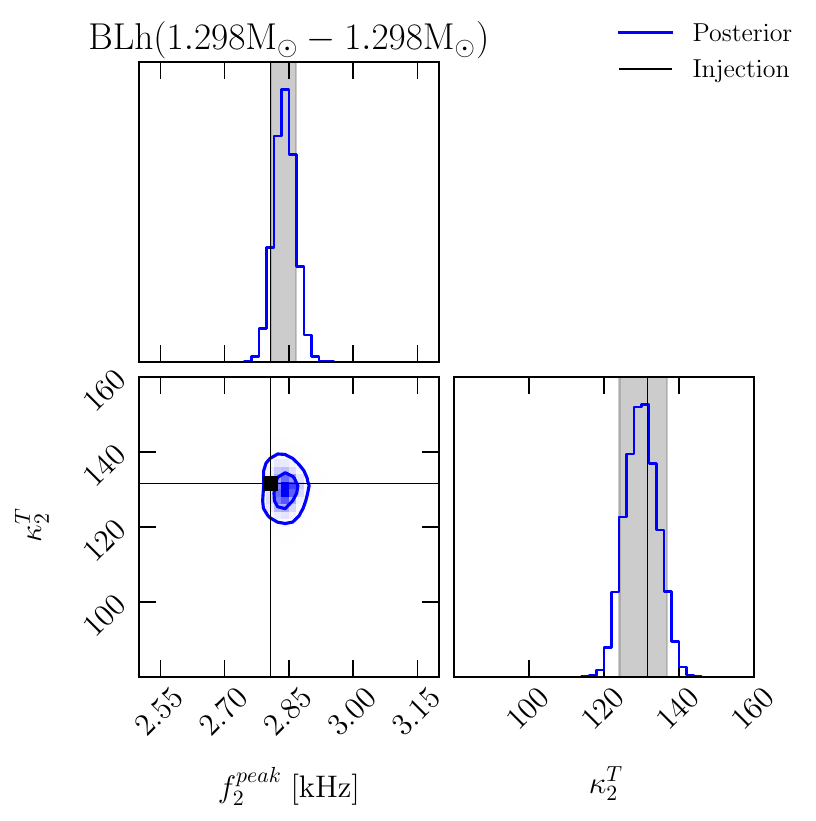}
	\includegraphics[width=0.49\textwidth]{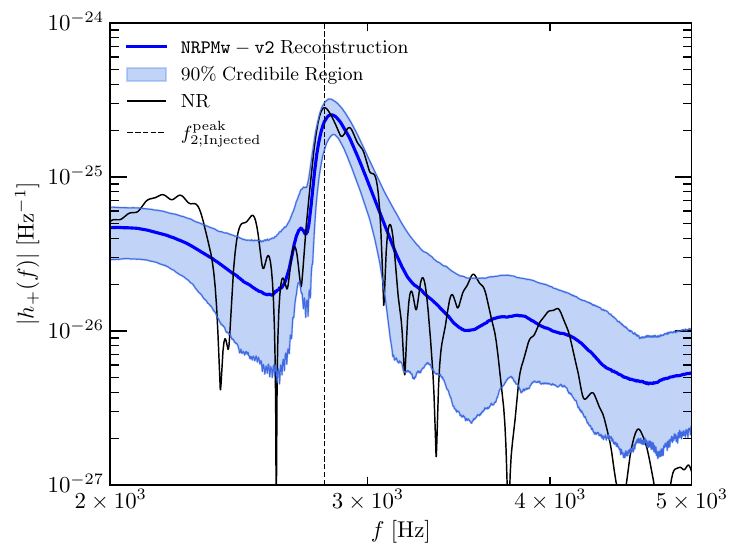}
	
	\caption{\textit{Left Panel}: The posterior distributions of the postmerger peak frequency $f_2^{\mathrm{peak}}$ and the tidal polarizability $\kappa_2^T$ corresponding to the $1.298 M_{\odot} - 1.298 M_{\odot}$ binary with the BLh EOS. Also shown is the lack of covariance of $f_2^{\mathrm{peak}}$ with $\kappa_2^T$ owing to the corresponding QUR being not used. The contours correspond to the 50 \% and 90\% CIs of the joint PDF. \textit{Right Panel}: The median reconstructed waveform from $\tt{NRPMw\_v2}$ shown along with the NR waveform.} 
	\label{fig:f2f0_free_recovery}
	\end{figure*}

\begin{figure*}[t]
		\includegraphics[width=0.49\textwidth]{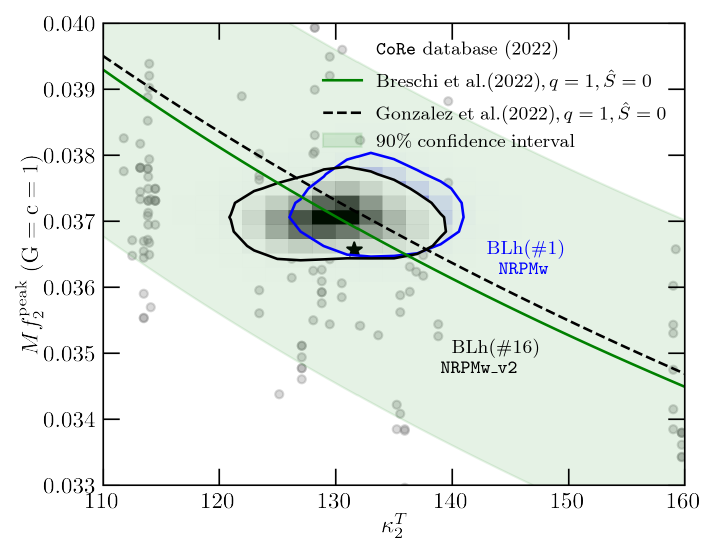}
		\caption{A comparison of the joint $f_2^{\rm peak} - \kappa_2^T$ posterior for the $1.298 M_{\odot} - 1.298 M_{\odot}$ binary between the $\tt{NRPMw}$ and $\tt{NRPMw\_v2}$ model configurations. We see that the 90\% contours of the joint posteriors for both the model configuratios contain the injection. }
		\label{fig:QUR_f2f0_free}
		\end{figure*}

In this appendix, we attempt to mitigate the source of bias in our hadronic models namely multiple amplitude modulations. We have seen in \ref{subsubsec:multi_amplitude_modulations} that $\tt{NRPMw}$ can only capture the first two peaks of the postmerger amplitude modulations which leads to an overestimation of the $f_2^{\rm peak}$ frequency. We test a new model configuration, in which we attempt to increase the flexibility of the model by freeing from universal relations not just the postmerger peak frequency parameter $f_2$, but also the parameter that models the radial pulsation modes of the remnant, i.e., $f_0$. We call this model configuration as $\tt{NRPMw\_v2}$ to distinguish from the other configurations employed in this work. This means that we do not use the recalibration parameter $\delta f_0$ which provided flexibility to the inference of $f_0$ when constrained from the universal relations instead, we set uniform priors on $f_0$ ranging between 0.1 to 2.5 kHz. The expectation is that making $f_0$ unconstrained can perhaps push the $\tilde{W}{\rm pul}$ wavelet that models amplitude modulations as defined in \cite{Breschi:2022xnc}, to include more of the amplitude modulations. 

We report however that this approach leads to only marginal improvements. We take the case of the $1.298 M_{\odot} - 1.298 M_{\odot}$ binary where the bias in measurement of $f_2^{\rm peak}$ is the largest. In figure \ref{fig:f2f0_free_recovery}, we report the same injection now being recovered from the modified $\tt{NRPMw\_v2}$ model configuration. We report that with the $\tt{NRPMw}$ model, 99.5\% CIs of the $f_2^{\rm peak}$ posterior contained the injection, which is now marginally improved to 97\% CIs containing the injection with $\tt{NRPMw\_v2}$. Nevertheless, the model configuration still reliably reconstructs the postmerger signal with a recovered SNR of 8.6 corresponding to an injected SNR of 10. 

Finally, in figure \ref{fig:QUR_f2f0_free}, we show a comparison of the joint $f_2^{\rm peak} - \kappa_2^T$ posterior between the $\tt{NRPMw}$ and the $\tt{NRPMw\_v2}$ model configurations, with reference to the $f_2 - \kappa_2^T$ universal relation. For both the configurations, the 90\% contours of the joint posterior capture the injection. We compute the Bayes' factor between the two models and find that $\rm{log}\;\mathcal{B.F.}^{\tt{NRPMw}}_{\tt{NRPMw\_v2}} = 0.06^{+0.28}_{-0.28}$ indicating that there is no preference to either models at a postmerger SNR of 10. 
 
\bibliography{references.bib}

\begin{thebibliography}{128}%
\makeatletter
\providecommand \@ifxundefined [1]{%
 \@ifx{#1\undefined}
}%
\providecommand \@ifnum [1]{%
 \ifnum #1\expandafter \@firstoftwo
 \else \expandafter \@secondoftwo
 \fi
}%
\providecommand \@ifx [1]{%
 \ifx #1\expandafter \@firstoftwo
 \else \expandafter \@secondoftwo
 \fi
}%
\providecommand \natexlab [1]{#1}%
\providecommand \enquote  [1]{``#1''}%
\providecommand \bibnamefont  [1]{#1}%
\providecommand \bibfnamefont [1]{#1}%
\providecommand \citenamefont [1]{#1}%
\providecommand \href@noop [0]{\@secondoftwo}%
\providecommand \href [0]{\begingroup \@sanitize@url \@href}%
\providecommand \@href[1]{\@@startlink{#1}\@@href}%
\providecommand \@@href[1]{\endgroup#1\@@endlink}%
\providecommand \@sanitize@url [0]{\catcode `\\12\catcode `\$12\catcode
  `\&12\catcode `\#12\catcode `\^12\catcode `\_12\catcode `\%12\relax}%
\providecommand \@@startlink[1]{}%
\providecommand \@@endlink[0]{}%
\providecommand \url  [0]{\begingroup\@sanitize@url \@url }%
\providecommand \@url [1]{\endgroup\@href {#1}{\urlprefix }}%
\providecommand \urlprefix  [0]{URL }%
\providecommand \Eprint [0]{\href }%
\providecommand \doibase [0]{https://doi.org/}%
\providecommand \selectlanguage [0]{\@gobble}%
\providecommand \bibinfo  [0]{\@secondoftwo}%
\providecommand \bibfield  [0]{\@secondoftwo}%
\providecommand \translation [1]{[#1]}%
\providecommand \BibitemOpen [0]{}%
\providecommand \bibitemStop [0]{}%
\providecommand \bibitemNoStop [0]{.\EOS\space}%
\providecommand \EOS [0]{\spacefactor3000\relax}%
\providecommand \BibitemShut  [1]{\csname bibitem#1\endcsname}%
\let\auto@bib@innerbib\@empty
%</preamble>
\bibitem [{\citenamefont {Abbott}\ \emph
  {et~al.}(2017{\natexlab{a}})\citenamefont {Abbott} \emph
  {et~al.}}]{LIGOScientific:2017vwq}%
  \BibitemOpen
  \bibfield  {author} {\bibinfo {author} {\bibfnamefont {B.~P.}\ \bibnamefont
  {Abbott}} \emph {et~al.} (\bibinfo {collaboration} {LIGO Scientific,
  Virgo}),\ }\bibfield  {title} {\bibinfo {title} {{GW170817: Observation of
  Gravitational Waves from a Binary Neutron Star Inspiral}},\ }\href
  {https://doi.org/10.1103/PhysRevLett.119.161101} {\bibfield  {journal}
  {\bibinfo  {journal} {Phys. Rev. Lett.}\ }\textbf {\bibinfo {volume} {119}},\
  \bibinfo {pages} {161101} (\bibinfo {year} {2017}{\natexlab{a}})},\ \Eprint
  {https://arxiv.org/abs/1710.05832} {arXiv:1710.05832 [gr-qc]} \BibitemShut
  {NoStop}%
\bibitem [{\citenamefont {Abbott}\ \emph
  {et~al.}(2017{\natexlab{b}})\citenamefont {Abbott} \emph
  {et~al.}}]{LIGOScientific:2017ync}%
  \BibitemOpen
  \bibfield  {author} {\bibinfo {author} {\bibfnamefont {B.~P.}\ \bibnamefont
  {Abbott}} \emph {et~al.} (\bibinfo {collaboration} {LIGO Scientific, Virgo,
  Fermi GBM, INTEGRAL, IceCube, AstroSat Cadmium Zinc Telluride Imager Team,
  IPN, Insight-Hxmt, ANTARES, Swift, AGILE Team, 1M2H Team, Dark Energy Camera
  GW-EM, DES, DLT40, GRAWITA, Fermi-LAT, ATCA, ASKAP, Las Cumbres Observatory
  Group, OzGrav, DWF (Deeper Wider Faster Program), AST3, CAASTRO, VINROUGE,
  MASTER, J-GEM, GROWTH, JAGWAR, CaltechNRAO, TTU-NRAO, NuSTAR, Pan-STARRS,
  MAXI Team, TZAC Consortium, KU, Nordic Optical Telescope, ePESSTO, GROND,
  Texas Tech University, SALT Group, TOROS, BOOTES, MWA, CALET, IKI-GW
  Follow-up, H.E.S.S., LOFAR, LWA, HAWC, Pierre Auger, ALMA, Euro VLBI Team, Pi
  of Sky, Chandra Team at McGill University, DFN, ATLAS Telescopes, High Time
  Resolution Universe Survey, RIMAS, RATIR, SKA South Africa/MeerKAT}),\
  }\bibfield  {title} {\bibinfo {title} {{Multi-messenger Observations of a
  Binary Neutron Star Merger}},\ }\href
  {https://doi.org/10.3847/2041-8213/aa91c9} {\bibfield  {journal} {\bibinfo
  {journal} {Astrophys. J. Lett.}\ }\textbf {\bibinfo {volume} {848}},\
  \bibinfo {pages} {L12} (\bibinfo {year} {2017}{\natexlab{b}})},\ \Eprint
  {https://arxiv.org/abs/1710.05833} {arXiv:1710.05833 [astro-ph.HE]}
  \BibitemShut {NoStop}%
\bibitem [{\citenamefont {Radice}\ \emph {et~al.}(2020)\citenamefont {Radice},
  \citenamefont {Bernuzzi},\ and\ \citenamefont {Perego}}]{Radice:2020ddv}%
  \BibitemOpen
  \bibfield  {author} {\bibinfo {author} {\bibfnamefont {D.}~\bibnamefont
  {Radice}}, \bibinfo {author} {\bibfnamefont {S.}~\bibnamefont {Bernuzzi}},\
  and\ \bibinfo {author} {\bibfnamefont {A.}~\bibnamefont {Perego}},\
  }\bibfield  {title} {\bibinfo {title} {{The Dynamics of Binary Neutron Star
  Mergers and GW170817}},\ }\href
  {https://doi.org/10.1146/annurev-nucl-013120-114541} {\bibfield  {journal}
  {\bibinfo  {journal} {Ann. Rev. Nucl. Part. Sci.}\ }\textbf {\bibinfo
  {volume} {70}},\ \bibinfo {pages} {95} (\bibinfo {year} {2020})},\ \Eprint
  {https://arxiv.org/abs/2002.03863} {arXiv:2002.03863 [astro-ph.HE]}
  \BibitemShut {NoStop}%
\bibitem [{\citenamefont {Bernuzzi}(2020)}]{Bernuzzi:2020tgt}%
  \BibitemOpen
  \bibfield  {author} {\bibinfo {author} {\bibfnamefont {S.}~\bibnamefont
  {Bernuzzi}},\ }\bibfield  {title} {\bibinfo {title} {{Neutron Star Merger
  Remnants}},\ }\href {https://doi.org/10.1007/s10714-020-02752-5} {\bibfield
  {journal} {\bibinfo  {journal} {Gen. Rel. Grav.}\ }\textbf {\bibinfo {volume}
  {52}},\ \bibinfo {pages} {108} (\bibinfo {year} {2020})},\ \Eprint
  {https://arxiv.org/abs/2004.06419} {arXiv:2004.06419 [astro-ph.HE]}
  \BibitemShut {NoStop}%
\bibitem [{\citenamefont {Punturo}\ \emph {et~al.}(2010)\citenamefont {Punturo}
  \emph {et~al.}}]{Punturo:2010zz}%
  \BibitemOpen
  \bibfield  {author} {\bibinfo {author} {\bibfnamefont {M.}~\bibnamefont
  {Punturo}} \emph {et~al.},\ }\bibfield  {title} {\bibinfo {title} {{The
  Einstein Telescope: A third-generation gravitational wave observatory}},\
  }\href {https://doi.org/10.1088/0264-9381/27/19/194002} {\bibfield  {journal}
  {\bibinfo  {journal} {Class. Quant. Grav.}\ }\textbf {\bibinfo {volume}
  {27}},\ \bibinfo {pages} {194002} (\bibinfo {year} {2010})}\BibitemShut
  {NoStop}%
\bibitem [{\citenamefont {Hild}\ \emph {et~al.}(2011)\citenamefont {Hild} \emph
  {et~al.}}]{Hild:2010id}%
  \BibitemOpen
  \bibfield  {author} {\bibinfo {author} {\bibfnamefont {S.}~\bibnamefont
  {Hild}} \emph {et~al.},\ }\bibfield  {title} {\bibinfo {title} {{Sensitivity
  Studies for Third-Generation Gravitational Wave Observatories}},\ }\href
  {https://doi.org/10.1088/0264-9381/28/9/094013} {\bibfield  {journal}
  {\bibinfo  {journal} {Class. Quant. Grav.}\ }\textbf {\bibinfo {volume}
  {28}},\ \bibinfo {pages} {094013} (\bibinfo {year} {2011})},\ \Eprint
  {https://arxiv.org/abs/1012.0908} {arXiv:1012.0908 [gr-qc]} \BibitemShut
  {NoStop}%
\bibitem [{\citenamefont {Abbott}\ \emph
  {et~al.}(2017{\natexlab{c}})\citenamefont {Abbott} \emph
  {et~al.}}]{LIGOScientific:2016wof}%
  \BibitemOpen
  \bibfield  {author} {\bibinfo {author} {\bibfnamefont {B.~P.}\ \bibnamefont
  {Abbott}} \emph {et~al.} (\bibinfo {collaboration} {LIGO Scientific}),\
  }\bibfield  {title} {\bibinfo {title} {{Exploring the Sensitivity of Next
  Generation Gravitational Wave Detectors}},\ }\href
  {https://doi.org/10.1088/1361-6382/aa51f4} {\bibfield  {journal} {\bibinfo
  {journal} {Class. Quant. Grav.}\ }\textbf {\bibinfo {volume} {34}},\ \bibinfo
  {pages} {044001} (\bibinfo {year} {2017}{\natexlab{c}})},\ \Eprint
  {https://arxiv.org/abs/1607.08697} {arXiv:1607.08697 [astro-ph.IM]}
  \BibitemShut {NoStop}%
\bibitem [{\citenamefont {Reitze}\ \emph {et~al.}(2019)\citenamefont {Reitze}
  \emph {et~al.}}]{Reitze:2019iox}%
  \BibitemOpen
  \bibfield  {author} {\bibinfo {author} {\bibfnamefont {D.}~\bibnamefont
  {Reitze}} \emph {et~al.},\ }\bibfield  {title} {\bibinfo {title} {{Cosmic
  Explorer: The U.S. Contribution to Gravitational-Wave Astronomy beyond
  LIGO}},\ }\href@noop {} {\bibfield  {journal} {\bibinfo  {journal} {Bull. Am.
  Astron. Soc.}\ }\textbf {\bibinfo {volume} {51}},\ \bibinfo {pages} {035}
  (\bibinfo {year} {2019})},\ \Eprint {https://arxiv.org/abs/1907.04833}
  {arXiv:1907.04833 [astro-ph.IM]} \BibitemShut {NoStop}%
\bibitem [{\citenamefont {Evans}\ \emph {et~al.}(2021)\citenamefont {Evans}
  \emph {et~al.}}]{Evans:2021gyd}%
  \BibitemOpen
  \bibfield  {author} {\bibinfo {author} {\bibfnamefont {M.}~\bibnamefont
  {Evans}} \emph {et~al.},\ }\bibfield  {title} {\bibinfo {title} {{A Horizon
  Study for Cosmic Explorer: Science, Observatories, and Community}},\
  }\href@noop {} {\  (\bibinfo {year} {2021})},\ \Eprint
  {https://arxiv.org/abs/2109.09882} {arXiv:2109.09882 [astro-ph.IM]}
  \BibitemShut {NoStop}%
\bibitem [{\citenamefont {Evans}\ \emph {et~al.}(2023)\citenamefont {Evans}
  \emph {et~al.}}]{Evans:2023euw}%
  \BibitemOpen
  \bibfield  {author} {\bibinfo {author} {\bibfnamefont {M.}~\bibnamefont
  {Evans}} \emph {et~al.},\ }\bibfield  {title} {\bibinfo {title} {{Cosmic
  Explorer: A Submission to the NSF MPSAC ngGW Subcommittee}},\ }\href@noop {}
  {\  (\bibinfo {year} {2023})},\ \Eprint {https://arxiv.org/abs/2306.13745}
  {arXiv:2306.13745 [astro-ph.IM]} \BibitemShut {NoStop}%
\bibitem [{\citenamefont {Branchesi}\ \emph {et~al.}(2023)\citenamefont
  {Branchesi} \emph {et~al.}}]{Branchesi:2023mws}%
  \BibitemOpen
  \bibfield  {author} {\bibinfo {author} {\bibfnamefont {M.}~\bibnamefont
  {Branchesi}} \emph {et~al.},\ }\bibfield  {title} {\bibinfo {title} {{Science
  with the Einstein Telescope: a comparison of different designs}},\ }\href
  {https://doi.org/10.1088/1475-7516/2023/07/068} {\bibfield  {journal}
  {\bibinfo  {journal} {JCAP}\ }\textbf {\bibinfo {volume} {07}},\ \bibinfo
  {pages} {068}},\ \Eprint {https://arxiv.org/abs/2303.15923} {arXiv:2303.15923
  [gr-qc]} \BibitemShut {NoStop}%
\bibitem [{\citenamefont {Gupta}\ \emph {et~al.}(2023)\citenamefont {Gupta}
  \emph {et~al.}}]{Gupta:2023lga}%
  \BibitemOpen
  \bibfield  {author} {\bibinfo {author} {\bibfnamefont {I.}~\bibnamefont
  {Gupta}} \emph {et~al.},\ }\bibfield  {title} {\bibinfo {title}
  {{Characterizing Gravitational Wave Detector Networks: From A$^\sharp$ to
  Cosmic Explorer}},\ }\href@noop {} {\  (\bibinfo {year} {2023})},\ \Eprint
  {https://arxiv.org/abs/2307.10421} {arXiv:2307.10421 [gr-qc]} \BibitemShut
  {NoStop}%
\bibitem [{\citenamefont {Perego}\ \emph {et~al.}(2019)\citenamefont {Perego},
  \citenamefont {Bernuzzi},\ and\ \citenamefont {Radice}}]{Perego:2019adq}%
  \BibitemOpen
  \bibfield  {author} {\bibinfo {author} {\bibfnamefont {A.}~\bibnamefont
  {Perego}}, \bibinfo {author} {\bibfnamefont {S.}~\bibnamefont {Bernuzzi}},\
  and\ \bibinfo {author} {\bibfnamefont {D.}~\bibnamefont {Radice}},\
  }\bibfield  {title} {\bibinfo {title} {{Thermodynamics conditions of matter
  in neutron star mergers}},\ }\href
  {https://doi.org/10.1140/epja/i2019-12810-7} {\bibfield  {journal} {\bibinfo
  {journal} {Eur. Phys. J. A}\ }\textbf {\bibinfo {volume} {55}},\ \bibinfo
  {pages} {124} (\bibinfo {year} {2019})},\ \Eprint
  {https://arxiv.org/abs/1903.07898} {arXiv:1903.07898 [gr-qc]} \BibitemShut
  {NoStop}%
\bibitem [{\citenamefont {Hammond}\ \emph {et~al.}(2021)\citenamefont
  {Hammond}, \citenamefont {Hawke},\ and\ \citenamefont
  {Andersson}}]{Hammond:2021vtv}%
  \BibitemOpen
  \bibfield  {author} {\bibinfo {author} {\bibfnamefont {P.}~\bibnamefont
  {Hammond}}, \bibinfo {author} {\bibfnamefont {I.}~\bibnamefont {Hawke}},\
  and\ \bibinfo {author} {\bibfnamefont {N.}~\bibnamefont {Andersson}},\
  }\bibfield  {title} {\bibinfo {title} {{Thermal aspects of neutron star
  mergers}},\ }\href {https://doi.org/10.1103/PhysRevD.104.103006} {\bibfield
  {journal} {\bibinfo  {journal} {Phys. Rev. D}\ }\textbf {\bibinfo {volume}
  {104}},\ \bibinfo {pages} {103006} (\bibinfo {year} {2021})},\ \Eprint
  {https://arxiv.org/abs/2108.08649} {arXiv:2108.08649 [astro-ph.HE]}
  \BibitemShut {NoStop}%
\bibitem [{\citenamefont {Blacker}\ \emph {et~al.}(2023)\citenamefont
  {Blacker}, \citenamefont {Bauswein},\ and\ \citenamefont
  {Typel}}]{Blacker:2023onp}%
  \BibitemOpen
  \bibfield  {author} {\bibinfo {author} {\bibfnamefont {S.}~\bibnamefont
  {Blacker}}, \bibinfo {author} {\bibfnamefont {A.}~\bibnamefont {Bauswein}},\
  and\ \bibinfo {author} {\bibfnamefont {S.}~\bibnamefont {Typel}},\ }\bibfield
   {title} {\bibinfo {title} {{Exploring thermal effects of the hadron-quark
  matter transition in neutron star mergers}},\ }\href@noop {} {\  (\bibinfo
  {year} {2023})},\ \Eprint {https://arxiv.org/abs/2304.01971}
  {arXiv:2304.01971 [astro-ph.HE]} \BibitemShut {NoStop}%
\bibitem [{\citenamefont {Most}\ and\ \citenamefont
  {Raithel}(2021)}]{Most:2021ktk}%
  \BibitemOpen
  \bibfield  {author} {\bibinfo {author} {\bibfnamefont {E.~R.}\ \bibnamefont
  {Most}}\ and\ \bibinfo {author} {\bibfnamefont {C.~A.}\ \bibnamefont
  {Raithel}},\ }\bibfield  {title} {\bibinfo {title} {{Impact of the nuclear
  symmetry energy on the post-merger phase of a binary neutron star
  coalescence}},\ }\href {https://doi.org/10.1103/PhysRevD.104.124012}
  {\bibfield  {journal} {\bibinfo  {journal} {Phys. Rev. D}\ }\textbf {\bibinfo
  {volume} {104}},\ \bibinfo {pages} {124012} (\bibinfo {year} {2021})},\
  \Eprint {https://arxiv.org/abs/2107.06804} {arXiv:2107.06804 [astro-ph.HE]}
  \BibitemShut {NoStop}%
\bibitem [{\citenamefont {Fields}\ \emph {et~al.}(2023)\citenamefont {Fields},
  \citenamefont {Prakash}, \citenamefont {Breschi}, \citenamefont {Radice},
  \citenamefont {Bernuzzi},\ and\ \citenamefont {Schneider}}]{Fields:2023bhs}%
  \BibitemOpen
  \bibfield  {author} {\bibinfo {author} {\bibfnamefont {J.}~\bibnamefont
  {Fields}}, \bibinfo {author} {\bibfnamefont {A.}~\bibnamefont {Prakash}},
  \bibinfo {author} {\bibfnamefont {M.}~\bibnamefont {Breschi}}, \bibinfo
  {author} {\bibfnamefont {D.}~\bibnamefont {Radice}}, \bibinfo {author}
  {\bibfnamefont {S.}~\bibnamefont {Bernuzzi}},\ and\ \bibinfo {author}
  {\bibfnamefont {A.~d.~S.}\ \bibnamefont {Schneider}},\ }\bibfield  {title}
  {\bibinfo {title} {{Thermal Effects in Binary Neutron Star Mergers}},\
  }\href@noop {} {\  (\bibinfo {year} {2023})},\ \Eprint
  {https://arxiv.org/abs/2302.11359} {arXiv:2302.11359 [astro-ph.HE]}
  \BibitemShut {NoStop}%
\bibitem [{\citenamefont {Radice}\ \emph {et~al.}(2016)\citenamefont {Radice},
  \citenamefont {Galeazzi}, \citenamefont {Lippuner}, \citenamefont {Roberts},
  \citenamefont {Ott},\ and\ \citenamefont {Rezzolla}}]{Radice:2016dwd}%
  \BibitemOpen
  \bibfield  {author} {\bibinfo {author} {\bibfnamefont {D.}~\bibnamefont
  {Radice}}, \bibinfo {author} {\bibfnamefont {F.}~\bibnamefont {Galeazzi}},
  \bibinfo {author} {\bibfnamefont {J.}~\bibnamefont {Lippuner}}, \bibinfo
  {author} {\bibfnamefont {L.~F.}\ \bibnamefont {Roberts}}, \bibinfo {author}
  {\bibfnamefont {C.~D.}\ \bibnamefont {Ott}},\ and\ \bibinfo {author}
  {\bibfnamefont {L.}~\bibnamefont {Rezzolla}},\ }\bibfield  {title} {\bibinfo
  {title} {{Dynamical Mass Ejection from Binary Neutron Star Mergers}},\ }\href
  {https://doi.org/10.1093/mnras/stw1227} {\bibfield  {journal} {\bibinfo
  {journal} {Mon. Not. Roy. Astron. Soc.}\ }\textbf {\bibinfo {volume} {460}},\
  \bibinfo {pages} {3255} (\bibinfo {year} {2016})},\ \Eprint
  {https://arxiv.org/abs/1601.02426} {arXiv:1601.02426 [astro-ph.HE]}
  \BibitemShut {NoStop}%
\bibitem [{\citenamefont {Radice}\ \emph {et~al.}(2018)\citenamefont {Radice},
  \citenamefont {Perego}, \citenamefont {Hotokezaka}, \citenamefont {Fromm},
  \citenamefont {Bernuzzi},\ and\ \citenamefont {Roberts}}]{Radice:2018pdn}%
  \BibitemOpen
  \bibfield  {author} {\bibinfo {author} {\bibfnamefont {D.}~\bibnamefont
  {Radice}}, \bibinfo {author} {\bibfnamefont {A.}~\bibnamefont {Perego}},
  \bibinfo {author} {\bibfnamefont {K.}~\bibnamefont {Hotokezaka}}, \bibinfo
  {author} {\bibfnamefont {S.~A.}\ \bibnamefont {Fromm}}, \bibinfo {author}
  {\bibfnamefont {S.}~\bibnamefont {Bernuzzi}},\ and\ \bibinfo {author}
  {\bibfnamefont {L.~F.}\ \bibnamefont {Roberts}},\ }\bibfield  {title}
  {\bibinfo {title} {{Binary Neutron Star Mergers: Mass Ejection,
  Electromagnetic Counterparts and Nucleosynthesis}},\ }\href
  {https://doi.org/10.3847/1538-4357/aaf054} {\bibfield  {journal} {\bibinfo
  {journal} {Astrophys. J.}\ }\textbf {\bibinfo {volume} {869}},\ \bibinfo
  {pages} {130} (\bibinfo {year} {2018})},\ \Eprint
  {https://arxiv.org/abs/1809.11161} {arXiv:1809.11161 [astro-ph.HE]}
  \BibitemShut {NoStop}%
\bibitem [{\citenamefont {Radice}\ \emph {et~al.}(2022)\citenamefont {Radice},
  \citenamefont {Bernuzzi}, \citenamefont {Perego},\ and\ \citenamefont
  {Haas}}]{Radice:2021jtw}%
  \BibitemOpen
  \bibfield  {author} {\bibinfo {author} {\bibfnamefont {D.}~\bibnamefont
  {Radice}}, \bibinfo {author} {\bibfnamefont {S.}~\bibnamefont {Bernuzzi}},
  \bibinfo {author} {\bibfnamefont {A.}~\bibnamefont {Perego}},\ and\ \bibinfo
  {author} {\bibfnamefont {R.}~\bibnamefont {Haas}},\ }\bibfield  {title}
  {\bibinfo {title} {{A new moment-based general-relativistic
  neutrino-radiation transport code: Methods and first applications to neutron
  star mergers}},\ }\href {https://doi.org/10.1093/mnras/stac589} {\bibfield
  {journal} {\bibinfo  {journal} {Mon. Not. Roy. Astron. Soc.}\ }\textbf
  {\bibinfo {volume} {512}},\ \bibinfo {pages} {1499} (\bibinfo {year}
  {2022})},\ \Eprint {https://arxiv.org/abs/2111.14858} {arXiv:2111.14858
  [astro-ph.HE]} \BibitemShut {NoStop}%
\bibitem [{\citenamefont {Schianchi}\ \emph {et~al.}(2023)\citenamefont
  {Schianchi}, \citenamefont {Gieg}, \citenamefont {Nedora}, \citenamefont
  {Neuweiler}, \citenamefont {Ujevic}, \citenamefont {Bulla},\ and\
  \citenamefont {Dietrich}}]{Schianchi:2023uky}%
  \BibitemOpen
  \bibfield  {author} {\bibinfo {author} {\bibfnamefont {F.}~\bibnamefont
  {Schianchi}}, \bibinfo {author} {\bibfnamefont {H.}~\bibnamefont {Gieg}},
  \bibinfo {author} {\bibfnamefont {V.}~\bibnamefont {Nedora}}, \bibinfo
  {author} {\bibfnamefont {A.}~\bibnamefont {Neuweiler}}, \bibinfo {author}
  {\bibfnamefont {M.}~\bibnamefont {Ujevic}}, \bibinfo {author} {\bibfnamefont
  {M.}~\bibnamefont {Bulla}},\ and\ \bibinfo {author} {\bibfnamefont
  {T.}~\bibnamefont {Dietrich}},\ }\bibfield  {title} {\bibinfo {title} {{M1
  neutrino transport within the numerical-relativistic code BAM with
  application to low mass binary neutron star mergers}},\ }\href@noop {} {\
  (\bibinfo {year} {2023})},\ \Eprint {https://arxiv.org/abs/2307.04572}
  {arXiv:2307.04572 [gr-qc]} \BibitemShut {NoStop}%
\bibitem [{\citenamefont {Foucart}(2022)}]{Foucart:2022bth}%
  \BibitemOpen
  \bibfield  {author} {\bibinfo {author} {\bibfnamefont {F.}~\bibnamefont
  {Foucart}},\ }\bibfield  {title} {\bibinfo {title} {{Neutrino transport in
  general relativistic neutron star merger simulations}}\ }\href
  {https://doi.org/10.1007/s41115-023-00016-y} {10.1007/s41115-023-00016-y}
  (\bibinfo {year} {2022}),\ \Eprint {https://arxiv.org/abs/2209.02538}
  {arXiv:2209.02538 [astro-ph.HE]} \BibitemShut {NoStop}%
\bibitem [{\citenamefont {Radice}\ and\ \citenamefont
  {Bernuzzi}(2023)}]{Radice:2023zlw}%
  \BibitemOpen
  \bibfield  {author} {\bibinfo {author} {\bibfnamefont {D.}~\bibnamefont
  {Radice}}\ and\ \bibinfo {author} {\bibfnamefont {S.}~\bibnamefont
  {Bernuzzi}},\ }\bibfield  {title} {\bibinfo {title} {{Ab-Initio
  General-Relativistic Neutrino-Radiation Hydrodynamics Simulations of
  Long-Lived Neutron Star Merger Remnants to Neutrino Cooling Timescales}},\
  }\href@noop {} {\  (\bibinfo {year} {2023})},\ \Eprint
  {https://arxiv.org/abs/2306.13709} {arXiv:2306.13709 [astro-ph.HE]}
  \BibitemShut {NoStop}%
\bibitem [{\citenamefont {Zappa}\ \emph {et~al.}(2022)\citenamefont {Zappa},
  \citenamefont {Bernuzzi}, \citenamefont {Radice},\ and\ \citenamefont
  {Perego}}]{Zappa:2022rpd}%
  \BibitemOpen
  \bibfield  {author} {\bibinfo {author} {\bibfnamefont {F.}~\bibnamefont
  {Zappa}}, \bibinfo {author} {\bibfnamefont {S.}~\bibnamefont {Bernuzzi}},
  \bibinfo {author} {\bibfnamefont {D.}~\bibnamefont {Radice}},\ and\ \bibinfo
  {author} {\bibfnamefont {A.}~\bibnamefont {Perego}},\ }\bibfield  {title}
  {\bibinfo {title} {{Binary neutron star merger simulations with neutrino
  transport and turbulent viscosity: impact of different schemes and grid
  resolution}}\ }\href {https://doi.org/10.1093/mnras/stad107}
  {10.1093/mnras/stad107} (\bibinfo {year} {2022}),\ \Eprint
  {https://arxiv.org/abs/2210.11491} {arXiv:2210.11491 [astro-ph.HE]}
  \BibitemShut {NoStop}%
\bibitem [{\citenamefont {Loffredo}\ \emph {et~al.}(2023)\citenamefont
  {Loffredo}, \citenamefont {Perego}, \citenamefont {Logoteta},\ and\
  \citenamefont {Branchesi}}]{Loffredo:2022prq}%
  \BibitemOpen
  \bibfield  {author} {\bibinfo {author} {\bibfnamefont {E.}~\bibnamefont
  {Loffredo}}, \bibinfo {author} {\bibfnamefont {A.}~\bibnamefont {Perego}},
  \bibinfo {author} {\bibfnamefont {D.}~\bibnamefont {Logoteta}},\ and\
  \bibinfo {author} {\bibfnamefont {M.}~\bibnamefont {Branchesi}},\ }\bibfield
  {title} {\bibinfo {title} {{Muons in the aftermath of neutron star mergers
  and their impact on trapped neutrinos}},\ }\href
  {https://doi.org/10.1051/0004-6361/202244927} {\bibfield  {journal} {\bibinfo
   {journal} {Astron. Astrophys.}\ }\textbf {\bibinfo {volume} {672}},\
  \bibinfo {pages} {A124} (\bibinfo {year} {2023})},\ \Eprint
  {https://arxiv.org/abs/2209.04458} {arXiv:2209.04458 [astro-ph.HE]}
  \BibitemShut {NoStop}%
\bibitem [{\citenamefont {Camilletti}\ \emph {et~al.}(2022)\citenamefont
  {Camilletti}, \citenamefont {Chiesa}, \citenamefont {Ricigliano},
  \citenamefont {Perego}, \citenamefont {Lippold}, \citenamefont {Padamata},
  \citenamefont {Bernuzzi}, \citenamefont {Radice}, \citenamefont {Logoteta},\
  and\ \citenamefont {Guercilena}}]{Camilletti:2022jms}%
  \BibitemOpen
  \bibfield  {author} {\bibinfo {author} {\bibfnamefont {A.}~\bibnamefont
  {Camilletti}}, \bibinfo {author} {\bibfnamefont {L.}~\bibnamefont {Chiesa}},
  \bibinfo {author} {\bibfnamefont {G.}~\bibnamefont {Ricigliano}}, \bibinfo
  {author} {\bibfnamefont {A.}~\bibnamefont {Perego}}, \bibinfo {author}
  {\bibfnamefont {L.~C.}\ \bibnamefont {Lippold}}, \bibinfo {author}
  {\bibfnamefont {S.}~\bibnamefont {Padamata}}, \bibinfo {author}
  {\bibfnamefont {S.}~\bibnamefont {Bernuzzi}}, \bibinfo {author}
  {\bibfnamefont {D.}~\bibnamefont {Radice}}, \bibinfo {author} {\bibfnamefont
  {D.}~\bibnamefont {Logoteta}},\ and\ \bibinfo {author} {\bibfnamefont
  {F.~M.}\ \bibnamefont {Guercilena}},\ }\bibfield  {title} {\bibinfo {title}
  {{Numerical relativity simulations of the neutron star merger GW190425:
  microphysics and mass ratio effects}}\ }\href
  {https://doi.org/10.1093/mnras/stac2333} {10.1093/mnras/stac2333} (\bibinfo
  {year} {2022}),\ \Eprint {https://arxiv.org/abs/2204.05336} {arXiv:2204.05336
  [astro-ph.HE]} \BibitemShut {NoStop}%
\bibitem [{\citenamefont {Most}\ \emph {et~al.}(2021)\citenamefont {Most},
  \citenamefont {Harris}, \citenamefont {Plumberg}, \citenamefont {Alford},
  \citenamefont {Noronha}, \citenamefont {Noronha-Hostler}, \citenamefont
  {Pretorius}, \citenamefont {Witek},\ and\ \citenamefont
  {Yunes}}]{Most:2021zvc}%
  \BibitemOpen
  \bibfield  {author} {\bibinfo {author} {\bibfnamefont {E.~R.}\ \bibnamefont
  {Most}}, \bibinfo {author} {\bibfnamefont {S.~P.}\ \bibnamefont {Harris}},
  \bibinfo {author} {\bibfnamefont {C.}~\bibnamefont {Plumberg}}, \bibinfo
  {author} {\bibfnamefont {M.~G.}\ \bibnamefont {Alford}}, \bibinfo {author}
  {\bibfnamefont {J.}~\bibnamefont {Noronha}}, \bibinfo {author} {\bibfnamefont
  {J.}~\bibnamefont {Noronha-Hostler}}, \bibinfo {author} {\bibfnamefont
  {F.}~\bibnamefont {Pretorius}}, \bibinfo {author} {\bibfnamefont
  {H.}~\bibnamefont {Witek}},\ and\ \bibinfo {author} {\bibfnamefont
  {N.}~\bibnamefont {Yunes}},\ }\bibfield  {title} {\bibinfo {title}
  {{Projecting the likely importance of weak-interaction-driven bulk viscosity
  in neutron star mergers}},\ }\href {https://doi.org/10.1093/mnras/stab2793}
  {\bibfield  {journal} {\bibinfo  {journal} {Mon. Not. Roy. Astron. Soc.}\
  }\textbf {\bibinfo {volume} {509}},\ \bibinfo {pages} {1096} (\bibinfo {year}
  {2021})},\ \Eprint {https://arxiv.org/abs/2107.05094} {arXiv:2107.05094
  [astro-ph.HE]} \BibitemShut {NoStop}%
\bibitem [{\citenamefont {Most}\ \emph {et~al.}(2022)\citenamefont {Most},
  \citenamefont {Haber}, \citenamefont {Harris}, \citenamefont {Zhang},
  \citenamefont {Alford},\ and\ \citenamefont {Noronha}}]{Most:2022yhe}%
  \BibitemOpen
  \bibfield  {author} {\bibinfo {author} {\bibfnamefont {E.~R.}\ \bibnamefont
  {Most}}, \bibinfo {author} {\bibfnamefont {A.}~\bibnamefont {Haber}},
  \bibinfo {author} {\bibfnamefont {S.~P.}\ \bibnamefont {Harris}}, \bibinfo
  {author} {\bibfnamefont {Z.}~\bibnamefont {Zhang}}, \bibinfo {author}
  {\bibfnamefont {M.~G.}\ \bibnamefont {Alford}},\ and\ \bibinfo {author}
  {\bibfnamefont {J.}~\bibnamefont {Noronha}},\ }\bibfield  {title} {\bibinfo
  {title} {{Emergence of microphysical viscosity in binary neutron star
  post-merger dynamics}},\ }\href@noop {} {\  (\bibinfo {year} {2022})},\
  \Eprint {https://arxiv.org/abs/2207.00442} {arXiv:2207.00442 [astro-ph.HE]}
  \BibitemShut {NoStop}%
\bibitem [{\citenamefont {Combi}\ and\ \citenamefont
  {Siegel}(2023)}]{Combi:2022nhg}%
  \BibitemOpen
  \bibfield  {author} {\bibinfo {author} {\bibfnamefont {L.}~\bibnamefont
  {Combi}}\ and\ \bibinfo {author} {\bibfnamefont {D.~M.}\ \bibnamefont
  {Siegel}},\ }\bibfield  {title} {\bibinfo {title} {{GRMHD Simulations of
  Neutron-star Mergers with Weak Interactions: r-process Nucleosynthesis and
  Electromagnetic Signatures of Dynamical Ejecta}},\ }\href
  {https://doi.org/10.3847/1538-4357/acac29} {\bibfield  {journal} {\bibinfo
  {journal} {Astrophys. J.}\ }\textbf {\bibinfo {volume} {944}},\ \bibinfo
  {pages} {28} (\bibinfo {year} {2023})},\ \Eprint
  {https://arxiv.org/abs/2206.03618} {arXiv:2206.03618 [astro-ph.HE]}
  \BibitemShut {NoStop}%
\bibitem [{\citenamefont {George}\ \emph {et~al.}(2020)\citenamefont {George},
  \citenamefont {Wu}, \citenamefont {Tamborra}, \citenamefont
  {Ardevol-Pulpillo},\ and\ \citenamefont {Janka}}]{George:2020veu}%
  \BibitemOpen
  \bibfield  {author} {\bibinfo {author} {\bibfnamefont {M.}~\bibnamefont
  {George}}, \bibinfo {author} {\bibfnamefont {M.-R.}\ \bibnamefont {Wu}},
  \bibinfo {author} {\bibfnamefont {I.}~\bibnamefont {Tamborra}}, \bibinfo
  {author} {\bibfnamefont {R.}~\bibnamefont {Ardevol-Pulpillo}},\ and\ \bibinfo
  {author} {\bibfnamefont {H.-T.}\ \bibnamefont {Janka}},\ }\bibfield  {title}
  {\bibinfo {title} {{Fast neutrino flavor conversion, ejecta properties, and
  nucleosynthesis in newly-formed hypermassive remnants of neutron-star
  mergers}},\ }\href {https://doi.org/10.1103/PhysRevD.102.103015} {\bibfield
  {journal} {\bibinfo  {journal} {Phys. Rev. D}\ }\textbf {\bibinfo {volume}
  {102}},\ \bibinfo {pages} {103015} (\bibinfo {year} {2020})},\ \Eprint
  {https://arxiv.org/abs/2009.04046} {arXiv:2009.04046 [astro-ph.HE]}
  \BibitemShut {NoStop}%
\bibitem [{\citenamefont {Siegel}\ and\ \citenamefont
  {Metzger}(2018)}]{Siegel:2017jug}%
  \BibitemOpen
  \bibfield  {author} {\bibinfo {author} {\bibfnamefont {D.~M.}\ \bibnamefont
  {Siegel}}\ and\ \bibinfo {author} {\bibfnamefont {B.~D.}\ \bibnamefont
  {Metzger}},\ }\bibfield  {title} {\bibinfo {title} {{Three-dimensional GRMHD
  simulations of neutrino-cooled accretion disks from neutron star mergers}},\
  }\href {https://doi.org/10.3847/1538-4357/aabaec} {\bibfield  {journal}
  {\bibinfo  {journal} {Astrophys. J.}\ }\textbf {\bibinfo {volume} {858}},\
  \bibinfo {pages} {52} (\bibinfo {year} {2018})},\ \Eprint
  {https://arxiv.org/abs/1711.00868} {arXiv:1711.00868 [astro-ph.HE]}
  \BibitemShut {NoStop}%
\bibitem [{\citenamefont {Martin}\ \emph {et~al.}(2018)\citenamefont {Martin},
  \citenamefont {Perego}, \citenamefont {Kastaun},\ and\ \citenamefont
  {Arcones}}]{Martin:2017dhc}%
  \BibitemOpen
  \bibfield  {author} {\bibinfo {author} {\bibfnamefont {D.}~\bibnamefont
  {Martin}}, \bibinfo {author} {\bibfnamefont {A.}~\bibnamefont {Perego}},
  \bibinfo {author} {\bibfnamefont {W.}~\bibnamefont {Kastaun}},\ and\ \bibinfo
  {author} {\bibfnamefont {A.}~\bibnamefont {Arcones}},\ }\bibfield  {title}
  {\bibinfo {title} {{The role of weak interactions in dynamic ejecta from
  binary neutron star mergers}},\ }\href
  {https://doi.org/10.1088/1361-6382/aa9f5a} {\bibfield  {journal} {\bibinfo
  {journal} {Class. Quant. Grav.}\ }\textbf {\bibinfo {volume} {35}},\ \bibinfo
  {pages} {034001} (\bibinfo {year} {2018})},\ \Eprint
  {https://arxiv.org/abs/1710.04900} {arXiv:1710.04900 [astro-ph.HE]}
  \BibitemShut {NoStop}%
\bibitem [{\citenamefont {Fujibayashi}\ \emph {et~al.}(2020)\citenamefont
  {Fujibayashi}, \citenamefont {Shibata}, \citenamefont {Wanajo}, \citenamefont
  {Kiuchi}, \citenamefont {Kyutoku},\ and\ \citenamefont
  {Sekiguchi}}]{Fujibayashi:2020qda}%
  \BibitemOpen
  \bibfield  {author} {\bibinfo {author} {\bibfnamefont {S.}~\bibnamefont
  {Fujibayashi}}, \bibinfo {author} {\bibfnamefont {M.}~\bibnamefont
  {Shibata}}, \bibinfo {author} {\bibfnamefont {S.}~\bibnamefont {Wanajo}},
  \bibinfo {author} {\bibfnamefont {K.}~\bibnamefont {Kiuchi}}, \bibinfo
  {author} {\bibfnamefont {K.}~\bibnamefont {Kyutoku}},\ and\ \bibinfo {author}
  {\bibfnamefont {Y.}~\bibnamefont {Sekiguchi}},\ }\bibfield  {title} {\bibinfo
  {title} {{Mass ejection from disks surrounding a low-mass black hole: Viscous
  neutrino-radiation hydrodynamics simulation in full general relativity}},\
  }\href {https://doi.org/10.1103/PhysRevD.101.083029} {\bibfield  {journal}
  {\bibinfo  {journal} {Phys. Rev. D}\ }\textbf {\bibinfo {volume} {101}},\
  \bibinfo {pages} {083029} (\bibinfo {year} {2020})},\ \Eprint
  {https://arxiv.org/abs/2001.04467} {arXiv:2001.04467 [astro-ph.HE]}
  \BibitemShut {NoStop}%
\bibitem [{\citenamefont {Grohs}\ \emph {et~al.}(2022)\citenamefont {Grohs},
  \citenamefont {Richers}, \citenamefont {Couch}, \citenamefont {Foucart},
  \citenamefont {Kneller},\ and\ \citenamefont {McLaughlin}}]{Grohs:2022fyq}%
  \BibitemOpen
  \bibfield  {author} {\bibinfo {author} {\bibfnamefont {E.}~\bibnamefont
  {Grohs}}, \bibinfo {author} {\bibfnamefont {S.}~\bibnamefont {Richers}},
  \bibinfo {author} {\bibfnamefont {S.~M.}\ \bibnamefont {Couch}}, \bibinfo
  {author} {\bibfnamefont {F.}~\bibnamefont {Foucart}}, \bibinfo {author}
  {\bibfnamefont {J.~P.}\ \bibnamefont {Kneller}},\ and\ \bibinfo {author}
  {\bibfnamefont {G.~C.}\ \bibnamefont {McLaughlin}},\ }\bibfield  {title}
  {\bibinfo {title} {{Neutrino Fast Flavor Instability in three dimensions for
  a Neutron Star Merger}},\ }\href@noop {} {\  (\bibinfo {year} {2022})},\
  \Eprint {https://arxiv.org/abs/2207.02214} {arXiv:2207.02214 [hep-ph]}
  \BibitemShut {NoStop}%
\bibitem [{\citenamefont {Grohs}\ \emph {et~al.}(2023)\citenamefont {Grohs},
  \citenamefont {Richers}, \citenamefont {Couch}, \citenamefont {Foucart},
  \citenamefont {Froustey}, \citenamefont {Kneller},\ and\ \citenamefont
  {McLaughlin}}]{Grohs:2023pgq}%
  \BibitemOpen
  \bibfield  {author} {\bibinfo {author} {\bibfnamefont {E.}~\bibnamefont
  {Grohs}}, \bibinfo {author} {\bibfnamefont {S.}~\bibnamefont {Richers}},
  \bibinfo {author} {\bibfnamefont {S.~M.}\ \bibnamefont {Couch}}, \bibinfo
  {author} {\bibfnamefont {F.}~\bibnamefont {Foucart}}, \bibinfo {author}
  {\bibfnamefont {J.}~\bibnamefont {Froustey}}, \bibinfo {author}
  {\bibfnamefont {J.}~\bibnamefont {Kneller}},\ and\ \bibinfo {author}
  {\bibfnamefont {G.}~\bibnamefont {McLaughlin}},\ }\bibfield  {title}
  {\bibinfo {title} {{Two-Moment Neutrino Flavor Transformation with
  applications to the Fast Flavor Instability in Neutron Star Mergers}},\
  }\href@noop {} {\  (\bibinfo {year} {2023})},\ \Eprint
  {https://arxiv.org/abs/2309.00972} {arXiv:2309.00972 [astro-ph.HE]}
  \BibitemShut {NoStop}%
\bibitem [{\citenamefont {Richers}\ \emph {et~al.}(2019)\citenamefont
  {Richers}, \citenamefont {McLaughlin}, \citenamefont {Kneller},\ and\
  \citenamefont {Vlasenko}}]{Richers:2019grc}%
  \BibitemOpen
  \bibfield  {author} {\bibinfo {author} {\bibfnamefont {S.~A.}\ \bibnamefont
  {Richers}}, \bibinfo {author} {\bibfnamefont {G.~C.}\ \bibnamefont
  {McLaughlin}}, \bibinfo {author} {\bibfnamefont {J.~P.}\ \bibnamefont
  {Kneller}},\ and\ \bibinfo {author} {\bibfnamefont {A.}~\bibnamefont
  {Vlasenko}},\ }\bibfield  {title} {\bibinfo {title} {{Neutrino Quantum
  Kinetics in Compact Objects}},\ }\href
  {https://doi.org/10.1103/PhysRevD.99.123014} {\bibfield  {journal} {\bibinfo
  {journal} {Phys. Rev. D}\ }\textbf {\bibinfo {volume} {99}},\ \bibinfo
  {pages} {123014} (\bibinfo {year} {2019})},\ \Eprint
  {https://arxiv.org/abs/1903.00022} {arXiv:1903.00022 [astro-ph.HE]}
  \BibitemShut {NoStop}%
\bibitem [{\citenamefont {Sekiguchi}\ \emph {et~al.}(2011)\citenamefont
  {Sekiguchi}, \citenamefont {Kiuchi}, \citenamefont {Kyutoku},\ and\
  \citenamefont {Shibata}}]{Sekiguchi:2011mc}%
  \BibitemOpen
  \bibfield  {author} {\bibinfo {author} {\bibfnamefont {Y.}~\bibnamefont
  {Sekiguchi}}, \bibinfo {author} {\bibfnamefont {K.}~\bibnamefont {Kiuchi}},
  \bibinfo {author} {\bibfnamefont {K.}~\bibnamefont {Kyutoku}},\ and\ \bibinfo
  {author} {\bibfnamefont {M.}~\bibnamefont {Shibata}},\ }\bibfield  {title}
  {\bibinfo {title} {{Effects of hyperons in binary neutron star mergers}},\
  }\href {https://doi.org/10.1103/PhysRevLett.107.211101} {\bibfield  {journal}
  {\bibinfo  {journal} {Phys. Rev. Lett.}\ }\textbf {\bibinfo {volume} {107}},\
  \bibinfo {pages} {211101} (\bibinfo {year} {2011})},\ \Eprint
  {https://arxiv.org/abs/1110.4442} {arXiv:1110.4442 [astro-ph.HE]}
  \BibitemShut {NoStop}%
\bibitem [{\citenamefont {Radice}\ \emph {et~al.}(2017)\citenamefont {Radice},
  \citenamefont {Bernuzzi}, \citenamefont {Del~Pozzo}, \citenamefont
  {Roberts},\ and\ \citenamefont {Ott}}]{Radice:2016rys}%
  \BibitemOpen
  \bibfield  {author} {\bibinfo {author} {\bibfnamefont {D.}~\bibnamefont
  {Radice}}, \bibinfo {author} {\bibfnamefont {S.}~\bibnamefont {Bernuzzi}},
  \bibinfo {author} {\bibfnamefont {W.}~\bibnamefont {Del~Pozzo}}, \bibinfo
  {author} {\bibfnamefont {L.~F.}\ \bibnamefont {Roberts}},\ and\ \bibinfo
  {author} {\bibfnamefont {C.~D.}\ \bibnamefont {Ott}},\ }\bibfield  {title}
  {\bibinfo {title} {{Probing Extreme-Density Matter with Gravitational Wave
  Observations of Binary Neutron Star Merger Remnants}},\ }\href
  {https://doi.org/10.3847/2041-8213/aa775f} {\bibfield  {journal} {\bibinfo
  {journal} {Astrophys. J. Lett.}\ }\textbf {\bibinfo {volume} {842}},\
  \bibinfo {pages} {L10} (\bibinfo {year} {2017})},\ \Eprint
  {https://arxiv.org/abs/1612.06429} {arXiv:1612.06429 [astro-ph.HE]}
  \BibitemShut {NoStop}%
\bibitem [{\citenamefont {Most}\ \emph {et~al.}(2019)\citenamefont {Most},
  \citenamefont {Papenfort}, \citenamefont {Dexheimer}, \citenamefont
  {Hanauske}, \citenamefont {Schramm}, \citenamefont {St\"ocker},\ and\
  \citenamefont {Rezzolla}}]{Most:2018eaw}%
  \BibitemOpen
  \bibfield  {author} {\bibinfo {author} {\bibfnamefont {E.~R.}\ \bibnamefont
  {Most}}, \bibinfo {author} {\bibfnamefont {L.~J.}\ \bibnamefont {Papenfort}},
  \bibinfo {author} {\bibfnamefont {V.}~\bibnamefont {Dexheimer}}, \bibinfo
  {author} {\bibfnamefont {M.}~\bibnamefont {Hanauske}}, \bibinfo {author}
  {\bibfnamefont {S.}~\bibnamefont {Schramm}}, \bibinfo {author} {\bibfnamefont
  {H.}~\bibnamefont {St\"ocker}},\ and\ \bibinfo {author} {\bibfnamefont
  {L.}~\bibnamefont {Rezzolla}},\ }\bibfield  {title} {\bibinfo {title}
  {{Signatures of quark-hadron phase transitions in general-relativistic
  neutron-star mergers}},\ }\href
  {https://doi.org/10.1103/PhysRevLett.122.061101} {\bibfield  {journal}
  {\bibinfo  {journal} {Phys. Rev. Lett.}\ }\textbf {\bibinfo {volume} {122}},\
  \bibinfo {pages} {061101} (\bibinfo {year} {2019})},\ \Eprint
  {https://arxiv.org/abs/1807.03684} {arXiv:1807.03684 [astro-ph.HE]}
  \BibitemShut {NoStop}%
\bibitem [{\citenamefont {Most}\ \emph {et~al.}(2020)\citenamefont {Most},
  \citenamefont {Jens~Papenfort}, \citenamefont {Dexheimer}, \citenamefont
  {Hanauske}, \citenamefont {Stoecker},\ and\ \citenamefont
  {Rezzolla}}]{Most:2019onn}%
  \BibitemOpen
  \bibfield  {author} {\bibinfo {author} {\bibfnamefont {E.~R.}\ \bibnamefont
  {Most}}, \bibinfo {author} {\bibfnamefont {L.}~\bibnamefont
  {Jens~Papenfort}}, \bibinfo {author} {\bibfnamefont {V.}~\bibnamefont
  {Dexheimer}}, \bibinfo {author} {\bibfnamefont {M.}~\bibnamefont {Hanauske}},
  \bibinfo {author} {\bibfnamefont {H.}~\bibnamefont {Stoecker}},\ and\
  \bibinfo {author} {\bibfnamefont {L.}~\bibnamefont {Rezzolla}},\ }\bibfield
  {title} {\bibinfo {title} {{On the deconfinement phase transition in
  neutron-star mergers}},\ }\href
  {https://doi.org/10.1140/epja/s10050-020-00073-4} {\bibfield  {journal}
  {\bibinfo  {journal} {Eur. Phys. J. A}\ }\textbf {\bibinfo {volume} {56}},\
  \bibinfo {pages} {59} (\bibinfo {year} {2020})},\ \Eprint
  {https://arxiv.org/abs/1910.13893} {arXiv:1910.13893 [astro-ph.HE]}
  \BibitemShut {NoStop}%
\bibitem [{\citenamefont {Bauswein}\ \emph {et~al.}(2019)\citenamefont
  {Bauswein}, \citenamefont {Bastian}, \citenamefont {Blaschke}, \citenamefont
  {Chatziioannou}, \citenamefont {Clark}, \citenamefont {Fischer},\ and\
  \citenamefont {Oertel}}]{Bauswein:2018bma}%
  \BibitemOpen
  \bibfield  {author} {\bibinfo {author} {\bibfnamefont {A.}~\bibnamefont
  {Bauswein}}, \bibinfo {author} {\bibfnamefont {N.-U.~F.}\ \bibnamefont
  {Bastian}}, \bibinfo {author} {\bibfnamefont {D.~B.}\ \bibnamefont
  {Blaschke}}, \bibinfo {author} {\bibfnamefont {K.}~\bibnamefont
  {Chatziioannou}}, \bibinfo {author} {\bibfnamefont {J.~A.}\ \bibnamefont
  {Clark}}, \bibinfo {author} {\bibfnamefont {T.}~\bibnamefont {Fischer}},\
  and\ \bibinfo {author} {\bibfnamefont {M.}~\bibnamefont {Oertel}},\
  }\bibfield  {title} {\bibinfo {title} {{Identifying a first-order phase
  transition in neutron star mergers through gravitational waves}},\ }\href
  {https://doi.org/10.1103/PhysRevLett.122.061102} {\bibfield  {journal}
  {\bibinfo  {journal} {Phys. Rev. Lett.}\ }\textbf {\bibinfo {volume} {122}},\
  \bibinfo {pages} {061102} (\bibinfo {year} {2019})},\ \Eprint
  {https://arxiv.org/abs/1809.01116} {arXiv:1809.01116 [astro-ph.HE]}
  \BibitemShut {NoStop}%
\bibitem [{\citenamefont {Bauswein}\ and\ \citenamefont
  {Blacker}(2020)}]{Bauswein:2020ggy}%
  \BibitemOpen
  \bibfield  {author} {\bibinfo {author} {\bibfnamefont {A.}~\bibnamefont
  {Bauswein}}\ and\ \bibinfo {author} {\bibfnamefont {S.}~\bibnamefont
  {Blacker}},\ }\bibfield  {title} {\bibinfo {title} {{Impact of quark
  deconfinement in neutron star mergers and hybrid star mergers}},\ }\href
  {https://doi.org/10.1140/epjst/e2020-000138-7} {\bibfield  {journal}
  {\bibinfo  {journal} {Eur. Phys. J. ST}\ }\textbf {\bibinfo {volume} {229}},\
  \bibinfo {pages} {3595} (\bibinfo {year} {2020})},\ \Eprint
  {https://arxiv.org/abs/2006.16183} {arXiv:2006.16183 [astro-ph.HE]}
  \BibitemShut {NoStop}%
\bibitem [{\citenamefont {Blacker}\ \emph {et~al.}(2020)\citenamefont
  {Blacker}, \citenamefont {Bastian}, \citenamefont {Bauswein}, \citenamefont
  {Blaschke}, \citenamefont {Fischer}, \citenamefont {Oertel}, \citenamefont
  {Soultanis},\ and\ \citenamefont {Typel}}]{Blacker:2020nlq}%
  \BibitemOpen
  \bibfield  {author} {\bibinfo {author} {\bibfnamefont {S.}~\bibnamefont
  {Blacker}}, \bibinfo {author} {\bibfnamefont {N.-U.~F.}\ \bibnamefont
  {Bastian}}, \bibinfo {author} {\bibfnamefont {A.}~\bibnamefont {Bauswein}},
  \bibinfo {author} {\bibfnamefont {D.~B.}\ \bibnamefont {Blaschke}}, \bibinfo
  {author} {\bibfnamefont {T.}~\bibnamefont {Fischer}}, \bibinfo {author}
  {\bibfnamefont {M.}~\bibnamefont {Oertel}}, \bibinfo {author} {\bibfnamefont
  {T.}~\bibnamefont {Soultanis}},\ and\ \bibinfo {author} {\bibfnamefont
  {S.}~\bibnamefont {Typel}},\ }\bibfield  {title} {\bibinfo {title}
  {{Constraining the onset density of the hadron-quark phase transition with
  gravitational-wave observations}},\ }\href
  {https://doi.org/10.1103/PhysRevD.102.123023} {\bibfield  {journal} {\bibinfo
   {journal} {Phys. Rev. D}\ }\textbf {\bibinfo {volume} {102}},\ \bibinfo
  {pages} {123023} (\bibinfo {year} {2020})},\ \Eprint
  {https://arxiv.org/abs/2006.03789} {arXiv:2006.03789 [astro-ph.HE]}
  \BibitemShut {NoStop}%
\bibitem [{\citenamefont {Weih}\ \emph {et~al.}(2020)\citenamefont {Weih},
  \citenamefont {Hanauske},\ and\ \citenamefont {Rezzolla}}]{Weih:2019xvw}%
  \BibitemOpen
  \bibfield  {author} {\bibinfo {author} {\bibfnamefont {L.~R.}\ \bibnamefont
  {Weih}}, \bibinfo {author} {\bibfnamefont {M.}~\bibnamefont {Hanauske}},\
  and\ \bibinfo {author} {\bibfnamefont {L.}~\bibnamefont {Rezzolla}},\
  }\bibfield  {title} {\bibinfo {title} {{Postmerger Gravitational-Wave
  Signatures of Phase Transitions in Binary Mergers}},\ }\href
  {https://doi.org/10.1103/PhysRevLett.124.171103} {\bibfield  {journal}
  {\bibinfo  {journal} {Phys. Rev. Lett.}\ }\textbf {\bibinfo {volume} {124}},\
  \bibinfo {pages} {171103} (\bibinfo {year} {2020})},\ \Eprint
  {https://arxiv.org/abs/1912.09340} {arXiv:1912.09340 [gr-qc]} \BibitemShut
  {NoStop}%
\bibitem [{\citenamefont {Prakash}\ \emph {et~al.}(2021)\citenamefont
  {Prakash}, \citenamefont {Radice}, \citenamefont {Logoteta}, \citenamefont
  {Perego}, \citenamefont {Nedora}, \citenamefont {Bombaci}, \citenamefont
  {Kashyap}, \citenamefont {Bernuzzi},\ and\ \citenamefont
  {Endrizzi}}]{Prakash:2021wpz}%
  \BibitemOpen
  \bibfield  {author} {\bibinfo {author} {\bibfnamefont {A.}~\bibnamefont
  {Prakash}}, \bibinfo {author} {\bibfnamefont {D.}~\bibnamefont {Radice}},
  \bibinfo {author} {\bibfnamefont {D.}~\bibnamefont {Logoteta}}, \bibinfo
  {author} {\bibfnamefont {A.}~\bibnamefont {Perego}}, \bibinfo {author}
  {\bibfnamefont {V.}~\bibnamefont {Nedora}}, \bibinfo {author} {\bibfnamefont
  {I.}~\bibnamefont {Bombaci}}, \bibinfo {author} {\bibfnamefont
  {R.}~\bibnamefont {Kashyap}}, \bibinfo {author} {\bibfnamefont
  {S.}~\bibnamefont {Bernuzzi}},\ and\ \bibinfo {author} {\bibfnamefont
  {A.}~\bibnamefont {Endrizzi}},\ }\bibfield  {title} {\bibinfo {title}
  {{Signatures of deconfined quark phases in binary neutron star mergers}},\
  }\href {https://doi.org/10.1103/PhysRevD.104.083029} {\bibfield  {journal}
  {\bibinfo  {journal} {Phys. Rev. D}\ }\textbf {\bibinfo {volume} {104}},\
  \bibinfo {pages} {083029} (\bibinfo {year} {2021})},\ \Eprint
  {https://arxiv.org/abs/2106.07885} {arXiv:2106.07885 [astro-ph.HE]}
  \BibitemShut {NoStop}%
\bibitem [{\citenamefont {Liebling}\ \emph {et~al.}(2021)\citenamefont
  {Liebling}, \citenamefont {Palenzuela},\ and\ \citenamefont
  {Lehner}}]{Liebling:2020dhf}%
  \BibitemOpen
  \bibfield  {author} {\bibinfo {author} {\bibfnamefont {S.~L.}\ \bibnamefont
  {Liebling}}, \bibinfo {author} {\bibfnamefont {C.}~\bibnamefont
  {Palenzuela}},\ and\ \bibinfo {author} {\bibfnamefont {L.}~\bibnamefont
  {Lehner}},\ }\bibfield  {title} {\bibinfo {title} {{Effects of High Density
  Phase Transitions on Neutron Star Dynamics}},\ }\href
  {https://doi.org/10.1088/1361-6382/abf898} {\bibfield  {journal} {\bibinfo
  {journal} {Class. Quant. Grav.}\ }\textbf {\bibinfo {volume} {38}},\ \bibinfo
  {pages} {115007} (\bibinfo {year} {2021})},\ \Eprint
  {https://arxiv.org/abs/2010.12567} {arXiv:2010.12567 [gr-qc]} \BibitemShut
  {NoStop}%
\bibitem [{\citenamefont {Kedia}\ \emph {et~al.}(2022)\citenamefont {Kedia},
  \citenamefont {Kim}, \citenamefont {Suh},\ and\ \citenamefont
  {Mathews}}]{Kedia:2022nns}%
  \BibitemOpen
  \bibfield  {author} {\bibinfo {author} {\bibfnamefont {A.}~\bibnamefont
  {Kedia}}, \bibinfo {author} {\bibfnamefont {H.~I.}\ \bibnamefont {Kim}},
  \bibinfo {author} {\bibfnamefont {I.-S.}\ \bibnamefont {Suh}},\ and\ \bibinfo
  {author} {\bibfnamefont {G.~J.}\ \bibnamefont {Mathews}},\ }\bibfield
  {title} {\bibinfo {title} {{Binary neutron star mergers as a probe of
  quark-hadron crossover equations of state}},\ }\href
  {https://doi.org/10.1103/PhysRevD.106.103027} {\bibfield  {journal} {\bibinfo
   {journal} {Phys. Rev. D}\ }\textbf {\bibinfo {volume} {106}},\ \bibinfo
  {pages} {103027} (\bibinfo {year} {2022})},\ \Eprint
  {https://arxiv.org/abs/2203.05461} {arXiv:2203.05461 [gr-qc]} \BibitemShut
  {NoStop}%
\bibitem [{\citenamefont {Mathews}\ \emph {et~al.}(2022)\citenamefont
  {Mathews}, \citenamefont {Kedia}, \citenamefont {Kim},\ and\ \citenamefont
  {Suh}}]{Mathews:2022ria}%
  \BibitemOpen
  \bibfield  {author} {\bibinfo {author} {\bibfnamefont {G.~J.}\ \bibnamefont
  {Mathews}}, \bibinfo {author} {\bibfnamefont {A.}~\bibnamefont {Kedia}},
  \bibinfo {author} {\bibfnamefont {H.~I.}\ \bibnamefont {Kim}},\ and\ \bibinfo
  {author} {\bibfnamefont {I.-S.}\ \bibnamefont {Suh}},\ }\bibfield  {title}
  {\bibinfo {title} {{Neutron Star Mergers and the Quark Matter Equation of
  State}},\ }\href {https://doi.org/10.1051/epjconf/202227401013} {\bibfield
  {journal} {\bibinfo  {journal} {EPJ Web Conf.}\ }\textbf {\bibinfo {volume}
  {274}},\ \bibinfo {pages} {01013} (\bibinfo {year} {2022})},\ \Eprint
  {https://arxiv.org/abs/2302.12897} {arXiv:2302.12897 [astro-ph.HE]}
  \BibitemShut {NoStop}%
\bibitem [{\citenamefont {Huang}\ \emph {et~al.}(2022)\citenamefont {Huang},
  \citenamefont {Baiotti}, \citenamefont {Kojo}, \citenamefont {Takami},
  \citenamefont {Sotani}, \citenamefont {Togashi}, \citenamefont {Hatsuda},
  \citenamefont {Nagataki},\ and\ \citenamefont {Fan}}]{Huang:2022mqp}%
  \BibitemOpen
  \bibfield  {author} {\bibinfo {author} {\bibfnamefont {Y.-J.}\ \bibnamefont
  {Huang}}, \bibinfo {author} {\bibfnamefont {L.}~\bibnamefont {Baiotti}},
  \bibinfo {author} {\bibfnamefont {T.}~\bibnamefont {Kojo}}, \bibinfo {author}
  {\bibfnamefont {K.}~\bibnamefont {Takami}}, \bibinfo {author} {\bibfnamefont
  {H.}~\bibnamefont {Sotani}}, \bibinfo {author} {\bibfnamefont
  {H.}~\bibnamefont {Togashi}}, \bibinfo {author} {\bibfnamefont
  {T.}~\bibnamefont {Hatsuda}}, \bibinfo {author} {\bibfnamefont
  {S.}~\bibnamefont {Nagataki}},\ and\ \bibinfo {author} {\bibfnamefont
  {Y.-Z.}\ \bibnamefont {Fan}},\ }\bibfield  {title} {\bibinfo {title} {{Merger
  and Postmerger of Binary Neutron Stars with a Quark-Hadron Crossover Equation
  of State}},\ }\href {https://doi.org/10.1103/PhysRevLett.129.181101}
  {\bibfield  {journal} {\bibinfo  {journal} {Phys. Rev. Lett.}\ }\textbf
  {\bibinfo {volume} {129}},\ \bibinfo {pages} {181101} (\bibinfo {year}
  {2022})},\ \Eprint {https://arxiv.org/abs/2203.04528} {arXiv:2203.04528
  [astro-ph.HE]} \BibitemShut {NoStop}%
\bibitem [{\citenamefont {Fujimoto}\ \emph {et~al.}(2023)\citenamefont
  {Fujimoto}, \citenamefont {Fukushima}, \citenamefont {Hotokezaka},\ and\
  \citenamefont {Kyutoku}}]{Fujimoto:2022xhv}%
  \BibitemOpen
  \bibfield  {author} {\bibinfo {author} {\bibfnamefont {Y.}~\bibnamefont
  {Fujimoto}}, \bibinfo {author} {\bibfnamefont {K.}~\bibnamefont {Fukushima}},
  \bibinfo {author} {\bibfnamefont {K.}~\bibnamefont {Hotokezaka}},\ and\
  \bibinfo {author} {\bibfnamefont {K.}~\bibnamefont {Kyutoku}},\ }\bibfield
  {title} {\bibinfo {title} {{Gravitational Wave Signal for Quark Matter with
  Realistic Phase Transition}},\ }\href
  {https://doi.org/10.1103/PhysRevLett.130.091404} {\bibfield  {journal}
  {\bibinfo  {journal} {Phys. Rev. Lett.}\ }\textbf {\bibinfo {volume} {130}},\
  \bibinfo {pages} {091404} (\bibinfo {year} {2023})},\ \Eprint
  {https://arxiv.org/abs/2205.03882} {arXiv:2205.03882 [astro-ph.HE]}
  \BibitemShut {NoStop}%
\bibitem [{\citenamefont {Tootle}\ \emph {et~al.}(2022)\citenamefont {Tootle},
  \citenamefont {Ecker}, \citenamefont {Topolski}, \citenamefont {Demircik},
  \citenamefont {J\"arvinen},\ and\ \citenamefont {Rezzolla}}]{Tootle:2022pvd}%
  \BibitemOpen
  \bibfield  {author} {\bibinfo {author} {\bibfnamefont {S.}~\bibnamefont
  {Tootle}}, \bibinfo {author} {\bibfnamefont {C.}~\bibnamefont {Ecker}},
  \bibinfo {author} {\bibfnamefont {K.}~\bibnamefont {Topolski}}, \bibinfo
  {author} {\bibfnamefont {T.}~\bibnamefont {Demircik}}, \bibinfo {author}
  {\bibfnamefont {M.}~\bibnamefont {J\"arvinen}},\ and\ \bibinfo {author}
  {\bibfnamefont {L.}~\bibnamefont {Rezzolla}},\ }\bibfield  {title} {\bibinfo
  {title} {{Quark formation and phenomenology in binary neutron-star mergers
  using V-QCD}},\ }\href {https://doi.org/10.21468/SciPostPhys.13.5.109}
  {\bibfield  {journal} {\bibinfo  {journal} {SciPost Phys.}\ }\textbf
  {\bibinfo {volume} {13}},\ \bibinfo {pages} {109} (\bibinfo {year} {2022})},\
  \Eprint {https://arxiv.org/abs/2205.05691} {arXiv:2205.05691 [astro-ph.HE]}
  \BibitemShut {NoStop}%
\bibitem [{\citenamefont {Demircik}\ \emph {et~al.}(2022)\citenamefont
  {Demircik}, \citenamefont {Ecker}, \citenamefont {J\"arvinen}, \citenamefont
  {Rezzolla}, \citenamefont {Tootle},\ and\ \citenamefont
  {Topolski}}]{Demircik:2022uol}%
  \BibitemOpen
  \bibfield  {author} {\bibinfo {author} {\bibfnamefont {T.}~\bibnamefont
  {Demircik}}, \bibinfo {author} {\bibfnamefont {C.}~\bibnamefont {Ecker}},
  \bibinfo {author} {\bibfnamefont {M.}~\bibnamefont {J\"arvinen}}, \bibinfo
  {author} {\bibfnamefont {L.}~\bibnamefont {Rezzolla}}, \bibinfo {author}
  {\bibfnamefont {S.}~\bibnamefont {Tootle}},\ and\ \bibinfo {author}
  {\bibfnamefont {K.}~\bibnamefont {Topolski}},\ }\bibfield  {title} {\bibinfo
  {title} {{Exploring the Phase Diagram of V-QCD with Neutron Star Merger
  Simulations}},\ }\href {https://doi.org/10.1051/epjconf/202227407006}
  {\bibfield  {journal} {\bibinfo  {journal} {EPJ Web Conf.}\ }\textbf
  {\bibinfo {volume} {274}},\ \bibinfo {pages} {07006} (\bibinfo {year}
  {2022})},\ \Eprint {https://arxiv.org/abs/2211.10118} {arXiv:2211.10118
  [astro-ph.HE]} \BibitemShut {NoStop}%
\bibitem [{\citenamefont {Espino}\ \emph
  {et~al.}(2023{\natexlab{a}})\citenamefont {Espino}, \citenamefont {Prakash},
  \citenamefont {Radice},\ and\ \citenamefont {Logoteta}}]{Espino:2023llj}%
  \BibitemOpen
  \bibfield  {author} {\bibinfo {author} {\bibfnamefont {P.~L.}\ \bibnamefont
  {Espino}}, \bibinfo {author} {\bibfnamefont {A.}~\bibnamefont {Prakash}},
  \bibinfo {author} {\bibfnamefont {D.}~\bibnamefont {Radice}},\ and\ \bibinfo
  {author} {\bibfnamefont {D.}~\bibnamefont {Logoteta}},\ }\bibfield  {title}
  {\bibinfo {title} {{Revealing Phase Transition in Dense Matter with
  Gravitational Wave Spectroscopy of Binary Neutron Star Mergers}},\
  }\href@noop {} {\  (\bibinfo {year} {2023}{\natexlab{a}})},\ \Eprint
  {https://arxiv.org/abs/2301.03619} {arXiv:2301.03619 [astro-ph.HE]}
  \BibitemShut {NoStop}%
\bibitem [{\citenamefont {Guo}\ \emph {et~al.}(2023)\citenamefont {Guo},
  \citenamefont {Yang}, \citenamefont {Ma},\ and\ \citenamefont
  {Wu}}]{Guo:2023som}%
  \BibitemOpen
  \bibfield  {author} {\bibinfo {author} {\bibfnamefont {L.-J.}\ \bibnamefont
  {Guo}}, \bibinfo {author} {\bibfnamefont {W.-C.}\ \bibnamefont {Yang}},
  \bibinfo {author} {\bibfnamefont {Y.-L.}\ \bibnamefont {Ma}},\ and\ \bibinfo
  {author} {\bibfnamefont {Y.-L.}\ \bibnamefont {Wu}},\ }\bibfield  {title}
  {\bibinfo {title} {{Probing hadron-quark transition through binary neutron
  star merger}},\ }\href@noop {} {\  (\bibinfo {year} {2023})},\ \Eprint
  {https://arxiv.org/abs/2308.01770} {arXiv:2308.01770 [astro-ph.HE]}
  \BibitemShut {NoStop}%
\bibitem [{\citenamefont {Haque}\ \emph {et~al.}(2022)\citenamefont {Haque},
  \citenamefont {Mallick},\ and\ \citenamefont {Thakur}}]{Haque:2022dsc}%
  \BibitemOpen
  \bibfield  {author} {\bibinfo {author} {\bibfnamefont {S.}~\bibnamefont
  {Haque}}, \bibinfo {author} {\bibfnamefont {R.}~\bibnamefont {Mallick}},\
  and\ \bibinfo {author} {\bibfnamefont {S.~K.}\ \bibnamefont {Thakur}},\
  }\bibfield  {title} {\bibinfo {title} {{Binary neutron star mergers and the
  effect of onset of phase transition on gravitational wave signals}},\
  }\href@noop {} {\  (\bibinfo {year} {2022})},\ \Eprint
  {https://arxiv.org/abs/2207.14485} {arXiv:2207.14485 [astro-ph.HE]}
  \BibitemShut {NoStop}%
\bibitem [{\citenamefont {Ciolfi}(2020)}]{Ciolfi:2020cpf}%
  \BibitemOpen
  \bibfield  {author} {\bibinfo {author} {\bibfnamefont {R.}~\bibnamefont
  {Ciolfi}},\ }\bibfield  {title} {\bibinfo {title} {{The key role of magnetic
  fields in binary neutron star mergers}},\ }\href
  {https://doi.org/10.1007/s10714-020-02714-x} {\bibfield  {journal} {\bibinfo
  {journal} {Gen. Rel. Grav.}\ }\textbf {\bibinfo {volume} {52}},\ \bibinfo
  {pages} {59} (\bibinfo {year} {2020})},\ \Eprint
  {https://arxiv.org/abs/2003.07572} {arXiv:2003.07572 [astro-ph.HE]}
  \BibitemShut {NoStop}%
\bibitem [{\citenamefont {Ciolfi}\ \emph {et~al.}(2017)\citenamefont {Ciolfi},
  \citenamefont {Kastaun}, \citenamefont {Giacomazzo}, \citenamefont
  {Endrizzi}, \citenamefont {Siegel},\ and\ \citenamefont
  {Perna}}]{Ciolfi:2017uak}%
  \BibitemOpen
  \bibfield  {author} {\bibinfo {author} {\bibfnamefont {R.}~\bibnamefont
  {Ciolfi}}, \bibinfo {author} {\bibfnamefont {W.}~\bibnamefont {Kastaun}},
  \bibinfo {author} {\bibfnamefont {B.}~\bibnamefont {Giacomazzo}}, \bibinfo
  {author} {\bibfnamefont {A.}~\bibnamefont {Endrizzi}}, \bibinfo {author}
  {\bibfnamefont {D.~M.}\ \bibnamefont {Siegel}},\ and\ \bibinfo {author}
  {\bibfnamefont {R.}~\bibnamefont {Perna}},\ }\bibfield  {title} {\bibinfo
  {title} {{General relativistic magnetohydrodynamic simulations of binary
  neutron star mergers forming a long-lived neutron star}},\ }\href
  {https://doi.org/10.1103/PhysRevD.95.063016} {\bibfield  {journal} {\bibinfo
  {journal} {Phys. Rev. D}\ }\textbf {\bibinfo {volume} {95}},\ \bibinfo
  {pages} {063016} (\bibinfo {year} {2017})},\ \Eprint
  {https://arxiv.org/abs/1701.08738} {arXiv:1701.08738 [astro-ph.HE]}
  \BibitemShut {NoStop}%
\bibitem [{\citenamefont {Radice}(2017)}]{Radice:2017zta}%
  \BibitemOpen
  \bibfield  {author} {\bibinfo {author} {\bibfnamefont {D.}~\bibnamefont
  {Radice}},\ }\bibfield  {title} {\bibinfo {title} {{General-Relativistic
  Large-Eddy Simulations of Binary Neutron Star Mergers}},\ }\href
  {https://doi.org/10.3847/2041-8213/aa6483} {\bibfield  {journal} {\bibinfo
  {journal} {Astrophys. J. Lett.}\ }\textbf {\bibinfo {volume} {838}},\
  \bibinfo {pages} {L2} (\bibinfo {year} {2017})},\ \Eprint
  {https://arxiv.org/abs/1703.02046} {arXiv:1703.02046 [astro-ph.HE]}
  \BibitemShut {NoStop}%
\bibitem [{\citenamefont {Shibata}\ and\ \citenamefont
  {Kiuchi}(2017)}]{Shibata:2017xht}%
  \BibitemOpen
  \bibfield  {author} {\bibinfo {author} {\bibfnamefont {M.}~\bibnamefont
  {Shibata}}\ and\ \bibinfo {author} {\bibfnamefont {K.}~\bibnamefont
  {Kiuchi}},\ }\bibfield  {title} {\bibinfo {title} {{Gravitational waves from
  remnant massive neutron stars of binary neutron star merger: Viscous
  hydrodynamics effects}},\ }\href {https://doi.org/10.1103/PhysRevD.95.123003}
  {\bibfield  {journal} {\bibinfo  {journal} {Phys. Rev. D}\ }\textbf {\bibinfo
  {volume} {95}},\ \bibinfo {pages} {123003} (\bibinfo {year} {2017})},\
  \Eprint {https://arxiv.org/abs/1705.06142} {arXiv:1705.06142 [astro-ph.HE]}
  \BibitemShut {NoStop}%
\bibitem [{\citenamefont {Margalit}\ \emph {et~al.}(2022)\citenamefont
  {Margalit}, \citenamefont {Jermyn}, \citenamefont {Metzger}, \citenamefont
  {Roberts},\ and\ \citenamefont {Quataert}}]{Margalit:2022rde}%
  \BibitemOpen
  \bibfield  {author} {\bibinfo {author} {\bibfnamefont {B.}~\bibnamefont
  {Margalit}}, \bibinfo {author} {\bibfnamefont {A.~S.}\ \bibnamefont
  {Jermyn}}, \bibinfo {author} {\bibfnamefont {B.~D.}\ \bibnamefont {Metzger}},
  \bibinfo {author} {\bibfnamefont {L.~F.}\ \bibnamefont {Roberts}},\ and\
  \bibinfo {author} {\bibfnamefont {E.}~\bibnamefont {Quataert}},\ }\bibfield
  {title} {\bibinfo {title} {{Angular-momentum Transport in Proto-neutron Stars
  and the Fate of Neutron Star Merger Remnants}},\ }\href
  {https://doi.org/10.3847/1538-4357/ac8b01} {\bibfield  {journal} {\bibinfo
  {journal} {Astrophys. J.}\ }\textbf {\bibinfo {volume} {939}},\ \bibinfo
  {pages} {51} (\bibinfo {year} {2022})},\ \Eprint
  {https://arxiv.org/abs/2206.10645} {arXiv:2206.10645 [astro-ph.HE]}
  \BibitemShut {NoStop}%
\bibitem [{\citenamefont {Hotokezaka}\ \emph {et~al.}(2013)\citenamefont
  {Hotokezaka}, \citenamefont {Kiuchi}, \citenamefont {Kyutoku}, \citenamefont
  {Muranushi}, \citenamefont {Sekiguchi}, \citenamefont {Shibata},\ and\
  \citenamefont {Taniguchi}}]{Hotokezaka:2013iia}%
  \BibitemOpen
  \bibfield  {author} {\bibinfo {author} {\bibfnamefont {K.}~\bibnamefont
  {Hotokezaka}}, \bibinfo {author} {\bibfnamefont {K.}~\bibnamefont {Kiuchi}},
  \bibinfo {author} {\bibfnamefont {K.}~\bibnamefont {Kyutoku}}, \bibinfo
  {author} {\bibfnamefont {T.}~\bibnamefont {Muranushi}}, \bibinfo {author}
  {\bibfnamefont {Y.-i.}\ \bibnamefont {Sekiguchi}}, \bibinfo {author}
  {\bibfnamefont {M.}~\bibnamefont {Shibata}},\ and\ \bibinfo {author}
  {\bibfnamefont {K.}~\bibnamefont {Taniguchi}},\ }\bibfield  {title} {\bibinfo
  {title} {{Remnant massive neutron stars of binary neutron star mergers:
  Evolution process and gravitational waveform}},\ }\href
  {https://doi.org/10.1103/PhysRevD.88.044026} {\bibfield  {journal} {\bibinfo
  {journal} {Phys. Rev. D}\ }\textbf {\bibinfo {volume} {88}},\ \bibinfo
  {pages} {044026} (\bibinfo {year} {2013})},\ \Eprint
  {https://arxiv.org/abs/1307.5888} {arXiv:1307.5888 [astro-ph.HE]}
  \BibitemShut {NoStop}%
\bibitem [{\citenamefont {Bauswein}\ \emph {et~al.}(2016)\citenamefont
  {Bauswein}, \citenamefont {Stergioulas},\ and\ \citenamefont
  {Janka}}]{Bauswein:2015vxa}%
  \BibitemOpen
  \bibfield  {author} {\bibinfo {author} {\bibfnamefont {A.}~\bibnamefont
  {Bauswein}}, \bibinfo {author} {\bibfnamefont {N.}~\bibnamefont
  {Stergioulas}},\ and\ \bibinfo {author} {\bibfnamefont {H.-T.}\ \bibnamefont
  {Janka}},\ }\bibfield  {title} {\bibinfo {title} {{Exploring properties of
  high-density matter through remnants of neutron-star mergers}},\ }\href
  {https://doi.org/10.1140/epja/i2016-16056-7} {\bibfield  {journal} {\bibinfo
  {journal} {Eur. Phys. J. A}\ }\textbf {\bibinfo {volume} {52}},\ \bibinfo
  {pages} {56} (\bibinfo {year} {2016})},\ \Eprint
  {https://arxiv.org/abs/1508.05493} {arXiv:1508.05493 [astro-ph.HE]}
  \BibitemShut {NoStop}%
\bibitem [{\citenamefont {Bose}\ \emph {et~al.}(2018)\citenamefont {Bose},
  \citenamefont {Chakravarti}, \citenamefont {Rezzolla}, \citenamefont
  {Sathyaprakash},\ and\ \citenamefont {Takami}}]{Bose:2017jvk}%
  \BibitemOpen
  \bibfield  {author} {\bibinfo {author} {\bibfnamefont {S.}~\bibnamefont
  {Bose}}, \bibinfo {author} {\bibfnamefont {K.}~\bibnamefont {Chakravarti}},
  \bibinfo {author} {\bibfnamefont {L.}~\bibnamefont {Rezzolla}}, \bibinfo
  {author} {\bibfnamefont {B.~S.}\ \bibnamefont {Sathyaprakash}},\ and\
  \bibinfo {author} {\bibfnamefont {K.}~\bibnamefont {Takami}},\ }\bibfield
  {title} {\bibinfo {title} {{Neutron-star Radius from a Population of Binary
  Neutron Star Mergers}},\ }\href
  {https://doi.org/10.1103/PhysRevLett.120.031102} {\bibfield  {journal}
  {\bibinfo  {journal} {Phys. Rev. Lett.}\ }\textbf {\bibinfo {volume} {120}},\
  \bibinfo {pages} {031102} (\bibinfo {year} {2018})},\ \Eprint
  {https://arxiv.org/abs/1705.10850} {arXiv:1705.10850 [gr-qc]} \BibitemShut
  {NoStop}%
\bibitem [{\citenamefont {Easter}\ \emph {et~al.}(2020)\citenamefont {Easter},
  \citenamefont {Ghonge}, \citenamefont {Lasky}, \citenamefont {Casey},
  \citenamefont {Clark}, \citenamefont {Vivanco},\ and\ \citenamefont
  {Chatziioannou}}]{Easter:2020ifj}%
  \BibitemOpen
  \bibfield  {author} {\bibinfo {author} {\bibfnamefont {P.~J.}\ \bibnamefont
  {Easter}}, \bibinfo {author} {\bibfnamefont {S.}~\bibnamefont {Ghonge}},
  \bibinfo {author} {\bibfnamefont {P.~D.}\ \bibnamefont {Lasky}}, \bibinfo
  {author} {\bibfnamefont {A.~R.}\ \bibnamefont {Casey}}, \bibinfo {author}
  {\bibfnamefont {J.~A.}\ \bibnamefont {Clark}}, \bibinfo {author}
  {\bibfnamefont {F.~H.}\ \bibnamefont {Vivanco}},\ and\ \bibinfo {author}
  {\bibfnamefont {K.}~\bibnamefont {Chatziioannou}},\ }\bibfield  {title}
  {\bibinfo {title} {{Detection and parameter estimation of binary neutron star
  merger remnants}},\ }\href {https://doi.org/10.1103/PhysRevD.102.043011}
  {\bibfield  {journal} {\bibinfo  {journal} {Phys. Rev. D}\ }\textbf {\bibinfo
  {volume} {102}},\ \bibinfo {pages} {043011} (\bibinfo {year} {2020})},\
  \Eprint {https://arxiv.org/abs/2006.04396} {arXiv:2006.04396 [astro-ph.HE]}
  \BibitemShut {NoStop}%
\bibitem [{\citenamefont {Soultanis}\ \emph {et~al.}(2022)\citenamefont
  {Soultanis}, \citenamefont {Bauswein},\ and\ \citenamefont
  {Stergioulas}}]{Soultanis:2021oia}%
  \BibitemOpen
  \bibfield  {author} {\bibinfo {author} {\bibfnamefont {T.}~\bibnamefont
  {Soultanis}}, \bibinfo {author} {\bibfnamefont {A.}~\bibnamefont
  {Bauswein}},\ and\ \bibinfo {author} {\bibfnamefont {N.}~\bibnamefont
  {Stergioulas}},\ }\bibfield  {title} {\bibinfo {title} {{Analytic models of
  the spectral properties of gravitational waves from neutron star merger
  remnants}},\ }\href {https://doi.org/10.1103/PhysRevD.105.043020} {\bibfield
  {journal} {\bibinfo  {journal} {Phys. Rev. D}\ }\textbf {\bibinfo {volume}
  {105}},\ \bibinfo {pages} {043020} (\bibinfo {year} {2022})},\ \Eprint
  {https://arxiv.org/abs/2111.08353} {arXiv:2111.08353 [astro-ph.HE]}
  \BibitemShut {NoStop}%
\bibitem [{\citenamefont {Tsang}\ \emph {et~al.}(2019)\citenamefont {Tsang},
  \citenamefont {Dietrich},\ and\ \citenamefont {Van
  Den~Broeck}}]{Tsang:2019esi}%
  \BibitemOpen
  \bibfield  {author} {\bibinfo {author} {\bibfnamefont {K.~W.}\ \bibnamefont
  {Tsang}}, \bibinfo {author} {\bibfnamefont {T.}~\bibnamefont {Dietrich}},\
  and\ \bibinfo {author} {\bibfnamefont {C.}~\bibnamefont {Van Den~Broeck}},\
  }\bibfield  {title} {\bibinfo {title} {{Modeling the postmerger gravitational
  wave signal and extracting binary properties from future binary neutron star
  detections}},\ }\href {https://doi.org/10.1103/PhysRevD.100.044047}
  {\bibfield  {journal} {\bibinfo  {journal} {Phys. Rev. D}\ }\textbf {\bibinfo
  {volume} {100}},\ \bibinfo {pages} {044047} (\bibinfo {year} {2019})},\
  \Eprint {https://arxiv.org/abs/1907.02424} {arXiv:1907.02424 [gr-qc]}
  \BibitemShut {NoStop}%
\bibitem [{\citenamefont {Breschi}\ \emph {et~al.}(2019)\citenamefont
  {Breschi}, \citenamefont {Bernuzzi}, \citenamefont {Zappa}, \citenamefont
  {Agathos}, \citenamefont {Perego}, \citenamefont {Radice},\ and\
  \citenamefont {Nagar}}]{Breschi:2019srl}%
  \BibitemOpen
  \bibfield  {author} {\bibinfo {author} {\bibfnamefont {M.}~\bibnamefont
  {Breschi}}, \bibinfo {author} {\bibfnamefont {S.}~\bibnamefont {Bernuzzi}},
  \bibinfo {author} {\bibfnamefont {F.}~\bibnamefont {Zappa}}, \bibinfo
  {author} {\bibfnamefont {M.}~\bibnamefont {Agathos}}, \bibinfo {author}
  {\bibfnamefont {A.}~\bibnamefont {Perego}}, \bibinfo {author} {\bibfnamefont
  {D.}~\bibnamefont {Radice}},\ and\ \bibinfo {author} {\bibfnamefont
  {A.}~\bibnamefont {Nagar}},\ }\bibfield  {title} {\bibinfo {title}
  {{kiloHertz gravitational waves from binary neutron star remnants:
  time-domain model and constraints on extreme matter}},\ }\href
  {https://doi.org/10.1103/PhysRevD.100.104029} {\bibfield  {journal} {\bibinfo
   {journal} {Phys. Rev. D}\ }\textbf {\bibinfo {volume} {100}},\ \bibinfo
  {pages} {104029} (\bibinfo {year} {2019})},\ \Eprint
  {https://arxiv.org/abs/1908.11418} {arXiv:1908.11418 [gr-qc]} \BibitemShut
  {NoStop}%
\bibitem [{\citenamefont {Breschi}\ \emph
  {et~al.}(2022{\natexlab{a}})\citenamefont {Breschi}, \citenamefont
  {Bernuzzi}, \citenamefont {Chakravarti}, \citenamefont {Camilletti},
  \citenamefont {Prakash},\ and\ \citenamefont {Perego}}]{Breschi:2022xnc}%
  \BibitemOpen
  \bibfield  {author} {\bibinfo {author} {\bibfnamefont {M.}~\bibnamefont
  {Breschi}}, \bibinfo {author} {\bibfnamefont {S.}~\bibnamefont {Bernuzzi}},
  \bibinfo {author} {\bibfnamefont {K.}~\bibnamefont {Chakravarti}}, \bibinfo
  {author} {\bibfnamefont {A.}~\bibnamefont {Camilletti}}, \bibinfo {author}
  {\bibfnamefont {A.}~\bibnamefont {Prakash}},\ and\ \bibinfo {author}
  {\bibfnamefont {A.}~\bibnamefont {Perego}},\ }\bibfield  {title} {\bibinfo
  {title} {{Kilohertz Gravitational Waves From Binary Neutron Star Mergers:
  Numerical-relativity Informed Postmerger Model}},\ }\href@noop {} {\
  (\bibinfo {year} {2022}{\natexlab{a}})},\ \Eprint
  {https://arxiv.org/abs/2205.09112} {arXiv:2205.09112 [gr-qc]} \BibitemShut
  {NoStop}%
\bibitem [{\citenamefont {Chatziioannou}\ \emph {et~al.}(2017)\citenamefont
  {Chatziioannou}, \citenamefont {Clark}, \citenamefont {Bauswein},
  \citenamefont {Millhouse}, \citenamefont {Littenberg},\ and\ \citenamefont
  {Cornish}}]{Chatziioannou:2017ixj}%
  \BibitemOpen
  \bibfield  {author} {\bibinfo {author} {\bibfnamefont {K.}~\bibnamefont
  {Chatziioannou}}, \bibinfo {author} {\bibfnamefont {J.~A.}\ \bibnamefont
  {Clark}}, \bibinfo {author} {\bibfnamefont {A.}~\bibnamefont {Bauswein}},
  \bibinfo {author} {\bibfnamefont {M.}~\bibnamefont {Millhouse}}, \bibinfo
  {author} {\bibfnamefont {T.~B.}\ \bibnamefont {Littenberg}},\ and\ \bibinfo
  {author} {\bibfnamefont {N.}~\bibnamefont {Cornish}},\ }\bibfield  {title}
  {\bibinfo {title} {{Inferring the post-merger gravitational wave emission
  from binary neutron star coalescences}},\ }\href
  {https://doi.org/10.1103/PhysRevD.96.124035} {\bibfield  {journal} {\bibinfo
  {journal} {Phys. Rev. D}\ }\textbf {\bibinfo {volume} {96}},\ \bibinfo
  {pages} {124035} (\bibinfo {year} {2017})},\ \Eprint
  {https://arxiv.org/abs/1711.00040} {arXiv:1711.00040 [gr-qc]} \BibitemShut
  {NoStop}%
\bibitem [{\citenamefont {Wijngaarden}\ \emph {et~al.}(2022)\citenamefont
  {Wijngaarden}, \citenamefont {Chatziioannou}, \citenamefont {Bauswein},
  \citenamefont {Clark},\ and\ \citenamefont {Cornish}}]{Wijngaarden:2022sah}%
  \BibitemOpen
  \bibfield  {author} {\bibinfo {author} {\bibfnamefont {M.}~\bibnamefont
  {Wijngaarden}}, \bibinfo {author} {\bibfnamefont {K.}~\bibnamefont
  {Chatziioannou}}, \bibinfo {author} {\bibfnamefont {A.}~\bibnamefont
  {Bauswein}}, \bibinfo {author} {\bibfnamefont {J.~A.}\ \bibnamefont
  {Clark}},\ and\ \bibinfo {author} {\bibfnamefont {N.~J.}\ \bibnamefont
  {Cornish}},\ }\bibfield  {title} {\bibinfo {title} {{Probing neutron stars
  with the full premerger and postmerger gravitational wave signal from binary
  coalescences}},\ }\href {https://doi.org/10.1103/PhysRevD.105.104019}
  {\bibfield  {journal} {\bibinfo  {journal} {Phys. Rev. D}\ }\textbf {\bibinfo
  {volume} {105}},\ \bibinfo {pages} {104019} (\bibinfo {year} {2022})},\
  \Eprint {https://arxiv.org/abs/2202.09382} {arXiv:2202.09382 [gr-qc]}
  \BibitemShut {NoStop}%
\bibitem [{\citenamefont {Breschi}\ \emph
  {et~al.}(2022{\natexlab{b}})\citenamefont {Breschi}, \citenamefont {Gamba},
  \citenamefont {Borhanian}, \citenamefont {Carullo},\ and\ \citenamefont
  {Bernuzzi}}]{Breschi:2022ens}%
  \BibitemOpen
  \bibfield  {author} {\bibinfo {author} {\bibfnamefont {M.}~\bibnamefont
  {Breschi}}, \bibinfo {author} {\bibfnamefont {R.}~\bibnamefont {Gamba}},
  \bibinfo {author} {\bibfnamefont {S.}~\bibnamefont {Borhanian}}, \bibinfo
  {author} {\bibfnamefont {G.}~\bibnamefont {Carullo}},\ and\ \bibinfo {author}
  {\bibfnamefont {S.}~\bibnamefont {Bernuzzi}},\ }\bibfield  {title} {\bibinfo
  {title} {{Kilohertz Gravitational Waves from Binary Neutron Star Mergers:
  Inference of Postmerger Signals with the Einstein Telescope}},\ }\href@noop
  {} {\  (\bibinfo {year} {2022}{\natexlab{b}})},\ \Eprint
  {https://arxiv.org/abs/2205.09979} {arXiv:2205.09979 [gr-qc]} \BibitemShut
  {NoStop}%
\bibitem [{\citenamefont {Breschi}\ \emph {et~al.}(2023)\citenamefont
  {Breschi}, \citenamefont {Carullo},\ and\ \citenamefont
  {Bernuzzi}}]{Breschi:2023mdj}%
  \BibitemOpen
  \bibfield  {author} {\bibinfo {author} {\bibfnamefont {M.}~\bibnamefont
  {Breschi}}, \bibinfo {author} {\bibfnamefont {G.}~\bibnamefont {Carullo}},\
  and\ \bibinfo {author} {\bibfnamefont {S.}~\bibnamefont {Bernuzzi}},\
  }\bibfield  {title} {\bibinfo {title} {{Pre/post-merger consistency test for
  gravitational signals from binary neutron star mergers}},\ }\href@noop {} {\
  (\bibinfo {year} {2023})},\ \Eprint {https://arxiv.org/abs/2301.09672}
  {arXiv:2301.09672 [gr-qc]} \BibitemShut {NoStop}%
\bibitem [{\citenamefont {Han}\ and\ \citenamefont
  {Steiner}(2019)}]{Han:2018mtj}%
  \BibitemOpen
  \bibfield  {author} {\bibinfo {author} {\bibfnamefont {S.}~\bibnamefont
  {Han}}\ and\ \bibinfo {author} {\bibfnamefont {A.~W.}\ \bibnamefont
  {Steiner}},\ }\bibfield  {title} {\bibinfo {title} {{Tidal deformability with
  sharp phase transitions in (binary) neutron stars}},\ }\href
  {https://doi.org/10.1103/PhysRevD.99.083014} {\bibfield  {journal} {\bibinfo
  {journal} {Phys. Rev. D}\ }\textbf {\bibinfo {volume} {99}},\ \bibinfo
  {pages} {083014} (\bibinfo {year} {2019})},\ \Eprint
  {https://arxiv.org/abs/1810.10967} {arXiv:1810.10967 [nucl-th]} \BibitemShut
  {NoStop}%
\bibitem [{\citenamefont {Sieniawska}\ \emph {et~al.}(2019)\citenamefont
  {Sieniawska}, \citenamefont {Turczanski}, \citenamefont {Bejger},\ and\
  \citenamefont {Zdunik}}]{Sieniawska:2018zzj}%
  \BibitemOpen
  \bibfield  {author} {\bibinfo {author} {\bibfnamefont {M.}~\bibnamefont
  {Sieniawska}}, \bibinfo {author} {\bibfnamefont {W.}~\bibnamefont
  {Turczanski}}, \bibinfo {author} {\bibfnamefont {M.}~\bibnamefont {Bejger}},\
  and\ \bibinfo {author} {\bibfnamefont {J.~L.}\ \bibnamefont {Zdunik}},\
  }\bibfield  {title} {\bibinfo {title} {{Tidal deformability and other global
  parameters of compact stars with strong phase transitions}},\ }\href
  {https://doi.org/10.1051/0004-6361/201833969} {\bibfield  {journal} {\bibinfo
   {journal} {Astron. Astrophys.}\ }\textbf {\bibinfo {volume} {622}},\
  \bibinfo {pages} {A174} (\bibinfo {year} {2019})},\ \Eprint
  {https://arxiv.org/abs/1807.11581} {arXiv:1807.11581 [astro-ph.HE]}
  \BibitemShut {NoStop}%
\bibitem [{\citenamefont {Raithel}\ and\ \citenamefont
  {Most}(2023{\natexlab{a}})}]{Raithel:2022efm}%
  \BibitemOpen
  \bibfield  {author} {\bibinfo {author} {\bibfnamefont {C.~A.}\ \bibnamefont
  {Raithel}}\ and\ \bibinfo {author} {\bibfnamefont {E.~R.}\ \bibnamefont
  {Most}},\ }\bibfield  {title} {\bibinfo {title} {{Degeneracy in the Inference
  of Phase Transitions in the Neutron Star Equation of State from Gravitational
  Wave Data}},\ }\href {https://doi.org/10.1103/PhysRevLett.130.201403}
  {\bibfield  {journal} {\bibinfo  {journal} {Phys. Rev. Lett.}\ }\textbf
  {\bibinfo {volume} {130}},\ \bibinfo {pages} {201403} (\bibinfo {year}
  {2023}{\natexlab{a}})},\ \Eprint {https://arxiv.org/abs/2208.04294}
  {arXiv:2208.04294 [astro-ph.HE]} \BibitemShut {NoStop}%
\bibitem [{\citenamefont {Essick}\ \emph {et~al.}(2023)\citenamefont {Essick},
  \citenamefont {Legred}, \citenamefont {Chatziioannou}, \citenamefont {Han},\
  and\ \citenamefont {Landry}}]{Essick:2023fso}%
  \BibitemOpen
  \bibfield  {author} {\bibinfo {author} {\bibfnamefont {R.}~\bibnamefont
  {Essick}}, \bibinfo {author} {\bibfnamefont {I.}~\bibnamefont {Legred}},
  \bibinfo {author} {\bibfnamefont {K.}~\bibnamefont {Chatziioannou}}, \bibinfo
  {author} {\bibfnamefont {S.}~\bibnamefont {Han}},\ and\ \bibinfo {author}
  {\bibfnamefont {P.}~\bibnamefont {Landry}},\ }\bibfield  {title} {\bibinfo
  {title} {{Phase transition phenomenology with nonparametric representations
  of the neutron star equation of state}},\ }\href
  {https://doi.org/10.1103/PhysRevD.108.043013} {\bibfield  {journal} {\bibinfo
   {journal} {Phys. Rev. D}\ }\textbf {\bibinfo {volume} {108}},\ \bibinfo
  {pages} {043013} (\bibinfo {year} {2023})},\ \Eprint
  {https://arxiv.org/abs/2305.07411} {arXiv:2305.07411 [astro-ph.HE]}
  \BibitemShut {NoStop}%
\bibitem [{\citenamefont {Mondal}\ \emph {et~al.}(2023)\citenamefont {Mondal},
  \citenamefont {Antonelli}, \citenamefont {Gulminelli}, \citenamefont
  {Mancini}, \citenamefont {Novak},\ and\ \citenamefont
  {Oertel}}]{Mondal:2023gbf}%
  \BibitemOpen
  \bibfield  {author} {\bibinfo {author} {\bibfnamefont {C.}~\bibnamefont
  {Mondal}}, \bibinfo {author} {\bibfnamefont {M.}~\bibnamefont {Antonelli}},
  \bibinfo {author} {\bibfnamefont {F.}~\bibnamefont {Gulminelli}}, \bibinfo
  {author} {\bibfnamefont {M.}~\bibnamefont {Mancini}}, \bibinfo {author}
  {\bibfnamefont {J.}~\bibnamefont {Novak}},\ and\ \bibinfo {author}
  {\bibfnamefont {M.}~\bibnamefont {Oertel}},\ }\bibfield  {title} {\bibinfo
  {title} {{Detectability of a phase transition in neutron star matter with
  third-generation gravitational wave interferometers}},\ }\href@noop {} {\
  (\bibinfo {year} {2023})},\ \Eprint {https://arxiv.org/abs/2305.05999}
  {arXiv:2305.05999 [astro-ph.HE]} \BibitemShut {NoStop}%
\bibitem [{\citenamefont {Raithel}\ and\ \citenamefont
  {Most}(2023{\natexlab{b}})}]{Raithel:2022aee}%
  \BibitemOpen
  \bibfield  {author} {\bibinfo {author} {\bibfnamefont {C.~A.}\ \bibnamefont
  {Raithel}}\ and\ \bibinfo {author} {\bibfnamefont {E.~R.}\ \bibnamefont
  {Most}},\ }\bibfield  {title} {\bibinfo {title} {{Tidal deformability
  doppelg\"anger: Implications of a low-density phase transition in the neutron
  star equation of state}},\ }\href
  {https://doi.org/10.1103/PhysRevD.108.023010} {\bibfield  {journal} {\bibinfo
   {journal} {Phys. Rev. D}\ }\textbf {\bibinfo {volume} {108}},\ \bibinfo
  {pages} {023010} (\bibinfo {year} {2023}{\natexlab{b}})},\ \Eprint
  {https://arxiv.org/abs/2208.04295} {arXiv:2208.04295 [astro-ph.HE]}
  \BibitemShut {NoStop}%
\bibitem [{\citenamefont {Pang}\ \emph {et~al.}(2020)\citenamefont {Pang},
  \citenamefont {Dietrich}, \citenamefont {Tews},\ and\ \citenamefont {Van
  Den~Broeck}}]{Pang:2020ilf}%
  \BibitemOpen
  \bibfield  {author} {\bibinfo {author} {\bibfnamefont {P.~T.~H.}\
  \bibnamefont {Pang}}, \bibinfo {author} {\bibfnamefont {T.}~\bibnamefont
  {Dietrich}}, \bibinfo {author} {\bibfnamefont {I.}~\bibnamefont {Tews}},\
  and\ \bibinfo {author} {\bibfnamefont {C.}~\bibnamefont {Van Den~Broeck}},\
  }\bibfield  {title} {\bibinfo {title} {{Parameter estimation for strong phase
  transitions in supranuclear matter using gravitational-wave astronomy}},\
  }\href {https://doi.org/10.1103/PhysRevResearch.2.033514} {\bibfield
  {journal} {\bibinfo  {journal} {Phys. Rev. Res.}\ }\textbf {\bibinfo {volume}
  {2}},\ \bibinfo {pages} {033514} (\bibinfo {year} {2020})},\ \Eprint
  {https://arxiv.org/abs/2006.14936} {arXiv:2006.14936 [astro-ph.HE]}
  \BibitemShut {NoStop}%
\bibitem [{\citenamefont {Bauswein}\ and\ \citenamefont
  {Janka}(2012)}]{Bauswein:2011tp}%
  \BibitemOpen
  \bibfield  {author} {\bibinfo {author} {\bibfnamefont {A.}~\bibnamefont
  {Bauswein}}\ and\ \bibinfo {author} {\bibfnamefont {H.~T.}\ \bibnamefont
  {Janka}},\ }\bibfield  {title} {\bibinfo {title} {{Measuring neutron-star
  properties via gravitational waves from binary mergers}},\ }\href
  {https://doi.org/10.1103/PhysRevLett.108.011101} {\bibfield  {journal}
  {\bibinfo  {journal} {Phys. Rev. Lett.}\ }\textbf {\bibinfo {volume} {108}},\
  \bibinfo {pages} {011101} (\bibinfo {year} {2012})},\ \Eprint
  {https://arxiv.org/abs/1106.1616} {arXiv:1106.1616 [astro-ph.SR]}
  \BibitemShut {NoStop}%
\bibitem [{\citenamefont {Bernuzzi}\ \emph {et~al.}(2014)\citenamefont
  {Bernuzzi}, \citenamefont {Nagar}, \citenamefont {Balmelli}, \citenamefont
  {Dietrich},\ and\ \citenamefont {Ujevic}}]{Bernuzzi:2014kca}%
  \BibitemOpen
  \bibfield  {author} {\bibinfo {author} {\bibfnamefont {S.}~\bibnamefont
  {Bernuzzi}}, \bibinfo {author} {\bibfnamefont {A.}~\bibnamefont {Nagar}},
  \bibinfo {author} {\bibfnamefont {S.}~\bibnamefont {Balmelli}}, \bibinfo
  {author} {\bibfnamefont {T.}~\bibnamefont {Dietrich}},\ and\ \bibinfo
  {author} {\bibfnamefont {M.}~\bibnamefont {Ujevic}},\ }\bibfield  {title}
  {\bibinfo {title} {{Quasiuniversal properties of neutron star mergers}},\
  }\href {https://doi.org/10.1103/PhysRevLett.112.201101} {\bibfield  {journal}
  {\bibinfo  {journal} {Phys. Rev. Lett.}\ }\textbf {\bibinfo {volume} {112}},\
  \bibinfo {pages} {201101} (\bibinfo {year} {2014})},\ \Eprint
  {https://arxiv.org/abs/1402.6244} {arXiv:1402.6244 [gr-qc]} \BibitemShut
  {NoStop}%
\bibitem [{\citenamefont {Bernuzzi}\ \emph {et~al.}(2015)\citenamefont
  {Bernuzzi}, \citenamefont {Dietrich},\ and\ \citenamefont
  {Nagar}}]{Bernuzzi:2015rla}%
  \BibitemOpen
  \bibfield  {author} {\bibinfo {author} {\bibfnamefont {S.}~\bibnamefont
  {Bernuzzi}}, \bibinfo {author} {\bibfnamefont {T.}~\bibnamefont {Dietrich}},\
  and\ \bibinfo {author} {\bibfnamefont {A.}~\bibnamefont {Nagar}},\ }\bibfield
   {title} {\bibinfo {title} {{Modeling the complete gravitational wave
  spectrum of neutron star mergers}},\ }\href
  {https://doi.org/10.1103/PhysRevLett.115.091101} {\bibfield  {journal}
  {\bibinfo  {journal} {Phys. Rev. Lett.}\ }\textbf {\bibinfo {volume} {115}},\
  \bibinfo {pages} {091101} (\bibinfo {year} {2015})},\ \Eprint
  {https://arxiv.org/abs/1504.01764} {arXiv:1504.01764 [gr-qc]} \BibitemShut
  {NoStop}%
\bibitem [{\citenamefont {Rezzolla}\ and\ \citenamefont
  {Takami}(2016)}]{Rezzolla:2016nxn}%
  \BibitemOpen
  \bibfield  {author} {\bibinfo {author} {\bibfnamefont {L.}~\bibnamefont
  {Rezzolla}}\ and\ \bibinfo {author} {\bibfnamefont {K.}~\bibnamefont
  {Takami}},\ }\bibfield  {title} {\bibinfo {title} {{Gravitational-wave signal
  from binary neutron stars: a systematic analysis of the spectral
  properties}},\ }\href {https://doi.org/10.1103/PhysRevD.93.124051} {\bibfield
   {journal} {\bibinfo  {journal} {Phys. Rev. D}\ }\textbf {\bibinfo {volume}
  {93}},\ \bibinfo {pages} {124051} (\bibinfo {year} {2016})},\ \Eprint
  {https://arxiv.org/abs/1604.00246} {arXiv:1604.00246 [gr-qc]} \BibitemShut
  {NoStop}%
\bibitem [{\citenamefont {Zappa}\ \emph {et~al.}(2018)\citenamefont {Zappa},
  \citenamefont {Bernuzzi}, \citenamefont {Radice}, \citenamefont {Perego},\
  and\ \citenamefont {Dietrich}}]{Zappa:2017xba}%
  \BibitemOpen
  \bibfield  {author} {\bibinfo {author} {\bibfnamefont {F.}~\bibnamefont
  {Zappa}}, \bibinfo {author} {\bibfnamefont {S.}~\bibnamefont {Bernuzzi}},
  \bibinfo {author} {\bibfnamefont {D.}~\bibnamefont {Radice}}, \bibinfo
  {author} {\bibfnamefont {A.}~\bibnamefont {Perego}},\ and\ \bibinfo {author}
  {\bibfnamefont {T.}~\bibnamefont {Dietrich}},\ }\bibfield  {title} {\bibinfo
  {title} {{Gravitational-wave luminosity of binary neutron stars mergers}},\
  }\href {https://doi.org/10.1103/PhysRevLett.120.111101} {\bibfield  {journal}
  {\bibinfo  {journal} {Phys. Rev. Lett.}\ }\textbf {\bibinfo {volume} {120}},\
  \bibinfo {pages} {111101} (\bibinfo {year} {2018})},\ \Eprint
  {https://arxiv.org/abs/1712.04267} {arXiv:1712.04267 [gr-qc]} \BibitemShut
  {NoStop}%
\bibitem [{\citenamefont {Bauswein}\ \emph {et~al.}(2012)\citenamefont
  {Bauswein}, \citenamefont {Janka}, \citenamefont {Hebeler},\ and\
  \citenamefont {Schwenk}}]{Bauswein:2012ya}%
  \BibitemOpen
  \bibfield  {author} {\bibinfo {author} {\bibfnamefont {A.}~\bibnamefont
  {Bauswein}}, \bibinfo {author} {\bibfnamefont {H.~T.}\ \bibnamefont {Janka}},
  \bibinfo {author} {\bibfnamefont {K.}~\bibnamefont {Hebeler}},\ and\ \bibinfo
  {author} {\bibfnamefont {A.}~\bibnamefont {Schwenk}},\ }\bibfield  {title}
  {\bibinfo {title} {{Equation-of-state dependence of the gravitational-wave
  signal from the ring-down phase of neutron-star mergers}},\ }\href
  {https://doi.org/10.1103/PhysRevD.86.063001} {\bibfield  {journal} {\bibinfo
  {journal} {Phys. Rev. D}\ }\textbf {\bibinfo {volume} {86}},\ \bibinfo
  {pages} {063001} (\bibinfo {year} {2012})},\ \Eprint
  {https://arxiv.org/abs/1204.1888} {arXiv:1204.1888 [astro-ph.SR]}
  \BibitemShut {NoStop}%
\bibitem [{\citenamefont {Lioutas}\ \emph {et~al.}(2021)\citenamefont
  {Lioutas}, \citenamefont {Bauswein},\ and\ \citenamefont
  {Stergioulas}}]{Lioutas:2021jbl}%
  \BibitemOpen
  \bibfield  {author} {\bibinfo {author} {\bibfnamefont {G.}~\bibnamefont
  {Lioutas}}, \bibinfo {author} {\bibfnamefont {A.}~\bibnamefont {Bauswein}},\
  and\ \bibinfo {author} {\bibfnamefont {N.}~\bibnamefont {Stergioulas}},\
  }\bibfield  {title} {\bibinfo {title} {{Frequency deviations in universal
  relations of isolated neutron stars and postmerger remnants}},\ }\href
  {https://doi.org/10.1103/PhysRevD.104.043011} {\bibfield  {journal} {\bibinfo
   {journal} {Phys. Rev. D}\ }\textbf {\bibinfo {volume} {104}},\ \bibinfo
  {pages} {043011} (\bibinfo {year} {2021})},\ \Eprint
  {https://arxiv.org/abs/2102.12455} {arXiv:2102.12455 [astro-ph.HE]}
  \BibitemShut {NoStop}%
\bibitem [{\citenamefont {Typel}\ \emph {et~al.}(2010)\citenamefont {Typel},
  \citenamefont {Ropke}, \citenamefont {Klahn}, \citenamefont {Blaschke},\ and\
  \citenamefont {Wolter}}]{Typel:2009sy}%
  \BibitemOpen
  \bibfield  {author} {\bibinfo {author} {\bibfnamefont {S.}~\bibnamefont
  {Typel}}, \bibinfo {author} {\bibfnamefont {G.}~\bibnamefont {Ropke}},
  \bibinfo {author} {\bibfnamefont {T.}~\bibnamefont {Klahn}}, \bibinfo
  {author} {\bibfnamefont {D.}~\bibnamefont {Blaschke}},\ and\ \bibinfo
  {author} {\bibfnamefont {H.~H.}\ \bibnamefont {Wolter}},\ }\bibfield  {title}
  {\bibinfo {title} {{Composition and thermodynamics of nuclear matter with
  light clusters}},\ }\href {https://doi.org/10.1103/PhysRevC.81.015803}
  {\bibfield  {journal} {\bibinfo  {journal} {Phys. Rev. C}\ }\textbf {\bibinfo
  {volume} {81}},\ \bibinfo {pages} {015803} (\bibinfo {year} {2010})},\
  \Eprint {https://arxiv.org/abs/0908.2344} {arXiv:0908.2344 [nucl-th]}
  \BibitemShut {NoStop}%
\bibitem [{\citenamefont {Alvarez-Castillo}\ \emph {et~al.}(2016)\citenamefont
  {Alvarez-Castillo}, \citenamefont {Ayriyan}, \citenamefont {Benic},
  \citenamefont {Blaschke}, \citenamefont {Grigorian},\ and\ \citenamefont
  {Typel}}]{Alvarez-Castillo:2016oln}%
  \BibitemOpen
  \bibfield  {author} {\bibinfo {author} {\bibfnamefont {D.}~\bibnamefont
  {Alvarez-Castillo}}, \bibinfo {author} {\bibfnamefont {A.}~\bibnamefont
  {Ayriyan}}, \bibinfo {author} {\bibfnamefont {S.}~\bibnamefont {Benic}},
  \bibinfo {author} {\bibfnamefont {D.}~\bibnamefont {Blaschke}}, \bibinfo
  {author} {\bibfnamefont {H.}~\bibnamefont {Grigorian}},\ and\ \bibinfo
  {author} {\bibfnamefont {S.}~\bibnamefont {Typel}},\ }\bibfield  {title}
  {\bibinfo {title} {{New class of hybrid EoS and Bayesian M-R data
  analysis}},\ }\href {https://doi.org/10.1140/epja/i2016-16069-2} {\bibfield
  {journal} {\bibinfo  {journal} {Eur. Phys. J. A}\ }\textbf {\bibinfo {volume}
  {52}},\ \bibinfo {pages} {69} (\bibinfo {year} {2016})},\ \Eprint
  {https://arxiv.org/abs/1603.03457} {arXiv:1603.03457 [nucl-th]} \BibitemShut
  {NoStop}%
\bibitem [{\citenamefont {Fischer}\ \emph {et~al.}(2018)\citenamefont
  {Fischer}, \citenamefont {Bastian}, \citenamefont {Wu}, \citenamefont
  {Baklanov}, \citenamefont {Sorokina}, \citenamefont {Blinnikov},
  \citenamefont {Typel}, \citenamefont {Kl\"ahn},\ and\ \citenamefont
  {Blaschke}}]{Fischer:2017lag}%
  \BibitemOpen
  \bibfield  {author} {\bibinfo {author} {\bibfnamefont {T.}~\bibnamefont
  {Fischer}}, \bibinfo {author} {\bibfnamefont {N.-U.~F.}\ \bibnamefont
  {Bastian}}, \bibinfo {author} {\bibfnamefont {M.-R.}\ \bibnamefont {Wu}},
  \bibinfo {author} {\bibfnamefont {P.}~\bibnamefont {Baklanov}}, \bibinfo
  {author} {\bibfnamefont {E.}~\bibnamefont {Sorokina}}, \bibinfo {author}
  {\bibfnamefont {S.}~\bibnamefont {Blinnikov}}, \bibinfo {author}
  {\bibfnamefont {S.}~\bibnamefont {Typel}}, \bibinfo {author} {\bibfnamefont
  {T.}~\bibnamefont {Kl\"ahn}},\ and\ \bibinfo {author} {\bibfnamefont {D.~B.}\
  \bibnamefont {Blaschke}},\ }\bibfield  {title} {\bibinfo {title} {{Quark
  deconfinement as a supernova explosion engine for massive blue supergiant
  stars}},\ }\href {https://doi.org/10.1038/s41550-018-0583-0} {\bibfield
  {journal} {\bibinfo  {journal} {Nature Astron.}\ }\textbf {\bibinfo {volume}
  {2}},\ \bibinfo {pages} {980} (\bibinfo {year} {2018})},\ \Eprint
  {https://arxiv.org/abs/1712.08788} {arXiv:1712.08788 [astro-ph.HE]}
  \BibitemShut {NoStop}%
\bibitem [{\citenamefont {Banyuls}\ \emph {et~al.}(1997)\citenamefont
  {Banyuls}, \citenamefont {Font}, \citenamefont {Ibanez}, \citenamefont
  {Marti},\ and\ \citenamefont {Miralles}}]{Banyuls:1997zz}%
  \BibitemOpen
  \bibfield  {author} {\bibinfo {author} {\bibfnamefont {F.}~\bibnamefont
  {Banyuls}}, \bibinfo {author} {\bibfnamefont {J.~A.}\ \bibnamefont {Font}},
  \bibinfo {author} {\bibfnamefont {J.~M.~A.}\ \bibnamefont {Ibanez}}, \bibinfo
  {author} {\bibfnamefont {J.~M.~A.}\ \bibnamefont {Marti}},\ and\ \bibinfo
  {author} {\bibfnamefont {J.~A.}\ \bibnamefont {Miralles}},\ }\bibfield
  {title} {\bibinfo {title} {{Numerical {3+1} General Relativistic
  Hydrodynamics: A Local Characteristic Approach}},\ }\href@noop {} {\bibfield
  {journal} {\bibinfo  {journal} {Astrophys. J.}\ }\textbf {\bibinfo {volume}
  {476}},\ \bibinfo {pages} {221} (\bibinfo {year} {1997})}\BibitemShut
  {NoStop}%
\bibitem [{\citenamefont {Radice}\ and\ \citenamefont
  {Rezzolla}(2012)}]{Radice:2012cu}%
  \BibitemOpen
  \bibfield  {author} {\bibinfo {author} {\bibfnamefont {D.}~\bibnamefont
  {Radice}}\ and\ \bibinfo {author} {\bibfnamefont {L.}~\bibnamefont
  {Rezzolla}},\ }\bibfield  {title} {\bibinfo {title} {{THC: a new high-order
  finite-difference high-resolution shock-capturing code for
  special-relativistic hydrodynamics}},\ }\href
  {https://doi.org/10.1051/0004-6361/201219735} {\bibfield  {journal} {\bibinfo
   {journal} {Astron. Astrophys.}\ }\textbf {\bibinfo {volume} {547}},\
  \bibinfo {pages} {A26} (\bibinfo {year} {2012})},\ \Eprint
  {https://arxiv.org/abs/1206.6502} {arXiv:1206.6502 [astro-ph.IM]}
  \BibitemShut {NoStop}%
\bibitem [{\citenamefont {Radice}\ \emph
  {et~al.}(2014{\natexlab{a}})\citenamefont {Radice}, \citenamefont
  {Rezzolla},\ and\ \citenamefont {Galeazzi}}]{Radice:2013hxh}%
  \BibitemOpen
  \bibfield  {author} {\bibinfo {author} {\bibfnamefont {D.}~\bibnamefont
  {Radice}}, \bibinfo {author} {\bibfnamefont {L.}~\bibnamefont {Rezzolla}},\
  and\ \bibinfo {author} {\bibfnamefont {F.}~\bibnamefont {Galeazzi}},\
  }\bibfield  {title} {\bibinfo {title} {{Beyond second-order convergence in
  simulations of binary neutron stars in full general-relativity}},\ }\href
  {https://doi.org/10.1093/mnrasl/slt137} {\bibfield  {journal} {\bibinfo
  {journal} {Mon. Not. Roy. Astron. Soc.}\ }\textbf {\bibinfo {volume} {437}},\
  \bibinfo {pages} {L46} (\bibinfo {year} {2014}{\natexlab{a}})},\ \Eprint
  {https://arxiv.org/abs/1306.6052} {arXiv:1306.6052 [gr-qc]} \BibitemShut
  {NoStop}%
\bibitem [{\citenamefont {Radice}\ \emph
  {et~al.}(2014{\natexlab{b}})\citenamefont {Radice}, \citenamefont
  {Rezzolla},\ and\ \citenamefont {Galeazzi}}]{Radice:2013xpa}%
  \BibitemOpen
  \bibfield  {author} {\bibinfo {author} {\bibfnamefont {D.}~\bibnamefont
  {Radice}}, \bibinfo {author} {\bibfnamefont {L.}~\bibnamefont {Rezzolla}},\
  and\ \bibinfo {author} {\bibfnamefont {F.}~\bibnamefont {Galeazzi}},\
  }\bibfield  {title} {\bibinfo {title} {{High-Order Fully General-Relativistic
  Hydrodynamics: new Approaches and Tests}},\ }\href
  {https://doi.org/10.1088/0264-9381/31/7/075012} {\bibfield  {journal}
  {\bibinfo  {journal} {Class. Quant. Grav.}\ }\textbf {\bibinfo {volume}
  {31}},\ \bibinfo {pages} {075012} (\bibinfo {year} {2014}{\natexlab{b}})},\
  \Eprint {https://arxiv.org/abs/1312.5004} {arXiv:1312.5004 [gr-qc]}
  \BibitemShut {NoStop}%
\bibitem [{\citenamefont {Pollney}\ \emph {et~al.}(2011)\citenamefont
  {Pollney}, \citenamefont {Reisswig}, \citenamefont {Schnetter}, \citenamefont
  {Dorband},\ and\ \citenamefont {Diener}}]{Pollney:2009yz}%
  \BibitemOpen
  \bibfield  {author} {\bibinfo {author} {\bibfnamefont {D.}~\bibnamefont
  {Pollney}}, \bibinfo {author} {\bibfnamefont {C.}~\bibnamefont {Reisswig}},
  \bibinfo {author} {\bibfnamefont {E.}~\bibnamefont {Schnetter}}, \bibinfo
  {author} {\bibfnamefont {N.}~\bibnamefont {Dorband}},\ and\ \bibinfo {author}
  {\bibfnamefont {P.}~\bibnamefont {Diener}},\ }\bibfield  {title} {\bibinfo
  {title} {{High accuracy binary black hole simulations with an extended wave
  zone}},\ }\href {https://doi.org/10.1103/PhysRevD.83.044045} {\bibfield
  {journal} {\bibinfo  {journal} {Phys. Rev. D}\ }\textbf {\bibinfo {volume}
  {83}},\ \bibinfo {pages} {044045} (\bibinfo {year} {2011})},\ \Eprint
  {https://arxiv.org/abs/0910.3803} {arXiv:0910.3803 [gr-qc]} \BibitemShut
  {NoStop}%
\bibitem [{\citenamefont {Reisswig}\ \emph
  {et~al.}(2013{\natexlab{a}})\citenamefont {Reisswig}, \citenamefont {Ott},
  \citenamefont {Abdikamalov}, \citenamefont {Haas}, \citenamefont {Moesta},\
  and\ \citenamefont {Schnetter}}]{Reisswig:2013sqa}%
  \BibitemOpen
  \bibfield  {author} {\bibinfo {author} {\bibfnamefont {C.}~\bibnamefont
  {Reisswig}}, \bibinfo {author} {\bibfnamefont {C.~D.}\ \bibnamefont {Ott}},
  \bibinfo {author} {\bibfnamefont {E.}~\bibnamefont {Abdikamalov}}, \bibinfo
  {author} {\bibfnamefont {R.}~\bibnamefont {Haas}}, \bibinfo {author}
  {\bibfnamefont {P.}~\bibnamefont {Moesta}},\ and\ \bibinfo {author}
  {\bibfnamefont {E.}~\bibnamefont {Schnetter}},\ }\bibfield  {title} {\bibinfo
  {title} {{Formation and Coalescence of Cosmological Supermassive Black Hole
  Binaries in Supermassive Star Collapse}},\ }\href
  {https://doi.org/10.1103/PhysRevLett.111.151101} {\bibfield  {journal}
  {\bibinfo  {journal} {Phys. Rev. Lett.}\ }\textbf {\bibinfo {volume} {111}},\
  \bibinfo {pages} {151101} (\bibinfo {year} {2013}{\natexlab{a}})},\ \Eprint
  {https://arxiv.org/abs/1304.7787} {arXiv:1304.7787 [astro-ph.CO]}
  \BibitemShut {NoStop}%
\bibitem [{\citenamefont {Werneck}\ \emph {et~al.}(2023)\citenamefont
  {Werneck}, \citenamefont {Cupp}, \citenamefont {Assumpção}, \citenamefont
  {Brandt}, \citenamefont {Cheng}, \citenamefont {Diener}, \citenamefont
  {Doherty}, \citenamefont {Etienne}, \citenamefont {Haas}, \citenamefont
  {Jacques}, \citenamefont {Karakaş}, \citenamefont {Topolski}, \citenamefont
  {Tsao}, \citenamefont {Alcubierre}, \citenamefont {Alic}, \citenamefont
  {Allen}, \citenamefont {Ansorg}, \citenamefont {Babiuc-Hamilton},
  \citenamefont {Baiotti}, \citenamefont {Benger}, \citenamefont {Bentivegna},
  \citenamefont {Bernuzzi}, \citenamefont {Bode}, \citenamefont {Bozzola},
  \citenamefont {Brendal}, \citenamefont {Bruegmann}, \citenamefont
  {Campanelli}, \citenamefont {Cipolletta}, \citenamefont {Corvino},
  \citenamefont {Pietri}, \citenamefont {Dima}, \citenamefont {Dimmelmeier},
  \citenamefont {Dooley}, \citenamefont {Dorband}, \citenamefont {Elley},
  \citenamefont {Khamra}, \citenamefont {Faber}, \citenamefont {Ficarra},
  \citenamefont {Font}, \citenamefont {Frieben}, \citenamefont {Giacomazzo},
  \citenamefont {Goodale}, \citenamefont {Gundlach}, \citenamefont {Hawke},
  \citenamefont {Hawley}, \citenamefont {Hinder}, \citenamefont {Huerta},
  \citenamefont {Husa}, \citenamefont {Ikeda}, \citenamefont {Iyer},
  \citenamefont {Ji}, \citenamefont {Johnson}, \citenamefont {Joshi},
  \citenamefont {Kalyanaraman}, \citenamefont {Kankani}, \citenamefont
  {Kastaun}, \citenamefont {Kellermann}, \citenamefont {Knapp}, \citenamefont
  {Koppitz}, \citenamefont {Kuo}, \citenamefont {Laguna}, \citenamefont
  {Lanferman}, \citenamefont {Lasky}, \citenamefont {Leung}, \citenamefont
  {L{\"o}ffler}, \citenamefont {Macpherson}, \citenamefont {Masso},
  \citenamefont {Menger}, \citenamefont {Merzky}, \citenamefont {Miller},
  \citenamefont {Miller}, \citenamefont {Moesta}, \citenamefont {Montero},
  \citenamefont {Mundim}, \citenamefont {Nelson}, \citenamefont {Nerozzi},
  \citenamefont {Noble}, \citenamefont {Ott}, \citenamefont {Papenfort},
  \citenamefont {Paruchuri}, \citenamefont {Pollney}, \citenamefont {Price},
  \citenamefont {Radice}, \citenamefont {Radke}, \citenamefont {Reisswig},
  \citenamefont {Rezzolla}, \citenamefont {Richards}, \citenamefont {Rideout},
  \citenamefont {Ripeanu}, \citenamefont {Sala}, \citenamefont {Schewtschenko},
  \citenamefont {Schnetter}, \citenamefont {Schutz}, \citenamefont {Seidel},
  \citenamefont {Seidel}, \citenamefont {Shalf}, \citenamefont {Sible},
  \citenamefont {Sperhake}, \citenamefont {Stergioulas}, \citenamefont {Suen},
  \citenamefont {Szilagyi}, \citenamefont {Takahashi}, \citenamefont {Thomas},
  \citenamefont {Thornburg}, \citenamefont {Tian}, \citenamefont {Tobias},
  \citenamefont {Tonita}, \citenamefont {Tootle}, \citenamefont {Walker},
  \citenamefont {Wan}, \citenamefont {Wardell}, \citenamefont {Wen},
  \citenamefont {Witek}, \citenamefont {Zilh{\~a}o}, \citenamefont {Zink},\
  and\ \citenamefont {Zlochower}}]{EinsteinToolkit:2023_05}%
  \BibitemOpen
  \bibfield  {author} {\bibinfo {author} {\bibfnamefont {L.}~\bibnamefont
  {Werneck}}, \bibinfo {author} {\bibfnamefont {S.}~\bibnamefont {Cupp}},
  \bibinfo {author} {\bibfnamefont {T.}~\bibnamefont {Assumpção}}, \bibinfo
  {author} {\bibfnamefont {S.~R.}\ \bibnamefont {Brandt}}, \bibinfo {author}
  {\bibfnamefont {C.-H.}\ \bibnamefont {Cheng}}, \bibinfo {author}
  {\bibfnamefont {P.}~\bibnamefont {Diener}}, \bibinfo {author} {\bibfnamefont
  {J.}~\bibnamefont {Doherty}}, \bibinfo {author} {\bibfnamefont
  {Z.}~\bibnamefont {Etienne}}, \bibinfo {author} {\bibfnamefont
  {R.}~\bibnamefont {Haas}}, \bibinfo {author} {\bibfnamefont {T.~P.}\
  \bibnamefont {Jacques}}, \bibinfo {author} {\bibfnamefont {B.}~\bibnamefont
  {Karakaş}}, \bibinfo {author} {\bibfnamefont {K.}~\bibnamefont {Topolski}},
  \bibinfo {author} {\bibfnamefont {B.-J.}\ \bibnamefont {Tsao}}, \bibinfo
  {author} {\bibfnamefont {M.}~\bibnamefont {Alcubierre}}, \bibinfo {author}
  {\bibfnamefont {D.}~\bibnamefont {Alic}}, \bibinfo {author} {\bibfnamefont
  {G.}~\bibnamefont {Allen}}, \bibinfo {author} {\bibfnamefont
  {M.}~\bibnamefont {Ansorg}}, \bibinfo {author} {\bibfnamefont
  {M.}~\bibnamefont {Babiuc-Hamilton}}, \bibinfo {author} {\bibfnamefont
  {L.}~\bibnamefont {Baiotti}}, \bibinfo {author} {\bibfnamefont
  {W.}~\bibnamefont {Benger}}, \bibinfo {author} {\bibfnamefont
  {E.}~\bibnamefont {Bentivegna}}, \bibinfo {author} {\bibfnamefont
  {S.}~\bibnamefont {Bernuzzi}}, \bibinfo {author} {\bibfnamefont
  {T.}~\bibnamefont {Bode}}, \bibinfo {author} {\bibfnamefont {G.}~\bibnamefont
  {Bozzola}}, \bibinfo {author} {\bibfnamefont {B.}~\bibnamefont {Brendal}},
  \bibinfo {author} {\bibfnamefont {B.}~\bibnamefont {Bruegmann}}, \bibinfo
  {author} {\bibfnamefont {M.}~\bibnamefont {Campanelli}}, \bibinfo {author}
  {\bibfnamefont {F.}~\bibnamefont {Cipolletta}}, \bibinfo {author}
  {\bibfnamefont {G.}~\bibnamefont {Corvino}}, \bibinfo {author} {\bibfnamefont
  {R.~D.}\ \bibnamefont {Pietri}}, \bibinfo {author} {\bibfnamefont
  {A.}~\bibnamefont {Dima}}, \bibinfo {author} {\bibfnamefont {H.}~\bibnamefont
  {Dimmelmeier}}, \bibinfo {author} {\bibfnamefont {R.}~\bibnamefont {Dooley}},
  \bibinfo {author} {\bibfnamefont {N.}~\bibnamefont {Dorband}}, \bibinfo
  {author} {\bibfnamefont {M.}~\bibnamefont {Elley}}, \bibinfo {author}
  {\bibfnamefont {Y.~E.}\ \bibnamefont {Khamra}}, \bibinfo {author}
  {\bibfnamefont {J.}~\bibnamefont {Faber}}, \bibinfo {author} {\bibfnamefont
  {G.}~\bibnamefont {Ficarra}}, \bibinfo {author} {\bibfnamefont
  {T.}~\bibnamefont {Font}}, \bibinfo {author} {\bibfnamefont {J.}~\bibnamefont
  {Frieben}}, \bibinfo {author} {\bibfnamefont {B.}~\bibnamefont {Giacomazzo}},
  \bibinfo {author} {\bibfnamefont {T.}~\bibnamefont {Goodale}}, \bibinfo
  {author} {\bibfnamefont {C.}~\bibnamefont {Gundlach}}, \bibinfo {author}
  {\bibfnamefont {I.}~\bibnamefont {Hawke}}, \bibinfo {author} {\bibfnamefont
  {S.}~\bibnamefont {Hawley}}, \bibinfo {author} {\bibfnamefont
  {I.}~\bibnamefont {Hinder}}, \bibinfo {author} {\bibfnamefont {E.~A.}\
  \bibnamefont {Huerta}}, \bibinfo {author} {\bibfnamefont {S.}~\bibnamefont
  {Husa}}, \bibinfo {author} {\bibfnamefont {T.}~\bibnamefont {Ikeda}},
  \bibinfo {author} {\bibfnamefont {S.}~\bibnamefont {Iyer}}, \bibinfo {author}
  {\bibfnamefont {L.}~\bibnamefont {Ji}}, \bibinfo {author} {\bibfnamefont
  {D.}~\bibnamefont {Johnson}}, \bibinfo {author} {\bibfnamefont {A.~V.}\
  \bibnamefont {Joshi}}, \bibinfo {author} {\bibfnamefont {H.}~\bibnamefont
  {Kalyanaraman}}, \bibinfo {author} {\bibfnamefont {A.}~\bibnamefont
  {Kankani}}, \bibinfo {author} {\bibfnamefont {W.}~\bibnamefont {Kastaun}},
  \bibinfo {author} {\bibfnamefont {T.}~\bibnamefont {Kellermann}}, \bibinfo
  {author} {\bibfnamefont {A.}~\bibnamefont {Knapp}}, \bibinfo {author}
  {\bibfnamefont {M.}~\bibnamefont {Koppitz}}, \bibinfo {author} {\bibfnamefont
  {N.}~\bibnamefont {Kuo}}, \bibinfo {author} {\bibfnamefont {P.}~\bibnamefont
  {Laguna}}, \bibinfo {author} {\bibfnamefont {G.}~\bibnamefont {Lanferman}},
  \bibinfo {author} {\bibfnamefont {P.}~\bibnamefont {Lasky}}, \bibinfo
  {author} {\bibfnamefont {L.}~\bibnamefont {Leung}}, \bibinfo {author}
  {\bibfnamefont {F.}~\bibnamefont {L{\"o}ffler}}, \bibinfo {author}
  {\bibfnamefont {H.}~\bibnamefont {Macpherson}}, \bibinfo {author}
  {\bibfnamefont {J.}~\bibnamefont {Masso}}, \bibinfo {author} {\bibfnamefont
  {L.}~\bibnamefont {Menger}}, \bibinfo {author} {\bibfnamefont
  {A.}~\bibnamefont {Merzky}}, \bibinfo {author} {\bibfnamefont {J.~M.}\
  \bibnamefont {Miller}}, \bibinfo {author} {\bibfnamefont {M.}~\bibnamefont
  {Miller}}, \bibinfo {author} {\bibfnamefont {P.}~\bibnamefont {Moesta}},
  \bibinfo {author} {\bibfnamefont {P.}~\bibnamefont {Montero}}, \bibinfo
  {author} {\bibfnamefont {B.}~\bibnamefont {Mundim}}, \bibinfo {author}
  {\bibfnamefont {P.}~\bibnamefont {Nelson}}, \bibinfo {author} {\bibfnamefont
  {A.}~\bibnamefont {Nerozzi}}, \bibinfo {author} {\bibfnamefont {S.~C.}\
  \bibnamefont {Noble}}, \bibinfo {author} {\bibfnamefont {C.}~\bibnamefont
  {Ott}}, \bibinfo {author} {\bibfnamefont {L.~J.}\ \bibnamefont {Papenfort}},
  \bibinfo {author} {\bibfnamefont {R.}~\bibnamefont {Paruchuri}}, \bibinfo
  {author} {\bibfnamefont {D.}~\bibnamefont {Pollney}}, \bibinfo {author}
  {\bibfnamefont {D.}~\bibnamefont {Price}}, \bibinfo {author} {\bibfnamefont
  {D.}~\bibnamefont {Radice}}, \bibinfo {author} {\bibfnamefont
  {T.}~\bibnamefont {Radke}}, \bibinfo {author} {\bibfnamefont
  {C.}~\bibnamefont {Reisswig}}, \bibinfo {author} {\bibfnamefont
  {L.}~\bibnamefont {Rezzolla}}, \bibinfo {author} {\bibfnamefont {C.~B.}\
  \bibnamefont {Richards}}, \bibinfo {author} {\bibfnamefont {D.}~\bibnamefont
  {Rideout}}, \bibinfo {author} {\bibfnamefont {M.}~\bibnamefont {Ripeanu}},
  \bibinfo {author} {\bibfnamefont {L.}~\bibnamefont {Sala}}, \bibinfo {author}
  {\bibfnamefont {J.~A.}\ \bibnamefont {Schewtschenko}}, \bibinfo {author}
  {\bibfnamefont {E.}~\bibnamefont {Schnetter}}, \bibinfo {author}
  {\bibfnamefont {B.}~\bibnamefont {Schutz}}, \bibinfo {author} {\bibfnamefont
  {E.}~\bibnamefont {Seidel}}, \bibinfo {author} {\bibfnamefont
  {E.}~\bibnamefont {Seidel}}, \bibinfo {author} {\bibfnamefont
  {J.}~\bibnamefont {Shalf}}, \bibinfo {author} {\bibfnamefont
  {K.}~\bibnamefont {Sible}}, \bibinfo {author} {\bibfnamefont
  {U.}~\bibnamefont {Sperhake}}, \bibinfo {author} {\bibfnamefont
  {N.}~\bibnamefont {Stergioulas}}, \bibinfo {author} {\bibfnamefont {W.-M.}\
  \bibnamefont {Suen}}, \bibinfo {author} {\bibfnamefont {B.}~\bibnamefont
  {Szilagyi}}, \bibinfo {author} {\bibfnamefont {R.}~\bibnamefont {Takahashi}},
  \bibinfo {author} {\bibfnamefont {M.}~\bibnamefont {Thomas}}, \bibinfo
  {author} {\bibfnamefont {J.}~\bibnamefont {Thornburg}}, \bibinfo {author}
  {\bibfnamefont {C.}~\bibnamefont {Tian}}, \bibinfo {author} {\bibfnamefont
  {M.}~\bibnamefont {Tobias}}, \bibinfo {author} {\bibfnamefont
  {A.}~\bibnamefont {Tonita}}, \bibinfo {author} {\bibfnamefont
  {S.}~\bibnamefont {Tootle}}, \bibinfo {author} {\bibfnamefont
  {P.}~\bibnamefont {Walker}}, \bibinfo {author} {\bibfnamefont {M.-B.}\
  \bibnamefont {Wan}}, \bibinfo {author} {\bibfnamefont {B.}~\bibnamefont
  {Wardell}}, \bibinfo {author} {\bibfnamefont {A.}~\bibnamefont {Wen}},
  \bibinfo {author} {\bibfnamefont {H.}~\bibnamefont {Witek}}, \bibinfo
  {author} {\bibfnamefont {M.}~\bibnamefont {Zilh{\~a}o}}, \bibinfo {author}
  {\bibfnamefont {B.}~\bibnamefont {Zink}},\ and\ \bibinfo {author}
  {\bibfnamefont {Y.}~\bibnamefont {Zlochower}},\ }\href
  {https://doi.org/10.5281/zenodo.7942541} {\bibinfo {title} {The einstein
  toolkit}} (\bibinfo {year} {2023}),\ \bibinfo {note} {to find out more, visit
  http://einsteintoolkit.org}\BibitemShut {NoStop}%
\bibitem [{\citenamefont {Bernuzzi}\ and\ \citenamefont
  {Hilditch}(2010)}]{Bernuzzi:2009ex}%
  \BibitemOpen
  \bibfield  {author} {\bibinfo {author} {\bibfnamefont {S.}~\bibnamefont
  {Bernuzzi}}\ and\ \bibinfo {author} {\bibfnamefont {D.}~\bibnamefont
  {Hilditch}},\ }\bibfield  {title} {\bibinfo {title} {{Constraint violation in
  free evolution schemes: Comparing BSSNOK with a conformal decomposition of
  Z4}},\ }\href {https://doi.org/10.1103/PhysRevD.81.084003} {\bibfield
  {journal} {\bibinfo  {journal} {Phys. Rev. D}\ }\textbf {\bibinfo {volume}
  {81}},\ \bibinfo {pages} {084003} (\bibinfo {year} {2010})},\ \Eprint
  {https://arxiv.org/abs/0912.2920} {arXiv:0912.2920 [gr-qc]} \BibitemShut
  {NoStop}%
\bibitem [{\citenamefont {Hilditch}\ \emph {et~al.}(2013)\citenamefont
  {Hilditch}, \citenamefont {Bernuzzi}, \citenamefont {Thierfelder},
  \citenamefont {Cao}, \citenamefont {Tichy},\ and\ \citenamefont
  {Bruegmann}}]{Hilditch:2012fp}%
  \BibitemOpen
  \bibfield  {author} {\bibinfo {author} {\bibfnamefont {D.}~\bibnamefont
  {Hilditch}}, \bibinfo {author} {\bibfnamefont {S.}~\bibnamefont {Bernuzzi}},
  \bibinfo {author} {\bibfnamefont {M.}~\bibnamefont {Thierfelder}}, \bibinfo
  {author} {\bibfnamefont {Z.}~\bibnamefont {Cao}}, \bibinfo {author}
  {\bibfnamefont {W.}~\bibnamefont {Tichy}},\ and\ \bibinfo {author}
  {\bibfnamefont {B.}~\bibnamefont {Bruegmann}},\ }\bibfield  {title} {\bibinfo
  {title} {{Compact binary evolutions with the Z4c formulation}},\ }\href
  {https://doi.org/10.1103/PhysRevD.88.084057} {\bibfield  {journal} {\bibinfo
  {journal} {Phys. Rev. D}\ }\textbf {\bibinfo {volume} {88}},\ \bibinfo
  {pages} {084057} (\bibinfo {year} {2013})},\ \Eprint
  {https://arxiv.org/abs/1212.2901} {arXiv:1212.2901 [gr-qc]} \BibitemShut
  {NoStop}%
\bibitem [{\citenamefont {Gourgoulhon}\ \emph {et~al.}(2001)\citenamefont
  {Gourgoulhon}, \citenamefont {Grandclement}, \citenamefont {Taniguchi},
  \citenamefont {Marck},\ and\ \citenamefont {Bonazzola}}]{Gourgoulhon:2000nn}%
  \BibitemOpen
  \bibfield  {author} {\bibinfo {author} {\bibfnamefont {E.}~\bibnamefont
  {Gourgoulhon}}, \bibinfo {author} {\bibfnamefont {P.}~\bibnamefont
  {Grandclement}}, \bibinfo {author} {\bibfnamefont {K.}~\bibnamefont
  {Taniguchi}}, \bibinfo {author} {\bibfnamefont {J.-A.}\ \bibnamefont
  {Marck}},\ and\ \bibinfo {author} {\bibfnamefont {S.}~\bibnamefont
  {Bonazzola}},\ }\bibfield  {title} {\bibinfo {title} {{Quasiequilibrium
  sequences of synchronized and irrotational binary neutron stars in general
  relativity: 1. Method and tests}},\ }\href
  {https://doi.org/10.1103/PhysRevD.63.064029} {\bibfield  {journal} {\bibinfo
  {journal} {Phys. Rev. D}\ }\textbf {\bibinfo {volume} {63}},\ \bibinfo
  {pages} {064029} (\bibinfo {year} {2001})},\ \Eprint
  {https://arxiv.org/abs/gr-qc/0007028} {arXiv:gr-qc/0007028} \BibitemShut
  {NoStop}%
\bibitem [{\citenamefont {Schnetter}\ \emph {et~al.}(2004)\citenamefont
  {Schnetter}, \citenamefont {Hawley},\ and\ \citenamefont
  {Hawke}}]{Schnetter:2003rb}%
  \BibitemOpen
  \bibfield  {author} {\bibinfo {author} {\bibfnamefont {E.}~\bibnamefont
  {Schnetter}}, \bibinfo {author} {\bibfnamefont {S.~H.}\ \bibnamefont
  {Hawley}},\ and\ \bibinfo {author} {\bibfnamefont {I.}~\bibnamefont
  {Hawke}},\ }\bibfield  {title} {\bibinfo {title} {{Evolutions in 3-D
  numerical relativity using fixed mesh refinement}},\ }\href
  {https://doi.org/10.1088/0264-9381/21/6/014} {\bibfield  {journal} {\bibinfo
  {journal} {Class. Quant. Grav.}\ }\textbf {\bibinfo {volume} {21}},\ \bibinfo
  {pages} {1465} (\bibinfo {year} {2004})},\ \Eprint
  {https://arxiv.org/abs/gr-qc/0310042} {arXiv:gr-qc/0310042} \BibitemShut
  {NoStop}%
\bibitem [{\citenamefont {Reisswig}\ \emph
  {et~al.}(2013{\natexlab{b}})\citenamefont {Reisswig}, \citenamefont {Haas},
  \citenamefont {Ott}, \citenamefont {Abdikamalov}, \citenamefont {M\"osta},
  \citenamefont {Pollney},\ and\ \citenamefont {Schnetter}}]{Reisswig:2012nc}%
  \BibitemOpen
  \bibfield  {author} {\bibinfo {author} {\bibfnamefont {C.}~\bibnamefont
  {Reisswig}}, \bibinfo {author} {\bibfnamefont {R.}~\bibnamefont {Haas}},
  \bibinfo {author} {\bibfnamefont {C.~D.}\ \bibnamefont {Ott}}, \bibinfo
  {author} {\bibfnamefont {E.}~\bibnamefont {Abdikamalov}}, \bibinfo {author}
  {\bibfnamefont {P.}~\bibnamefont {M\"osta}}, \bibinfo {author} {\bibfnamefont
  {D.}~\bibnamefont {Pollney}},\ and\ \bibinfo {author} {\bibfnamefont
  {E.}~\bibnamefont {Schnetter}},\ }\bibfield  {title} {\bibinfo {title}
  {{Three-Dimensional General-Relativistic Hydrodynamic Simulations of Binary
  Neutron Star Coalescence and Stellar Collapse with Multipatch Grids}},\
  }\href {https://doi.org/10.1103/PhysRevD.87.064023} {\bibfield  {journal}
  {\bibinfo  {journal} {Phys. Rev. D}\ }\textbf {\bibinfo {volume} {87}},\
  \bibinfo {pages} {064023} (\bibinfo {year} {2013}{\natexlab{b}})},\ \Eprint
  {https://arxiv.org/abs/1212.1191} {arXiv:1212.1191 [astro-ph.HE]}
  \BibitemShut {NoStop}%
\bibitem [{\citenamefont {Bombaci}\ and\ \citenamefont
  {Logoteta}(2018)}]{Bombaci:2018ksa}%
  \BibitemOpen
  \bibfield  {author} {\bibinfo {author} {\bibfnamefont {I.}~\bibnamefont
  {Bombaci}}\ and\ \bibinfo {author} {\bibfnamefont {D.}~\bibnamefont
  {Logoteta}},\ }\bibfield  {title} {\bibinfo {title} {{Equation of state of
  dense nuclear matter and neutron star structure from nuclear chiral
  interactions}},\ }\href {https://doi.org/10.1051/0004-6361/201731604}
  {\bibfield  {journal} {\bibinfo  {journal} {Astron. Astrophys.}\ }\textbf
  {\bibinfo {volume} {609}},\ \bibinfo {pages} {A128} (\bibinfo {year}
  {2018})},\ \Eprint {https://arxiv.org/abs/1805.11846} {arXiv:1805.11846
  [astro-ph.HE]} \BibitemShut {NoStop}%
\bibitem [{\citenamefont {Logoteta}\ \emph {et~al.}(2021)\citenamefont
  {Logoteta}, \citenamefont {Perego},\ and\ \citenamefont
  {Bombaci}}]{Logoteta:2020yxf}%
  \BibitemOpen
  \bibfield  {author} {\bibinfo {author} {\bibfnamefont {D.}~\bibnamefont
  {Logoteta}}, \bibinfo {author} {\bibfnamefont {A.}~\bibnamefont {Perego}},\
  and\ \bibinfo {author} {\bibfnamefont {I.}~\bibnamefont {Bombaci}},\
  }\bibfield  {title} {\bibinfo {title} {{Microscopic equation of state of hot
  nuclear matter for numerical relativity simulations}},\ }\href
  {https://doi.org/10.1051/0004-6361/202039457} {\bibfield  {journal} {\bibinfo
   {journal} {Astron. Astrophys.}\ }\textbf {\bibinfo {volume} {646}},\
  \bibinfo {pages} {A55} (\bibinfo {year} {2021})},\ \Eprint
  {https://arxiv.org/abs/2012.03599} {arXiv:2012.03599 [nucl-th]} \BibitemShut
  {NoStop}%
\bibitem [{\citenamefont {Kashyap}\ \emph {et~al.}(2022)\citenamefont {Kashyap}
  \emph {et~al.}}]{Kashyap:2021wzs}%
  \BibitemOpen
  \bibfield  {author} {\bibinfo {author} {\bibfnamefont {R.}~\bibnamefont
  {Kashyap}} \emph {et~al.},\ }\bibfield  {title} {\bibinfo {title} {{Numerical
  relativity simulations of prompt collapse mergers: Threshold mass and
  phenomenological constraints on neutron star properties after GW170817}},\
  }\href {https://doi.org/10.1103/PhysRevD.105.103022} {\bibfield  {journal}
  {\bibinfo  {journal} {Phys. Rev. D}\ }\textbf {\bibinfo {volume} {105}},\
  \bibinfo {pages} {103022} (\bibinfo {year} {2022})},\ \Eprint
  {https://arxiv.org/abs/2111.05183} {arXiv:2111.05183 [astro-ph.HE]}
  \BibitemShut {NoStop}%
\bibitem [{\citenamefont {Perego}\ \emph {et~al.}(2022)\citenamefont {Perego},
  \citenamefont {Logoteta}, \citenamefont {Radice}, \citenamefont {Bernuzzi},
  \citenamefont {Kashyap}, \citenamefont {Das}, \citenamefont {Padamata},\ and\
  \citenamefont {Prakash}}]{Perego:2021mkd}%
  \BibitemOpen
  \bibfield  {author} {\bibinfo {author} {\bibfnamefont {A.}~\bibnamefont
  {Perego}}, \bibinfo {author} {\bibfnamefont {D.}~\bibnamefont {Logoteta}},
  \bibinfo {author} {\bibfnamefont {D.}~\bibnamefont {Radice}}, \bibinfo
  {author} {\bibfnamefont {S.}~\bibnamefont {Bernuzzi}}, \bibinfo {author}
  {\bibfnamefont {R.}~\bibnamefont {Kashyap}}, \bibinfo {author} {\bibfnamefont
  {A.}~\bibnamefont {Das}}, \bibinfo {author} {\bibfnamefont {S.}~\bibnamefont
  {Padamata}},\ and\ \bibinfo {author} {\bibfnamefont {A.}~\bibnamefont
  {Prakash}},\ }\bibfield  {title} {\bibinfo {title} {{Probing the
  Incompressibility of Nuclear Matter at Ultrahigh Density through the Prompt
  Collapse of Asymmetric Neutron Star Binaries}},\ }\href
  {https://doi.org/10.1103/PhysRevLett.129.032701} {\bibfield  {journal}
  {\bibinfo  {journal} {Phys. Rev. Lett.}\ }\textbf {\bibinfo {volume} {129}},\
  \bibinfo {pages} {032701} (\bibinfo {year} {2022})},\ \Eprint
  {https://arxiv.org/abs/2112.05864} {arXiv:2112.05864 [astro-ph.HE]}
  \BibitemShut {NoStop}%
\bibitem [{\citenamefont {Reisswig}\ and\ \citenamefont
  {Pollney}(2011)}]{Reisswig:2010di}%
  \BibitemOpen
  \bibfield  {author} {\bibinfo {author} {\bibfnamefont {C.}~\bibnamefont
  {Reisswig}}\ and\ \bibinfo {author} {\bibfnamefont {D.}~\bibnamefont
  {Pollney}},\ }\bibfield  {title} {\bibinfo {title} {{Notes on the integration
  of numerical relativity waveforms}},\ }\href
  {https://doi.org/10.1088/0264-9381/28/19/195015} {\bibfield  {journal}
  {\bibinfo  {journal} {Class. Quant. Grav.}\ }\textbf {\bibinfo {volume}
  {28}},\ \bibinfo {pages} {195015} (\bibinfo {year} {2011})},\ \Eprint
  {https://arxiv.org/abs/1006.1632} {arXiv:1006.1632 [gr-qc]} \BibitemShut
  {NoStop}%
\bibitem [{\citenamefont {Harris}(1978)}]{1455106}%
  \BibitemOpen
  \bibfield  {author} {\bibinfo {author} {\bibfnamefont {F.}~\bibnamefont
  {Harris}},\ }\bibfield  {title} {\bibinfo {title} {On the use of windows for
  harmonic analysis with the discrete fourier transform},\ }\href
  {https://doi.org/10.1109/PROC.1978.10837} {\bibfield  {journal} {\bibinfo
  {journal} {Proceedings of the IEEE}\ }\textbf {\bibinfo {volume} {66}},\
  \bibinfo {pages} {51} (\bibinfo {year} {1978})}\BibitemShut {NoStop}%
\bibitem [{\citenamefont {Buchner}(2021)}]{johannes_buchner_2021_4636924}%
  \BibitemOpen
  \bibfield  {author} {\bibinfo {author} {\bibfnamefont {J.}~\bibnamefont
  {Buchner}},\ }\href {https://doi.org/10.5281/zenodo.4636924} {\bibinfo
  {title} {{UltraNest - a robust, general purpose Bayesian inference engine}}}
  (\bibinfo {year} {2021})\BibitemShut {NoStop}%
\bibitem [{\citenamefont {Breschi}\ \emph {et~al.}(2021)\citenamefont
  {Breschi}, \citenamefont {Gamba},\ and\ \citenamefont
  {Bernuzzi}}]{Breschi:2021wzr}%
  \BibitemOpen
  \bibfield  {author} {\bibinfo {author} {\bibfnamefont {M.}~\bibnamefont
  {Breschi}}, \bibinfo {author} {\bibfnamefont {R.}~\bibnamefont {Gamba}},\
  and\ \bibinfo {author} {\bibfnamefont {S.}~\bibnamefont {Bernuzzi}},\
  }\bibfield  {title} {\bibinfo {title} {{Bayesian inference of multimessenger
  astrophysical data: Methods and applications to gravitational waves}},\
  }\href {https://doi.org/10.1103/PhysRevD.104.042001} {\bibfield  {journal}
  {\bibinfo  {journal} {Phys. Rev. D}\ }\textbf {\bibinfo {volume} {104}},\
  \bibinfo {pages} {042001} (\bibinfo {year} {2021})},\ \Eprint
  {https://arxiv.org/abs/2102.00017} {arXiv:2102.00017 [gr-qc]} \BibitemShut
  {NoStop}%
\bibitem [{\citenamefont {Thrane}\ and\ \citenamefont
  {Talbot}(2019)}]{Thrane:2018qnx}%
  \BibitemOpen
  \bibfield  {author} {\bibinfo {author} {\bibfnamefont {E.}~\bibnamefont
  {Thrane}}\ and\ \bibinfo {author} {\bibfnamefont {C.}~\bibnamefont
  {Talbot}},\ }\bibfield  {title} {\bibinfo {title} {{An introduction to
  Bayesian inference in gravitational-wave astronomy: parameter estimation,
  model selection, and hierarchical models}},\ }\href
  {https://doi.org/10.1017/pasa.2019.2} {\bibfield  {journal} {\bibinfo
  {journal} {Publ. Astron. Soc. Austral.}\ }\textbf {\bibinfo {volume} {36}},\
  \bibinfo {pages} {e010} (\bibinfo {year} {2019})},\ \bibinfo {note}
  {[Erratum: Publ.Astron.Soc.Austral. 37, e036 (2020)]},\ \Eprint
  {https://arxiv.org/abs/1809.02293} {arXiv:1809.02293 [astro-ph.IM]}
  \BibitemShut {NoStop}%
\bibitem [{\citenamefont {Callister}(2021)}]{Callister:2021gxf}%
  \BibitemOpen
  \bibfield  {author} {\bibinfo {author} {\bibfnamefont {T.}~\bibnamefont
  {Callister}},\ }\bibfield  {title} {\bibinfo {title} {{A Thesaurus for Common
  Priors in Gravitational-Wave Astronomy}},\ }\href@noop {} {\  (\bibinfo
  {year} {2021})},\ \Eprint {https://arxiv.org/abs/2104.09508}
  {arXiv:2104.09508 [gr-qc]} \BibitemShut {NoStop}%
\bibitem [{\citenamefont {Sathyaprakash}\ and\ \citenamefont
  {Dhurandhar}(1991)}]{Sathyaprakash:1991mt}%
  \BibitemOpen
  \bibfield  {author} {\bibinfo {author} {\bibfnamefont {B.~S.}\ \bibnamefont
  {Sathyaprakash}}\ and\ \bibinfo {author} {\bibfnamefont {S.~V.}\ \bibnamefont
  {Dhurandhar}},\ }\bibfield  {title} {\bibinfo {title} {{Choice of filters for
  the detection of gravitational waves from coalescing binaries}},\ }\href
  {https://doi.org/10.1103/PhysRevD.44.3819} {\bibfield  {journal} {\bibinfo
  {journal} {Phys. Rev. D}\ }\textbf {\bibinfo {volume} {44}},\ \bibinfo
  {pages} {3819} (\bibinfo {year} {1991})}\BibitemShut {NoStop}%
\bibitem [{\citenamefont {Boh\'e}\ \emph {et~al.}(2013)\citenamefont {Boh\'e},
  \citenamefont {Marsat},\ and\ \citenamefont {Blanchet}}]{Bohe:2013cla}%
  \BibitemOpen
  \bibfield  {author} {\bibinfo {author} {\bibfnamefont {A.}~\bibnamefont
  {Boh\'e}}, \bibinfo {author} {\bibfnamefont {S.}~\bibnamefont {Marsat}},\
  and\ \bibinfo {author} {\bibfnamefont {L.}~\bibnamefont {Blanchet}},\
  }\bibfield  {title} {\bibinfo {title} {{Next-to-next-to-leading order
  spin\textendash{}orbit effects in the gravitational wave flux and orbital
  phasing of compact binaries}},\ }\href
  {https://doi.org/10.1088/0264-9381/30/13/135009} {\bibfield  {journal}
  {\bibinfo  {journal} {Class. Quant. Grav.}\ }\textbf {\bibinfo {volume}
  {30}},\ \bibinfo {pages} {135009} (\bibinfo {year} {2013})},\ \Eprint
  {https://arxiv.org/abs/1303.7412} {arXiv:1303.7412 [gr-qc]} \BibitemShut
  {NoStop}%
\bibitem [{\citenamefont {Arun}\ \emph {et~al.}(2009)\citenamefont {Arun},
  \citenamefont {Buonanno}, \citenamefont {Faye},\ and\ \citenamefont
  {Ochsner}}]{Arun:2008kb}%
  \BibitemOpen
  \bibfield  {author} {\bibinfo {author} {\bibfnamefont {K.~G.}\ \bibnamefont
  {Arun}}, \bibinfo {author} {\bibfnamefont {A.}~\bibnamefont {Buonanno}},
  \bibinfo {author} {\bibfnamefont {G.}~\bibnamefont {Faye}},\ and\ \bibinfo
  {author} {\bibfnamefont {E.}~\bibnamefont {Ochsner}},\ }\bibfield  {title}
  {\bibinfo {title} {{Higher-order spin effects in the amplitude and phase of
  gravitational waveforms emitted by inspiraling compact binaries: Ready-to-use
  gravitational waveforms}},\ }\href
  {https://doi.org/10.1103/PhysRevD.79.104023} {\bibfield  {journal} {\bibinfo
  {journal} {Phys. Rev. D}\ }\textbf {\bibinfo {volume} {79}},\ \bibinfo
  {pages} {104023} (\bibinfo {year} {2009})},\ \bibinfo {note} {[Erratum:
  Phys.Rev.D 84, 049901 (2011)]},\ \Eprint {https://arxiv.org/abs/0810.5336}
  {arXiv:0810.5336 [gr-qc]} \BibitemShut {NoStop}%
\bibitem [{\citenamefont {Mikoczi}\ \emph {et~al.}(2005)\citenamefont
  {Mikoczi}, \citenamefont {Vasuth},\ and\ \citenamefont
  {Gergely}}]{Mikoczi:2005dn}%
  \BibitemOpen
  \bibfield  {author} {\bibinfo {author} {\bibfnamefont {B.}~\bibnamefont
  {Mikoczi}}, \bibinfo {author} {\bibfnamefont {M.}~\bibnamefont {Vasuth}},\
  and\ \bibinfo {author} {\bibfnamefont {L.~A.}\ \bibnamefont {Gergely}},\
  }\bibfield  {title} {\bibinfo {title} {{Self-interaction spin effects in
  inspiralling compact binaries}},\ }\href
  {https://doi.org/10.1103/PhysRevD.71.124043} {\bibfield  {journal} {\bibinfo
  {journal} {Phys. Rev. D}\ }\textbf {\bibinfo {volume} {71}},\ \bibinfo
  {pages} {124043} (\bibinfo {year} {2005})},\ \Eprint
  {https://arxiv.org/abs/astro-ph/0504538} {arXiv:astro-ph/0504538}
  \BibitemShut {NoStop}%
\bibitem [{\citenamefont {Boh\'e}\ \emph {et~al.}(2015)\citenamefont {Boh\'e},
  \citenamefont {Faye}, \citenamefont {Marsat},\ and\ \citenamefont
  {Porter}}]{Bohe:2015ana}%
  \BibitemOpen
  \bibfield  {author} {\bibinfo {author} {\bibfnamefont {A.}~\bibnamefont
  {Boh\'e}}, \bibinfo {author} {\bibfnamefont {G.}~\bibnamefont {Faye}},
  \bibinfo {author} {\bibfnamefont {S.}~\bibnamefont {Marsat}},\ and\ \bibinfo
  {author} {\bibfnamefont {E.~K.}\ \bibnamefont {Porter}},\ }\bibfield  {title}
  {\bibinfo {title} {{Quadratic-in-spin effects in the orbital dynamics and
  gravitational-wave energy flux of compact binaries at the 3PN order}},\
  }\href {https://doi.org/10.1088/0264-9381/32/19/195010} {\bibfield  {journal}
  {\bibinfo  {journal} {Class. Quant. Grav.}\ }\textbf {\bibinfo {volume}
  {32}},\ \bibinfo {pages} {195010} (\bibinfo {year} {2015})},\ \Eprint
  {https://arxiv.org/abs/1501.01529} {arXiv:1501.01529 [gr-qc]} \BibitemShut
  {NoStop}%
\bibitem [{\citenamefont {Mishra}\ \emph {et~al.}(2016)\citenamefont {Mishra},
  \citenamefont {Kela}, \citenamefont {Arun},\ and\ \citenamefont
  {Faye}}]{Mishra:2016whh}%
  \BibitemOpen
  \bibfield  {author} {\bibinfo {author} {\bibfnamefont {C.~K.}\ \bibnamefont
  {Mishra}}, \bibinfo {author} {\bibfnamefont {A.}~\bibnamefont {Kela}},
  \bibinfo {author} {\bibfnamefont {K.~G.}\ \bibnamefont {Arun}},\ and\
  \bibinfo {author} {\bibfnamefont {G.}~\bibnamefont {Faye}},\ }\bibfield
  {title} {\bibinfo {title} {{Ready-to-use post-Newtonian gravitational
  waveforms for binary black holes with nonprecessing spins: An update}},\
  }\href {https://doi.org/10.1103/PhysRevD.93.084054} {\bibfield  {journal}
  {\bibinfo  {journal} {Phys. Rev. D}\ }\textbf {\bibinfo {volume} {93}},\
  \bibinfo {pages} {084054} (\bibinfo {year} {2016})},\ \Eprint
  {https://arxiv.org/abs/1601.05588} {arXiv:1601.05588 [gr-qc]} \BibitemShut
  {NoStop}%
\bibitem [{\citenamefont {Poisson}(1998)}]{Poisson:1997ha}%
  \BibitemOpen
  \bibfield  {author} {\bibinfo {author} {\bibfnamefont {E.}~\bibnamefont
  {Poisson}},\ }\bibfield  {title} {\bibinfo {title} {{Gravitational waves from
  inspiraling compact binaries: The Quadrupole moment term}},\ }\href
  {https://doi.org/10.1103/PhysRevD.57.5287} {\bibfield  {journal} {\bibinfo
  {journal} {Phys. Rev. D}\ }\textbf {\bibinfo {volume} {57}},\ \bibinfo
  {pages} {5287} (\bibinfo {year} {1998})},\ \Eprint
  {https://arxiv.org/abs/gr-qc/9709032} {arXiv:gr-qc/9709032} \BibitemShut
  {NoStop}%
\bibitem [{\citenamefont {Wade}\ \emph {et~al.}(2014)\citenamefont {Wade},
  \citenamefont {Creighton}, \citenamefont {Ochsner}, \citenamefont {Lackey},
  \citenamefont {Farr}, \citenamefont {Littenberg},\ and\ \citenamefont
  {Raymond}}]{Wade:2014vqa}%
  \BibitemOpen
  \bibfield  {author} {\bibinfo {author} {\bibfnamefont {L.}~\bibnamefont
  {Wade}}, \bibinfo {author} {\bibfnamefont {J.~D.~E.}\ \bibnamefont
  {Creighton}}, \bibinfo {author} {\bibfnamefont {E.}~\bibnamefont {Ochsner}},
  \bibinfo {author} {\bibfnamefont {B.~D.}\ \bibnamefont {Lackey}}, \bibinfo
  {author} {\bibfnamefont {B.~F.}\ \bibnamefont {Farr}}, \bibinfo {author}
  {\bibfnamefont {T.~B.}\ \bibnamefont {Littenberg}},\ and\ \bibinfo {author}
  {\bibfnamefont {V.}~\bibnamefont {Raymond}},\ }\bibfield  {title} {\bibinfo
  {title} {{Systematic and statistical errors in a bayesian approach to the
  estimation of the neutron-star equation of state using advanced gravitational
  wave detectors}},\ }\href {https://doi.org/10.1103/PhysRevD.89.103012}
  {\bibfield  {journal} {\bibinfo  {journal} {Phys. Rev. D}\ }\textbf {\bibinfo
  {volume} {89}},\ \bibinfo {pages} {103012} (\bibinfo {year} {2014})},\
  \Eprint {https://arxiv.org/abs/1402.5156} {arXiv:1402.5156 [gr-qc]}
  \BibitemShut {NoStop}%
\bibitem [{\citenamefont {Ashton}\ \emph {et~al.}(2019)\citenamefont {Ashton}
  \emph {et~al.}}]{Ashton:2018jfp}%
  \BibitemOpen
  \bibfield  {author} {\bibinfo {author} {\bibfnamefont {G.}~\bibnamefont
  {Ashton}} \emph {et~al.},\ }\bibfield  {title} {\bibinfo {title} {{BILBY: A
  user-friendly Bayesian inference library for gravitational-wave astronomy}},\
  }\href {https://doi.org/10.3847/1538-4365/ab06fc} {\bibfield  {journal}
  {\bibinfo  {journal} {Astrophys. J. Suppl.}\ }\textbf {\bibinfo {volume}
  {241}},\ \bibinfo {pages} {27} (\bibinfo {year} {2019})},\ \Eprint
  {https://arxiv.org/abs/1811.02042} {arXiv:1811.02042 [astro-ph.IM]}
  \BibitemShut {NoStop}%
\bibitem [{\citenamefont {Romero-Shaw}\ \emph {et~al.}(2020)\citenamefont
  {Romero-Shaw} \emph {et~al.}}]{Romero-Shaw:2020owr}%
  \BibitemOpen
  \bibfield  {author} {\bibinfo {author} {\bibfnamefont {I.~M.}\ \bibnamefont
  {Romero-Shaw}} \emph {et~al.},\ }\bibfield  {title} {\bibinfo {title}
  {{Bayesian inference for compact binary coalescences with bilby: validation
  and application to the first LIGO\textendash{}Virgo gravitational-wave
  transient catalogue}},\ }\href {https://doi.org/10.1093/mnras/staa2850}
  {\bibfield  {journal} {\bibinfo  {journal} {Mon. Not. Roy. Astron. Soc.}\
  }\textbf {\bibinfo {volume} {499}},\ \bibinfo {pages} {3295} (\bibinfo {year}
  {2020})},\ \Eprint {https://arxiv.org/abs/2006.00714} {arXiv:2006.00714
  [astro-ph.IM]} \BibitemShut {NoStop}%
\bibitem [{\citenamefont {Ashton}\ and\ \citenamefont
  {Talbot}(2021)}]{Ashton:2021anp}%
  \BibitemOpen
  \bibfield  {author} {\bibinfo {author} {\bibfnamefont {G.}~\bibnamefont
  {Ashton}}\ and\ \bibinfo {author} {\bibfnamefont {C.}~\bibnamefont
  {Talbot}},\ }\bibfield  {title} {\bibinfo {title} {{B\,ilby-MCMC: an MCMC
  sampler for gravitational-wave inference}},\ }\href
  {https://doi.org/10.1093/mnras/stab2236} {\bibfield  {journal} {\bibinfo
  {journal} {Mon. Not. Roy. Astron. Soc.}\ }\textbf {\bibinfo {volume} {507}},\
  \bibinfo {pages} {2037} (\bibinfo {year} {2021})},\ \Eprint
  {https://arxiv.org/abs/2106.08730} {arXiv:2106.08730 [gr-qc]} \BibitemShut
  {NoStop}%
\bibitem [{\citenamefont {Zackay}\ \emph {et~al.}(2018)\citenamefont {Zackay},
  \citenamefont {Dai},\ and\ \citenamefont {Venumadhav}}]{Zackay:2018qdy}%
  \BibitemOpen
  \bibfield  {author} {\bibinfo {author} {\bibfnamefont {B.}~\bibnamefont
  {Zackay}}, \bibinfo {author} {\bibfnamefont {L.}~\bibnamefont {Dai}},\ and\
  \bibinfo {author} {\bibfnamefont {T.}~\bibnamefont {Venumadhav}},\ }\bibfield
   {title} {\bibinfo {title} {{Relative Binning and Fast Likelihood Evaluation
  for Gravitational Wave Parameter Estimation}},\ }\href@noop {} {\  (\bibinfo
  {year} {2018})},\ \Eprint {https://arxiv.org/abs/1806.08792}
  {arXiv:1806.08792 [astro-ph.IM]} \BibitemShut {NoStop}%
\bibitem [{\citenamefont {Chatziioannou}\ \emph {et~al.}(2018)\citenamefont
  {Chatziioannou}, \citenamefont {Haster},\ and\ \citenamefont
  {Zimmerman}}]{Chatziioannou:2018vzf}%
  \BibitemOpen
  \bibfield  {author} {\bibinfo {author} {\bibfnamefont {K.}~\bibnamefont
  {Chatziioannou}}, \bibinfo {author} {\bibfnamefont {C.-J.}\ \bibnamefont
  {Haster}},\ and\ \bibinfo {author} {\bibfnamefont {A.}~\bibnamefont
  {Zimmerman}},\ }\bibfield  {title} {\bibinfo {title} {{Measuring the neutron
  star tidal deformability with equation-of-state-independent relations and
  gravitational waves}},\ }\href {https://doi.org/10.1103/PhysRevD.97.104036}
  {\bibfield  {journal} {\bibinfo  {journal} {Phys. Rev. D}\ }\textbf {\bibinfo
  {volume} {97}},\ \bibinfo {pages} {104036} (\bibinfo {year} {2018})},\
  \Eprint {https://arxiv.org/abs/1804.03221} {arXiv:1804.03221 [gr-qc]}
  \BibitemShut {NoStop}%
\bibitem [{\citenamefont {Kastaun}\ and\ \citenamefont
  {Ohme}(2019)}]{Kastaun:2019bxo}%
  \BibitemOpen
  \bibfield  {author} {\bibinfo {author} {\bibfnamefont {W.}~\bibnamefont
  {Kastaun}}\ and\ \bibinfo {author} {\bibfnamefont {F.}~\bibnamefont {Ohme}},\
  }\bibfield  {title} {\bibinfo {title} {{Finite tidal effects in GW170817:
  Observational evidence or model assumptions?}},\ }\href
  {https://doi.org/10.1103/PhysRevD.100.103023} {\bibfield  {journal} {\bibinfo
   {journal} {Phys. Rev. D}\ }\textbf {\bibinfo {volume} {100}},\ \bibinfo
  {pages} {103023} (\bibinfo {year} {2019})},\ \Eprint
  {https://arxiv.org/abs/1909.12718} {arXiv:1909.12718 [gr-qc]} \BibitemShut
  {NoStop}%
\bibitem [{\citenamefont {Gonzalez}\ \emph {et~al.}(2023)\citenamefont
  {Gonzalez} \emph {et~al.}}]{Gonzalez:2022mgo}%
  \BibitemOpen
  \bibfield  {author} {\bibinfo {author} {\bibfnamefont {A.}~\bibnamefont
  {Gonzalez}} \emph {et~al.},\ }\bibfield  {title} {\bibinfo {title} {{Second
  release of the CoRe database of binary neutron star merger waveforms}},\
  }\href {https://doi.org/10.1088/1361-6382/acc231} {\bibfield  {journal}
  {\bibinfo  {journal} {Class. Quant. Grav.}\ }\textbf {\bibinfo {volume}
  {40}},\ \bibinfo {pages} {085011} (\bibinfo {year} {2023})},\ \Eprint
  {https://arxiv.org/abs/2210.16366} {arXiv:2210.16366 [gr-qc]} \BibitemShut
  {NoStop}%
\bibitem [{\citenamefont {Bauswein}\ \emph {et~al.}(2020)\citenamefont
  {Bauswein}, \citenamefont {Blacker}, \citenamefont {Vijayan}, \citenamefont
  {Stergioulas}, \citenamefont {Chatziioannou}, \citenamefont {Clark},
  \citenamefont {Bastian}, \citenamefont {Blaschke}, \citenamefont {Cierniak},\
  and\ \citenamefont {Fischer}}]{Bauswein:2020aag}%
  \BibitemOpen
  \bibfield  {author} {\bibinfo {author} {\bibfnamefont {A.}~\bibnamefont
  {Bauswein}}, \bibinfo {author} {\bibfnamefont {S.}~\bibnamefont {Blacker}},
  \bibinfo {author} {\bibfnamefont {V.}~\bibnamefont {Vijayan}}, \bibinfo
  {author} {\bibfnamefont {N.}~\bibnamefont {Stergioulas}}, \bibinfo {author}
  {\bibfnamefont {K.}~\bibnamefont {Chatziioannou}}, \bibinfo {author}
  {\bibfnamefont {J.~A.}\ \bibnamefont {Clark}}, \bibinfo {author}
  {\bibfnamefont {N.-U.~F.}\ \bibnamefont {Bastian}}, \bibinfo {author}
  {\bibfnamefont {D.~B.}\ \bibnamefont {Blaschke}}, \bibinfo {author}
  {\bibfnamefont {M.}~\bibnamefont {Cierniak}},\ and\ \bibinfo {author}
  {\bibfnamefont {T.}~\bibnamefont {Fischer}},\ }\bibfield  {title} {\bibinfo
  {title} {{Equation of state constraints from the threshold binary mass for
  prompt collapse of neutron star mergers}},\ }\href
  {https://doi.org/10.1103/PhysRevLett.125.141103} {\bibfield  {journal}
  {\bibinfo  {journal} {Phys. Rev. Lett.}\ }\textbf {\bibinfo {volume} {125}},\
  \bibinfo {pages} {141103} (\bibinfo {year} {2020})},\ \Eprint
  {https://arxiv.org/abs/2004.00846} {arXiv:2004.00846 [astro-ph.HE]}
  \BibitemShut {NoStop}%
\bibitem [{\citenamefont {Espino}\ \emph
  {et~al.}(2023{\natexlab{b}})\citenamefont {Espino}, \citenamefont {Bozzola},\
  and\ \citenamefont {Paschalidis}}]{Espino:2022mtb}%
  \BibitemOpen
  \bibfield  {author} {\bibinfo {author} {\bibfnamefont {P.~L.}\ \bibnamefont
  {Espino}}, \bibinfo {author} {\bibfnamefont {G.}~\bibnamefont {Bozzola}},\
  and\ \bibinfo {author} {\bibfnamefont {V.}~\bibnamefont {Paschalidis}},\
  }\bibfield  {title} {\bibinfo {title} {{Quantifying uncertainties in general
  relativistic magnetohydrodynamic codes}},\ }\href
  {https://doi.org/10.1103/PhysRevD.107.104059} {\bibfield  {journal} {\bibinfo
   {journal} {Phys. Rev. D}\ }\textbf {\bibinfo {volume} {107}},\ \bibinfo
  {pages} {104059} (\bibinfo {year} {2023}{\natexlab{b}})},\ \Eprint
  {https://arxiv.org/abs/2210.13481} {arXiv:2210.13481 [gr-qc]} \BibitemShut
  {NoStop}%
\end{thebibliography}%
\acrodef{ADM}{Arnowitt-Deser-Misner}
\acrodef{AMR}{adaptive mesh-refinement}
\acrodef{BH}{black hole}
\acrodef{BBH}{binary black-hole}
\acrodef{BHNS}{black-hole neutron-star}
\acrodef{BNS}{binary neutron star}
\acrodef{CCSN}{core-collapse supernova}
\acrodefplural{CCSN}[CCSNe]{core-collapse supernovae}
\acrodef{CMA}{consistent multi-fluid advection}
\acrodef{CFL}{Courant-Friedrichs-Lewy}
\acrodef{DG}{discontinuous Galerkin}
\acrodef{HMNS}{hypermassive neutron star}
\acrodef{EM}{electromagnetic}
\acrodef{ET}{Einstein Telescope}
\acrodef{EOB}{effective-one-body}
\acrodef{EOS}{equation of state}
\acrodef{QUR}{quasi-universal relation}
\acrodef{FF}{fitting factor}
\acrodef{GR}{general-relativistic}
\acrodef{GRLES}{general-relativistic large-eddy simulation}
\acrodef{GRHD}{general-relativistic hydrodynamics}
\acrodef{GRMHD}{general-relativistic magnetohydrodynamics}
\acrodef{GW}{gravitational wave}
\acrodef{ILES}{implicit large-eddy simulations}
\acrodef{LIA}{linear interaction analysis}
\acrodef{LES}{large-eddy simulation}
\acrodefplural{LES}[LES]{large-eddy simulations}
\acrodef{MRI}{magnetorotational instability}
\acrodef{NR}{numerical relativity}
\acrodef{NS}{neutron star}
\acrodef{PNS}{protoneutron star}
\acrodef{RMNS}{remnant massive neutron star}
\acrodef{SASI}{standing accretion shock instability}
\acrodef{SGRB}{short $\gamma$-ray burst}
\acrodef{SPH}{smoothed particle hydrodynamics}
\acrodef{SN}{supernova}
\acrodefplural{SN}[SNe]{supernovae}
\acrodef{SNR}{signal-to-noise ratio}
\acrodef{ZAMS}{zero age main sequence}
\acrodef{PE}{parameter estimation}
\acrodef{PDF}{probability distribution function}
\acrodef{CIs}{Credible Intervals}
\acrodef{ET}{Einstein Telescope}
\acrodef{CE}{Cosmic Explorer}
\acrodef{MHD}{Magnetohydrodynamic}
\end{document}